\documentclass[11pt,a4paper]{article}
\usepackage{amsmath,amsfonts,amssymb,amsthm,amstext,amscd,array,bbold}
\DeclareMathOperator{\AdS}{AdS}

\DeclareMathOperator{\Sphere}{S}
\DeclareMathOperator{\Torus}{T}
\DeclareMathOperator{\NW}{NW}
\DeclareMathOperator{\CW}{CW}
\DeclareMathOperator{\Hyp}{H}
\let\S\Sphere

\newcommand{\NO}{\mbox{$\substack{\circ\\\circ}$}}      

\newcommand{\ee}[1]{{\rm e}^{#1}}
\newcommand{\ii}{{\rm i}}

\newcommand{\C}{\complex}

\newcommand{\mbf}[1]{{\boldsymbol {#1} }}
\def\ii{{\,{\rm i}\,}}
\def\dd{{\rm d}}

\def\P{{\sf P}}

\def\T{{\sf T}}

\def\Q{{\sf Q}}
\def\C{{\sf C}}
\def\J{{\sf J}}

\def\Pu{\underline{\sf P}{}}

\def\mz{{\mbf z}}
\def\mw{{\mbf w}}

\def\mm{{\mbf m}}
\def\mell{{\mbf\ell}}
\def\mdell{{\mbf\partial}}

\def\mcDp{{\mathcal D}^{p^+,p^-}}

\newcommand{\eq}{\begin{equation}}
\newcommand{\eqend}{\end{equation}}
\newcommand{\eqa}{\begin{eqnarray}}
\newcommand{\nonueqa}{\begin{eqnarray*}}
\newcommand{\eqaend}{\end{eqnarray}}
\newcommand{\nonueqaend}{\end{eqnarray*}}

\newcommand{\bma}[1]{\begin{array}{#1}}
\newcommand{\ema}{\end{array}}
\newcommand{\bc}{\begin{center}}
\newcommand{\ec}{\end{center}}

\newcommand{\R}{\real}

\renewcommand{\thefootnote}{\fnsymbol{footnote}}
\newcommand{\newsection}{\setcounter{equation}{0}\section}

\def\appendix#1{\addtocounter{section}{1}\setcounter{equation}{0}
\renewcommand{\thesection}{\Alph{section}}
\section*{Appendix \thesection\protect\indent \parbox[t]{11.715cm} {#1}}
\addcontentsline{toc}{section}{Appendix \thesection\ \ \ #1} }

\newcommand{\complex}{{\mathbb C}} 
\newcommand{\zed}{{\mathbb Z}} 
\newcommand{\nat}{{\mathbb N}} 
\newcommand{\real}{{\mathbb R}} 
\newcommand{\eucl}{{\mathbb E}}
\newcommand{\id}{{1\!\!1}} 

\newif\ifold             \oldtrue

\def\nn{\nonumber}

\newcommand{\Tr}[1]{\:{\rm Tr}\,#1}

\def\e{{\,\rm e}\,}

\def\be{\begin{equation}}
\def\ee{\end{equation}}
\def\bea{\begin{eqnarray}}
\def\eea{\end{eqnarray}}
\def\bd{\begin{displaymath}}
\def\ed{\end{displaymath}}

\newcommand{\beq}{\begin{eqnarray}}
\newcommand{\eeq}{\end{eqnarray}}

\makeatletter
\newdimen\normalarrayskip              
\newdimen\minarrayskip                 
\normalarrayskip\baselineskip
\minarrayskip\jot
\newif\ifold             \oldtrue            
\def\arraymode{\ifold\relax\else\displaystyle\fi} 
\def\@arrayskip{\ifold\baselineskip\z@\lineskip\z@
     \else
     \baselineskip\minarrayskip\lineskip2\minarrayskip\fi}
\def\@arrayclassz{\ifcase \@lastchclass \@acolampacol \or
\@ampacol \or \or \or \@addamp \or
   \@acolampacol \or \@firstampfalse \@acol \fi
\edef\@preamble{\@preamble
  \ifcase \@chnum
     \hfil$\relax\arraymode\@sharp$\hfil
     \or $\relax\arraymode\@sharp$\hfil
     \or \hfil$\relax\arraymode\@sharp$\fi}}
\def\@array[#1]#2{\setbox\@arstrutbox=\hbox{\vrule
     height\arraystretch \ht\strutbox
     depth\arraystretch \dp\strutbox
     width\z@}\@mkpream{#2}\edef\@preamble{\halign \noexpand\@halignto
\bgroup \tabskip\z@ \@arstrut \@preamble \tabskip\z@ \cr}%
\let\@startpbox\@@startpbox \let\@endpbox\@@endpbox
  \if #1t\vtop \else \if#1b\vbox \else \vcenter \fi\fi
  \bgroup \let\par\relax
  \let\@sharp##\let\protect\relax
  \@arrayskip\@preamble}
\makeatother

\begin{document}
\begin{titlepage}
\begin{flushright}

\baselineskip=12pt

HWM--05--01\\
EMPG--05--01\\
hep--th/0502054\\
\hfill{ }\\
February 2005
\end{flushright}

\begin{center}

\vspace{2cm}

\baselineskip=24pt

{\Large\bf Isometric Embeddings and Noncommutative Branes\\ in
  Homogeneous Gravitational Waves}

\baselineskip=14pt

\vspace{1cm}

{\bf Sam Halliday} and {\bf Richard J. Szabo}
\\[4mm]
{\it Department of Mathematics\\ School of Mathematical and Computer
  Sciences\\ Heriot-Watt University\\ Scott Russell Building,
  Riccarton, Edinburgh EH14 4AS, U.K.}
\\{\tt samuel@ma.hw.ac.uk} , {\tt R.J.Szabo@ma.hw.ac.uk}
\\[40mm]

\end{center}

\begin{abstract}

\baselineskip=12pt

We characterize the worldvolume theories on symmetric D-branes in a
six-dimensional Cahen-Wallach pp-wave supported by
a constant Neveu-Schwarz three-form flux. We find a class of flat
noncommutative euclidean D3-branes analogous to branes in a constant
magnetic field, as well as curved noncommutative lorentzian D3-branes
analogous to branes in an electric background. In the former case the
noncommutative field theory on the branes is constructed from first
principles, related to dynamics of fuzzy spheres in the worldvolumes,
and used to analyse the flat space limits of the string theory. The
worldvolume theories on all other symmetric branes in the background
are local field theories. The physical origins of all these theories are
described through the interplay between isometric embeddings of branes
in the spacetime and the Penrose-G\"uven limit of $\AdS_3\times\S^3$
with Neveu-Schwarz three-form flux. The noncommutative field theory of
a non-symmetric spacetime-filling D-brane is also constructed, giving
a spatially varying but time-independent noncommutativity analogous to
that of the Dolan-Nappi model.

\end{abstract}

\end{titlepage}
\setcounter{page}{2}

\newpage

\tableofcontents

\newpage

\renewcommand{\thefootnote}{\arabic{footnote}} \setcounter{footnote}{0}

\newsection{Introduction and Summary\label{Intro}}

The dynamics of strings in the backgrounds of plane-polarized
gravitational waves (pp-waves) has been of interest recently for a
variety of reasons. They provide explicit realizations of string
theory in time-dependent backgrounds which is necessary for
applications of string cosmology. They also provide scenarios in
which the AdS/CFT correspondence may be tested beyond the supergravity
approximation by taking the Penrose limit of an $\AdS_m\times\S^n$
background~\cite{BFP1} and the BMN limit of the dual superconformal
field theory~\cite{BMN1}. The property of these backgrounds that make them
appealing in these contexts is that string dynamics on them is
solvable in some instances, even in the presence of R--R fields or
non-trivial NS $B$-fields~\cite{BMN1,BOLPT1,Met1,PRT1,RT1}. The
spectrum of the theory can be studied in light-cone gauge wherein the
two-dimensional $\sigma$-models become free, while scattering
amplitudes can be analysed using light-cone string field theory.

When D-branes are added to such closed string backgrounds, in some cases
decoupling limits exist in which one can freeze out massive open
string modes and closed string excitations. The low-energy effective
theory governing the dynamics of open strings living on the branes is
non-gravitational and can be reformulated as a field theory. The
typical result is a noncommutative gauge theory with a spacetime dependent
noncommutativity parameter~\cite{CLO1,DRRS1,DN1,HS1,HT1,LNR1}. The role of
spacetime dependence in these worldvolume field theories leads to
interesting violations of energy-momentum conservation~\cite{BG1,RS1},
and their potential time-dependence is especially important for
cosmological applications. In some instances the decoupled open
strings also have a dual description in terms of a gravitational
theory via the AdS/CFT correspondence~\cite{HS1}. This suggests
that the holographic description of cosmological spacetimes may be
described by non-local field theories.

In this paper we will study the noncommutative gauge theories that
reside on some D-branes in the four-dimensional Nappi-Witten gravitational
wave~\cite{NW1} and its six-dimensional generalization~\cite{KM1}. We
will refer to both of these pp-waves in the following as Nappi-Witten
spacetimes and denote them respectively by $\NW_4$ and $\NW_6$. In the
full superstring setting the backgrounds we study are $\NW_4\times\Torus^6$ and
$\NW_6\times\Torus^4$, although we shall not write the toroidal factors
explicitly in what follows. These backgrounds can be
supported by either R--R flux or NS--NS flux, or by a combination of
both. We will only consider the noncommutative deformations of the
bosonic parts of these string theories, and hence only a non-trivial
NS background, as R--R fields simply add mass terms to the light-cone
$\sigma$-model action and do not contribute to the noncommutativity of
spacetime coordinates in the bosonic sector~\cite{CH1,Seiberg1}.

The interest in this particular class of pp-waves is that string
theory in these backgrounds can be solved completely and in a fully
covariant way~\cite{BAKZ1,CFS1,DAK1,FHHP1,KK1,KKL1,RT2}. They describe
homogeneous gravitational waves (Hpp-waves) and represent the
``minimal'' deformation of flat spacetime by $H$-flux. They may be
formulated as WZW models based on a twisted Heisenberg group, for
which the wave is an exact solution of the worldsheet
$\sigma$-model~\cite{KM1,NW1}. The $\NW_6$ spacetime already captures
the generic features of higher-dimensional Hpp-waves. It can be
regarded as the Penrose-G\"uven limit of the background
$\AdS_3\times\S^3\times\Torus^4$~\cite{BFP1,BFHP1} supported by an
NS--NS three-form flux, which describes the near horizon geometry of
an NS5/F1 bound state~\cite{GKS1}. The dual superconformal field
theory is believed to be the nonlinear $\sigma$-model with target
space the symmetric product orbifold ${\rm
  Sym}^N(\Torus^4)$. Similarly, the plane wave metric of $\NW_4$
arises from the Penrose limit of
$\AdS_2\times\S^2$~\cite{BFP1,BFHP1}. Both $\NW_4$ and the Penrose
limit of $\AdS_3\times\S^3$ are examples of non-dilatonic,
parallelizable pp-wave solutions of six-dimensional
supergravity~\cite{SS-J1}. However, as we discuss in detail
in this paper, the Penrose-G\"uven limit of $\AdS_2\times\S^2$ does
{\it not} induce the full NS-supported geometry of the $\NW_4$
spacetime. Therefore, contrary to some claims~\cite{DeKa1,SF1},
the four-dimensional Nappi-Witten spacetime cannot be studied as the
Penrose-G\"uven limit of $\AdS_2\times\S^2$. Instead, it arises as a
Penrose-G\"uven limit of the near horizon geometry of NS5-branes~\cite{GO1}, on
which string theory is dual to little string theory. This feature can be
understood by regarding $\AdS_2\times\S^2$ as the worldvolume of a
symmetric D-brane in the $\AdS_3\times\S^3$ spacetime, while $\NW_4$
may only be realized as the worldvolume of a non-symmetric D-brane in
$\NW_6$. In fact, we shall find that the most natural plane wave
limits of embedded $\AdS_2\times\S^2$ submanifolds of
$\AdS_3\times\S^3$ correspond to two classes of symmetric D-branes in
$\NW_6$. The first one is a {\it flat} euclidean D3-brane in a
constant magnetic field, which carries a noncommutative
worldvolume field theory with constant noncommutativity parameter
determined by the constant time slices of the plane wave
background. The second one is a lorentzian D3-brane isometric to
$\NW_4$ with vanishing NS fields but with a null worldvolume electric
field, which is described in the decoupling limit by a
non-gravitational theory of noncommutative open strings, rather than
by a noncommutative field theory. It is tempting to speculate that the
full $\NW_4$ deformation of this noncommutative open string theory
describes the dynamics of the dual little string theory.

This problem is not peculiar to the class of plane wave geometries
that we study, and it leads us into a detailed investigation of how
D-branes behave under the Penrose-G\"uven limit of a
spacetime. Similar analyses in some specific contexts are considered
in~\cite{DK2,SZ1,SF1}. We formulate and solve this problem in some
generality, and then apply it to our specific backgrounds of
interest. With this motivation at hand, we then proceed to reanalyse
the classification of the symmetric D-branes of $\NW_6$, elaborating
on the analysis initiated in~\cite{FS1,SF1} and extending it to a
detailed study of the worldvolume supergravity fields supported by
each of these branes. We also clarify some points which were missed
in the analysis of~\cite{SF1}. In each instance we identify the
$\AdS_3\times\S^3$ origin of the brane in question, and quantize its
worldvolume geometry using standard techniques and the representation
theory of the twisted Heisenberg
group~\cite{BAKZ1,CFS1,KK1,Streater1}. We will find that most of
these branes support {\it local} worldvolume effective field theories,
because on most of them the pertinent supergravity form fields
are trivial. In fact, we find that all symmetric D-branes in $\NW_6$
(both twisted and untwisted) have vanishing NS--NS three-form flux,
and only the two classes of branes mentioned above support a
non-vanishing gauge-invariant two-form field. The overall consistency
of these results, along with their agreement with the exact boundary
conformal field theory description of Cardy branes in
$\NW_4$~\cite{DK2}, provides an important check that the standard
techniques for quantization of worldvolume geometries in compact group
manifolds (see~\cite{Schom1} for a review) extend to these classes of
non-compact (and non-semisimple) Lie groups.

Somewhat surprisingly, even the spacetime filling symmetric D5-brane
in $\NW_6$ has trivial supergravity form fields. Motivated by this fact, we
systematically construct the noncommutative geometry underlying
the non-local field theory living on a non-symmetric D5-brane wrapping
$\NW_6$. The resulting noncommutativity is non-constant, but
independent of the plane wave time coordinate. This agrees with the
recent analysis in~\cite{HT1} of the Dolan-Nappi model~\cite{DN1}
which describes a time-dependent noncommutative geometry on the
worldvolume of a D3-brane wrapping $\NW_4$. However, the background
used in~\cite{DN1} is not conformally invariant and hence not a closed string
background. Correctly reinstating conformal invariance~\cite{HT1}
gives a spatially dependent but time-independent noncommutativity
parameter. We elaborate on this noncommutativity somewhat and show
that it may be regarded as arising from a formal quantization of the
twisted Heisenberg algebra.

The organization of this paper is as follows. In Section~\ref{NWPW}
we review the definition and geometrical properties of the
four-dimensional Nappi-Witten spacetime. We show that its natural
pp-wave isometry group is isomorphic to the six-dimensional twisted
Heisenberg group, paving the way to an analysis of the isometric
embeddings $\NW_4\hookrightarrow\NW_6$. We also emphasize the
time-independent harmonic oscillator character of point-particle
dynamics in these backgrounds, as it helps to clarify the nature of
the noncommutative worldvolume field theories constructed later on. In
Section~\ref{IsomEmb} we study the interplay between isometric
embeddings and Penrose-G\"uven limits of branes, first in generality
and then to the particular instances of Nappi-Witten spacetimes. From
this analysis it becomes clear that both the $\NW_4$ and $\NW_6$
gravitational waves are necessarily wrapped by non-symmetric D-branes.
In Section~\ref{NCBranes} we begin our analysis of the symmetric
branes in $\NW_6$, beginning with those described by conjugacy classes
of the twisted Heisenberg group. We identify classes of null branes
(with degenerate worldvolume metrics), and show that their quantized
geometries are commutative but generically differ from those of the
classical conjugacy classes due to a unitary rotational
symmetry of the background. We also find a class of euclidean
D3-branes and show, directly from the representation theory of the
twisted Heisenberg group, that their worldvolumes carry a Moyal-type
noncommutativity akin to that induced on branes in constant magnetic
fields~\cite{DNek1,SW1,Sz1,Sz2}. This sort of noncommutativity is
natural from the point of view of the time-independent harmonic
oscillator dynamics. These D3-branes are also naturally related to a
foliation by fuzzy two-spheres through a noncommutative version of the
Hopf fibration $\S^3\to\S^2$. This provides a geometrical worldvolume
picture of the accidental ${\rm SU}(2)$ transverse space symmetry of
our $\NW_6$ background~\cite{BAKZ1}, acting by outer automorphisms of
the twisted Heisenberg group. It also implies that there exists a flat
space limit of the string theory in which the low-energy dynamics will
still be described by a noncommutative gauge theory with constant
noncommutativity. In Section~\ref{TwistedNCBranes} we analyse
symmetric branes in $\NW_6$ which are described by twisted conjugacy
classes. We show, again through explicit quantization via
representation theory and analysis of the worldvolume supergravity
fields, that the low-energy effective field theories on {\it all}
twisted D-branes are local. Finally, in Section~\ref{TDNC} we quantize
a non-symmetric D5-brane wrapping $\NW_6$ by foliating its geometry in
terms of the noncommutative euclidean D3-branes, and compare the
resulting noncommutativity with that of the Dolan-Nappi model. Again
the result is natural from the perspective of time-independent
harmonic oscillator dynamics and the quantization of the coadjoint
orbits.

\newsection{The Nappi-Witten Plane Wave \label{NWPW}}

In this section we will define and analyse the geometry of the
Nappi-Witten spacetime NW$_4$~\cite{NW1}. It is a four-dimensional
homogeneous spacetime of Minkowski signature which defines
a monochromatic plane wave. It is further equipped with a supergravity
NS $B$-field of constant flux, which in the presence of D-branes is
responsible for the spacetime noncommutativity of the
pp-wave. We will emphasize the simple, time-independent harmonic
oscillator form of the dynamics in this background, as it will play a
crucial role in subsequent sections.

\subsection{Definitions \label{Defs}}

The spacetime NW$_4$ is defined as the group manifold of
the Nappi-Witten group, the universal central extension of the
two-dimensional euclidean group ${\rm ISO}(2)={\rm
  SO}(2)\ltimes\real^2$. The corresponding simply connected group
$\mathcal N_4$ is homeomorphic to four-dimensional Minkowski space
$\eucl^{1,3}$. Its non-semisimple Lie algebra $\mathfrak n_4$ is
generated by elements $\P^\pm$, $\J$, $\T$ obeying the commutation
relations
\bea
\left[\P^+\,,\,\P^-\right]&=&2\ii\T \ , \nn\\
\left[\J\,,\,\P^\pm\right]&=&\pm\ii\P^\pm \ , \nn\\
\left[\T\,,\,\J\right]&=&\left[\T\,,\,\P^\pm\right]~=~0 \ .
\label{NW4algdef}\eea
This is just the three-dimensional Heisenberg algebra extended by an
outer automorphism which rotates the noncommuting coordinates. The
twisted Heisenberg algebra
may be regarded as defining the harmonic oscillator algebra of a
particle moving in one-dimension, with the additional generator $\J$
playing the role of the number operator (or equivalently the
oscillator hamiltonian). It is a solvable algebra whose properties are
much more tractable than, for instance, those of the
${\rm su}(2)$ or ${\rm sl}(2,\real)$ Lie algebras which are
at the opposite extreme.

The center of the universal enveloping algebra $U(\mathfrak n_4)$
contains the central element $\T$ of the Lie algebra $\mathfrak{n}_4$
and also the quadratic Casimir element
\beq
\C_4=2\,\J\,\T+\mbox{$\frac12$}\,\left(\P^+\,\P^-+\P^-\,\P^+\right) \ .
\label{NW4Casimir}\eeq
The most general invariant, non-degenerate symmetric bilinear form
$\langle\,\cdot\,,\,\cdot\,\rangle:\mathfrak{n}_4\times\mathfrak{n}_4\to\real$
is defined by~\cite{NW1}
\bea
\left\langle\P^+\,,\,\P^-\right\rangle&=&2\,
\left\langle\J\,,\,\T\right\rangle~=~2 \ , \nn\\
\left\langle\J\,,\,\J\right\rangle&=&b \ , \nn\\
\left\langle\P^\pm\,,\,\P^\pm\right\rangle&=&
\left\langle\T\,,\,\T\right\rangle~=~0 \ , \nn\\
\left\langle\J\,,\,\P^\pm\right\rangle&=&\left\langle\T\,,\,
\P^\pm\right\rangle~=~0
\label{NW4innerprod}\eea
for any $b\in\real$. This inner product has Minkowski signature, so that the
group manifold of $\mathcal N_4$ possesses a
homogeneous, bi-invariant lorentzian metric defined by the pairing of
the Cartan-Maurer left-invariant, $\mathfrak n_4$-valued one-forms
$g^{-1}~\dd g$ for $g\in\mathcal N_4$ as
\beq
\dd s_4^2=\left\langle g^{-1}~\dd g\,,\,g^{-1}~\dd g\right\rangle \ .
\label{NW4CM}\eeq
A generic group element $g\in\mathcal N_4$ may be parametrized as
\beq
g(u,v,a,\overline{a}\,)=\e^{a\,\P^++\overline{a}\,\P^-}~
\e^{\mu\,u\,\J}~\e^{\mu^{-1}\,v\,\T}
\label{NW4coords}\eeq
where $u,v\in\real$, $a\in\complex$, and the parameter $\mu\in\real$
controls the strength of the NS $B$-field background. For definiteness
we will take $\mu>0$ in the following. In these
global coordinates, the Cartan-Maurer one-form is given by
\beq
g^{-1}~\dd g=\e^{-\ii\mu\,u}~\dd a~\P^++\e^{\ii\mu\,u}~
\dd\overline{a}~\P^-+\mu~\dd u~\J+\left(\mu^{-1}~\dd v+\ii
a~\dd\overline{a}-\ii\overline{a}~\dd a\right)~\T
\label{NW4CMform}\eeq
so that the metric (\ref{NW4CM}) reads
\beq
\dd s_4^2=2~\dd u~\dd v+|\dd a|^2+2\ii\mu\,\left(a~
\dd\overline{a}-\overline{a}~\dd a\right)~\dd u+b\,\mu^2~\dd u^2 \ .
\label{NW4metricNW}\eeq

The metric (\ref{NW4metricNW}) assumes the standard form of the plane
wave metric for a conformally flat, indecomposable Cahen-Wallach
lorentzian symmetric spacetime CW$_4$ in four dimensions~\cite{CW1} upon
introduction of Brinkman harmonic coordinates
$(x^+,x^-,z)$~\cite{Brink1} defined by rotating the transverse plane at
a Larmor frequency as $u=x^+$, $v=x^-$ and
$a=\e^{\frac{\ii\mu}2\,x^+}\,z$. In these coordinates the metric
assumes the stationary form
\beq
\dd s_4^2=2~\dd x^+~\dd x^-+|\dd z|^2+\mu^2\,
\left(b-\mbox{$\frac14$}\,|z|^2\right)~
\left(\dd x^+\right)^2 \ ,
\label{NW4metricBrink}\eeq
revealing the pp-wave nature of the geometry for $b=0$. The physical
meaning of the arbitrary parameter $b$ will be elucidated below. It
may be set to zero by exploiting the translational symmetry of the
geometry in $x^-$ to shift $x^-\to x^--\frac{\mu^2\,b}2\,x^+$, which
corresponds to a Lie algebra automorphism of $\mathfrak n_4$. Note
that on the null planes of constant $u=x^+$, the geometry becomes that of
flat two-dimensional euclidean space $\eucl^2$. This is the geometry
appropriate to the Heisenberg subgroup of $\mathcal{N}_4$, where the
effects of the twisting generator $\J$ are turned off.

Thus far, the Nappi-Witten spacetime has been described geometrically
as a four-dimensional Cahen-Wallach space CW$_4$. The spacetime NW$_4$
is further supported by a Neveu-Schwarz two-form field $B_4$ of constant
field strength
\beq
H_4&=&-\mbox{$\frac13$}\,\bigl\langle g^{-1}~\dd g\,,\,
\left[g^{-1}~\dd g\,,\,g^{-1}~\dd g\right]\bigl\rangle
{}~=~2\ii\mu~\dd x^+\wedge\dd z\wedge\dd\overline{z}~=~\dd B_4 \ ,
\label{NS3formBrink}\eeq
where
\beq
B_4&=&-\mbox{$\frac12$}\,\bigl\langle g^{-1}~\dd g\,,\,
\frac{\id+{\rm Ad}_g}{\id-{\rm Ad}_g}\,g^{-1}~\dd g\bigl\rangle~=~
2\ii\mu\,x^+~\dd z\wedge\dd\overline{z}
\label{NS2formBrink}\eeq
is defined to be non-zero only on those vector fields lying in the
range of the operator $\id-{\rm Ad}_g$ on $T_g\mathcal{N}_4$, i.e. on
vectors tangent to the conjugacy class containing
$g\in\mathcal{N}_4$. The corresponding contracted two-form $H_4^2$
compensates exactly the constant Riemann curvature of the metric
(\ref{NW4metricBrink}), so that NW$_4$ provides a viable supergravity
background. In fact, in this case the cancellation is exact at the
level of the full string equations of motion, so that the plane wave
is an exact background of string theory~\cite{NW1}. It is the presence of
this $B$-field that induces noncommutativity of the string background
in the presence of D-branes.

\subsection{Isometries \label{Isoms}}

The realization of the geometry of NW$_4$ as a standard plane wave of
Cahen-Wallach type enables us to study its isometry group using the
standard classification~\cite{BOL1}. Writing $\partial_\pm:=\partial/\partial
x^\pm$, the metric (\ref{NW4metricBrink}) has the obvious null Killing
vector
\beq
T=\mu\,\partial_-
\label{ZKilling}\eeq
generating translations in $x^-$ and characterizing a pp-wave, and
also the null Killing vector
\beq
J=\mu^{-1}\,\partial_+
\label{HKilling}\eeq
generating translations in $x^+$. An analysis of the Killing
equations~\cite{BOL1} shows that there are also four extra
Killing vectors $P^{(k)}$, $P^{\prime\,(k)}$, $k=1,2$ which generate
twisted translations in the transverse plane $z\in\complex$ to
the motion of the plane wave. Denoting $\partial:=\partial/\partial
z$, they are given in the form
\bea
P^{(k)}&=&c^{(k)}(x^+)\,\partial+\overline{c}^{\,(k)}(x^+)\,
\overline{\partial}-\mu^{-1}\,\left(\dot c^{(k)}(x^+)\,\overline{z}+
\dot{\overline{c}}{}^{\,(k)}(x^+)\,z\right)\,\partial_- \ , \nn\\
P^{\prime\,(k)}&=&c^{\prime\,(k)}(x^+)\,\partial+\overline{c}^{\,
\prime\,(k)}(x^+)\,
\overline{\partial}-\mu^{-1}\,\left(\dot c^{\prime\,(k)}(x^+)\,\overline{z}+
\dot{\overline{c}}{}^{\,\prime\,(k)}(x^+)\,z\right)\,\partial_- \ ,
\label{XkXprimegen}\eea
where the dots denote differentiation with respect to the light-cone
time coordinate $u=x^+$, and the complex-valued coefficient functions in
(\ref{XkXprimegen}) solve the harmonic oscillator equation of motion
\beq
\ddot c(x^+)=-\mbox{$\frac{\mu^2}4$}\,c(x^+) \ .
\label{HODE}\eeq
The four linearly independent solutions of (\ref{HODE}) are
characterized by their initial conditions on the null surface $x^+=0$
as
\bea
c^{(k)}(0)~=~\delta_{k1}+\ii\delta_{k2} ~~~~ &,& ~~~~ \dot c^{(k)}(0)~=~0
\ , \nn\\c^{\prime\,(k)}(0)~=~0 ~~~~ &,& ~~~~ \dot
c^{\prime\,(k)}(0)~=~\mu\,\left(\delta_{k1}+\ii\delta_{k2}\right) \ .
\label{cinitialconds}\eea

The solutions of (\ref{HODE}) and (\ref{cinitialconds}) are given by
\bea
c^{(1)}(x^+)~=~\cos\mbox{$\frac{\mu\,x^+}2$} ~~~~ &,& ~~~~
c^{(2)}(x^+)~=~\ii\cos\mbox{$\frac{\mu\,x^+}2$} \ , \nn\\
c^{\prime\,(1)}(x^+)~=~2\sin\mbox{$\frac{\mu\,x^+}2$} ~~~~ &,& ~~~~
c^{\prime\,(2)}(x^+)~=~2\ii\sin\mbox{$\frac{\mu\,x^+}2$} \ .
\label{cexplsoln}\eea
An interesting feature of these functions is that they generate the
Rosen form~\cite{Rosen1} of the plane wave metric
(\ref{NW4metricBrink}). It is defined by the transformation to local
coordinates $(u,v,y^1,y^2)$ given by
\bea
u&=&x^+ \ , \nn\\v&=&x^--\mbox{$\frac\mu4$}\,(z+\overline{z}\,)^2\,
\tan\mbox{$\frac{\mu\,x^+}2$}-\mbox{$\frac\mu4$}\,(z-\overline{z}\,)^2\,
\cot\mbox{$\frac{\mu\,x^+}2$} \ , \nn\\y^1&=&\mbox{$\frac12$}\,
\left((z+\overline{z}\,)\sec\mbox{$\frac{\mu\,x^+}2$}+\ii(z-\overline{z}\,)
\,\csc\mbox{$\frac{\mu\,x^+}2$}\right) \ , \nn\\y^2&=&\mbox{$\frac12$}\,
\left((z+\overline{z}\,)\sec\mbox{$\frac{\mu\,x^+}2$}-\ii(z-\overline{z}\,)
\,\csc\mbox{$\frac{\mu\,x^+}2$}\right) \ ,
\label{Rosen}\eeq
under which the metric becomes
\beq
\dd s_4^2=2~\dd u~\dd v+C_{ij}(u)~\dd y^i~\dd y^j+b\,\mu^2~\dd u^2
\label{NW4metricRosen}\eeq
where
\beq
C(u)=\bigl(C_{ij}(u)\bigr)=\begin{pmatrix}1&\cos\mu\,u\\\cos\mu\,
  u&1\end{pmatrix} \ .
\label{Cumatrix}\eeq
This form of the metric is degenerate at the conjugate points where
$\cos\mu\,u=\pm\,1$. The harmonic oscillator solutions
(\ref{cexplsoln}) then generate an orthonormal frame for the
transverse plane metric~(\ref{Cumatrix}),
\beq
C(u)=E(u)\,E^\top(u) \ ,
\label{CQrel}\eeq
with
\beq
E=\frac12\,\begin{pmatrix}c^{(1)}+\mbox{$\frac12$}\,c^{\prime\,(2)}+
\overline{c}^{\,(1)}+\mbox{$\frac12$}\,\overline{c}^{\,\prime
\,(2)}& &-\left(c^{(2)}-\mbox{$\frac12$}\,c^{\prime\,(1)}+
\overline{c}^{\,(2)}-\mbox{$\frac12$}\,\overline{c}^{\,\prime
\,(1)}\right)\\c^{(1)}+\mbox{$\frac12$}\,c^{\prime\,(2)}+
\overline{c}^{\,(1)}+\mbox{$\frac12$}\,\overline{c}^{\,\prime
\,(2)}& &c^{(2)}-\mbox{$\frac12$}\,c^{\prime\,(1)}+
\overline{c}^{\,(2)}-\mbox{$\frac12$}\,\overline{c}^{\,\prime
\,(1)}\end{pmatrix}
\label{Qvielbein}\eeq
satisfying the symmetry condition
\beq
\dot E(u)\,E^\top(u)=E(u)\,\dot E^\top(u) \ .
\label{Esymcond}\eeq
Note that in contrast to the Brinkman coordinate system, in the Rosen
form (\ref{NW4metricRosen}) two extra commuting translational
symmetries in the transverse plane $(y^1,y^2)$ are manifest, while
time translation symmetry is lost.

By defining $P^\pm:=P^{\prime\,(1)}\pm\ii P^{(1)}$ and
$ Q{}^{\pm}:=P^{\prime\,(2)}\pm\ii P^{(2)}$, the six
Killing vectors generated by the basic Cahen-Wallach structure of the
plane wave may be summarized as
\bea
T&=&\mu\,\partial_- \ , \nn\\J&=&\mu^{-1}\,\partial_+ \ , \nn\\
P^\pm&=&\left(\sin\mbox{$\frac{\mu\,x^+}2$}\pm\ii\e^{\mp\,
\frac{\ii\mu}2\,x^+}\right)
\left(\partial+\overline{\partial}\,\right)-\mu~\e^{\mp\,
\frac{\ii\mu}2\,x^+}\,\left(
z+\overline{z}\,\right)~\partial_- \ , \nn
\\ Q{}^{\pm}&=&\left(\ii\sin\mbox{$\frac{\mu\,x^+}2$}\mp
\e^{\mp\,\frac{\ii\mu}2\,x^+}\right)
\left(\partial-\overline{\partial}\,\right)+\ii\mu~\e^{\mp\,
\frac{\ii\mu}2\,x^+}\,\left(
z-\overline{z}\,\right)~\partial_- \ .
\label{6CWKilling}\eea
Together, they generate the harmonic oscillator algebra
$\mathfrak{n}_6$ of a particle moving in {\it two} dimensions,
\bea
\left[P^\alpha\,,\, Q{}^{\beta}\right]&=&0 \ , ~~
\alpha,\beta=\pm \ , \nn\\\left[T\,,P^\pm\right]&=&\left[T
\,,\, Q{}^{\pm}\right]~=~\left[T\,,\,J\right]~=~0 \ ,
\nn\\\left[P^+\,,\,P^-\right]&=&\left[Q{}^{+}\,,\,
 Q{}^{-}\right]~=~2\ii T \ , \nn\\\left[J\,,\,
P^\pm\right]&=&\pm\ii P^\pm \ , \nn\\\left[J\,,\, Q{}^{
\pm}\right]&=&\pm\ii Q{}^{\pm} \ .
\label{NW4isomalg}\eea
This isometry algebra acts transitively on the null planes of constant
  $x^+$ and it generates a central extension ${\mathcal
  N}_6$ of the subgroup
\beq
\mathcal{S}_5={\rm SO}(2)\ltimes\real^4
\label{S5subgp}\eeq
of the four-dimensional euclidean group ${\rm
  ISO}(4)={\rm SO}(4)\ltimes\real^4$, where ${\rm SO}(2)$ is the
diagonal subgroup of ${\rm SO}(2)\times{\rm SO}(2)\subset{\rm SO}(4)$.
It is defined by extending the commutation relations (\ref{NW4algdef})
by generators $\Q^{\pm}$ obeying relations as in
(\ref{NW4isomalg}). The quadratic Casimir element $\C_6\in
U(\mathfrak{n}_6)$ and inner product on $\mathfrak{n}_6$ are defined
in the obvious way by extending (\ref{NW4Casimir}) and
(\ref{NW4innerprod}) symmetrically under
$\P^\pm\leftrightarrow\Q^\pm$.

Following the analysis of the previous section, one can show that the
group manifold of ${\mathcal N}_6$ is a six-dimensional Cahen-Wallach
space CW$_6$, with Brinkman metric
\beq
\dd s_6^2=2~\dd x^+~\dd x^-+|\dd\mz|^2+\mu^2\,\left(b-\mbox{$\frac14$}\,
|\mz|^2\right)~\left(\dd x^+\right)^2
\label{NW6metricBrink}\eeq
where $\mz^\top=(z,w)\in\complex^2$, which carries a constant
Neveu-Schwarz three-form flux
\beq
H_6=-2\ii\mu~\dd x^+\wedge\dd\overline{\mz}{}^{\,\top}
\wedge\dd\mz=\dd B_6 \ , ~~
B_6=-2\ii\mu\,x^+~\dd\overline{\mz}{}^{\,\top}\wedge\dd\mz \ .
\label{NW6Bfield}\eeq
It thereby defines a six-dimensional version NW$_6$ of the
Nappi-Witten pp-wave~\cite{KM1}. This observation will be exploited in the
ensuing sections to view the Nappi-Witten wave as an isometrically
embedded D-submanifold ${\rm NW}_4\hookrightarrow{\rm NW}_6$. In this
setting, it corresponds to a symmetry-breaking D3-brane in a non-zero
$H$-flux.

However, for the Nappi-Witten wave this is not the end of the
story. Because of the bi-invariance of the metric (\ref{NW4CM}), the
actual isometry group is the direct product $\mathcal
N_4\times\overline{\mathcal N}_4$ acting by left and right
multiplication on the group $\mathcal N_4$ itself. Since the left
and right actions of the central generator $\T$ coincide, the isometry group is
seven-dimensional. In the present basis, the missing generator from the list
(\ref{6CWKilling}) is the left-moving copy $\overline{J}$ of the
oscillator hamiltonian with
\beq
\left[~\overline{J}\,,\,P^\pm\right]=\mp\ii
P^\pm+ Q{}^{\pm} \ , ~~ \left[~\overline{J}\,,\,
 Q{}^{\pm}\right]=\mp\ii Q{}^{\pm}-P^\pm \ ,
\label{JPpm}\eeq
and it is straightforward to compute that it is given by
\beq
\overline{J}=-\mu^{-1}\,\partial_+-\ii\left(z\,\partial-\overline{z}\,
\overline{\partial}\,\right) \ .
\label{extraHKilling}\eeq
The vector field $J+\overline{J}$ generates rigid rotations in the
transverse plane.

\subsection{Field Theory\label{Dynamics}}

Standard covariant quantization of a massless relativistic scalar
particle in NW$_4$ leads to the Klein-Gordon equation in the curved
background,
\beq
\Box_4\phi~=~0 \ ,
\label{KGeqn}\eeq
where
\beq
\Box_4=2\,\partial_+\,\partial_--\mu^2\,
\left(b-\mbox{$\frac14$}\,|z|^2\right)\,\partial_-^2+|\partial|^2
\label{BoxNW4def}\eeq
is the laplacian corresponding to the Brinkman metric
(\ref{NW4metricBrink}). It coincides with the Casimir
(\ref{NW4Casimir}) expressed in terms of left or right isometry
generators (\ref{6CWKilling},\ref{extraHKilling}). The dependence on
the light-cone coordinates $x^\pm$ drops out of the Klein-Gordon
equation because of the isometries generated by the Killing vectors
(\ref{ZKilling}) and (\ref{HKilling}).

By using a Fourier transformation of the covariant Klein-Gordon field
$\phi$ along the $x^-$ direction,
\beq
\phi\left(x^+,x^-,z,\overline{z}\,\right)=\int\limits_{-\infty}^\infty
\dd p^+~\psi\left(x^+,z,\overline{z};p^+\right)~\e^{\ii p^+x^-} \ ,
\label{KGphiFT}\eeq
we may write (\ref{KGeqn}) equivalently as
\beq
\left[|\partial|^2+2\ii p^+\,\partial_++\left(b-
\mbox{$\frac14$}\,|z|^2\right)\,\left(\mu\,p^+\right)^2\right]\psi\left(x^+,z,
\overline{z};p^+\right)=0 \ .
\label{KGmomsp}\eeq
Introducing the time parameter $\tau$ through
\beq
u=x^+=p^+\,\tau \ ,
\label{utaudef}\eeq
the differential equation (\ref{KGmomsp}) becomes the
Schr\"odinger wave equation
\beq
\ii\,\frac{\partial\psi\left(\tau,z,\overline{z};p^+\right)}{\partial\tau}=
\left[-\mbox{$\frac12$}\,|\partial|^2+\mbox{$\frac12$}\,\left(
\mbox{$\frac{\mu\,p^+}2$}\right)^2\,
|z|^2-\mbox{$\frac b2$}\,\left(\mu\,p^+\right)^2\right]\psi\left(\tau,z,
\overline{z};p^+\right)
\label{SchHO}\eeq
for the non-relativistic two-dimensional harmonic oscillator with a
time independent frequency given by the light-cone momentum and
$H$-flux as $\omega=|\mu\,p^+|/2$. The only role of the arbitrary
parameter $b$ is to shift the zero-point energy of the harmonic
oscillator, and it thereby carries no physical significance.

Let us remark that the same hamiltonian that appears in the
Schr\"odinger equation (\ref{SchHO}) could also have been derived in
light-cone gauge in the plane wave metric (\ref{NW4metricBrink})
starting from the massless relativistic particle lagrangian
\beq
L=\dot x^+\,\dot x^-+\mbox{$\frac{\mu^2}2$}\,\left(b-\mbox{$\frac14$}\,
|z|^2\right)\,\left(\dot x^+\right)^2+\mbox{$\frac12$}\,\left|\dot z
\right|^2
\label{masslessLag}\eeq
describing free geodesic motion in the Nappi-Witten spacetime. In the
light-cone gauge, the light-cone momentum is $p^+=p_-=\partial
L/\partial\dot x^-=\dot x^+=1$, while the hamiltonian is
$J=p_+=\partial L/\partial\dot x^+$. Imposing the mass-shell
constraint $L=0$ at $\dot x^+=1$ gives the equation of motion for
$x^-$, which when substituted into $J$ yields exactly the hamiltonian
appearing on the right-hand side of (\ref{SchHO}) with transverse
momentum $p^{~}_\perp=\dot z=-\ii\partial$.

The time-independence of the effective dynamics here follows from
homogeneity of the plane wave geometry, which prevents dispersion
along the light-cone time direction. These calculations give the quantization
of a particle in NW$_4$ only in the commutative geometry limit,
i.e. in the spacetime CW$_4$, because they do not incorporate the
supergravity $B$-field supported by the Nappi-Witten spacetime. In the
following we will describe how to incorporate the deformation of
CW$_4$ caused by the non-trivial NS-sector. Henceforth we will
drop the zero-point energy and set $b=0$.

\newsection{Isometric Embeddings of Branes\label{IsomEmb}}

A remarkable feature of the Nappi-Witten spacetime is the extent to
which it shares common features with many of the more ``standard''
curved spaces. It is formally similar to the spacetimes built on the
${\rm SL}(2,\real)$ and ${\rm SU}(2)$ group manifolds, but in many
ways is much simpler. As a twisted Heisenberg group, it lies somewhere
in between these curved spaces and the flat space based on the usual
Heisenberg algebra. One way to see this feature at
a quantitative level is by examining Penrose-G\"uven limits involving the
(universal covers of the) ${\rm SL}(2,\real)$ and ${\rm SU}(2)$ group
manifolds which produce the spacetime NW$_4$. This will provide an aid
in understanding various physical properties which arise in later
constructions.

In looking for D-submanifolds, we are primarily interested in those
D-embeddings which are NS-supported and thereby carry a noncommutative
geometry. As we will discuss, this involves certain
important subtleties that must be carefully taken into account. As the
Nappi-Witten spacetime can be viewed as a Cahen-Wallach space, i.e. as
a plane wave, its geometry will arise as Penrose limits of other
metrics. This opens up the possibility of extracting features of
$\NW_4$ by mapping them directly from properties of simpler, better
studied noncommutative spaces. In
this section we will begin with a thorough general analysis of the interplay
between Penrose-G\"uven limits and isometric embeddings of lorentzian
manifolds, and derive simple criteria for the limit and embedding to
commute. Then we apply these results to derive the possible limits
that can be used to describe the NS-supported D-embeddings of NW$_4$.

\subsection{Penrose-G\"uven Limits and Isometric Embeddings: \\ General
  Considerations \label{PGLIE}}

Any lorentzian spacetime has a limiting spacetime which is a plane
wave, whereby the limit can be thought of as a ``first order
approximation'' along a null geodesic in that
spacetime~\cite{Penrose1}. The limiting spacetime depends on the
choice of null geodesic, and hence any spacetime can have more than
one Penrose limit. Let $(M,G)$ be a $d$-dimensional lorentzian
spacetime. We can always introduce local Penrose coordinates
$(U,V,\mbf Y)$, $\mbf Y^\top=(Y^i)\in\real^{d-2}$ in the neighbourhood
of a segment of a null geodesic $\gamma\subset M$ which contains no
conjugate points, whereby the metric assumes the form~\cite{LP1}
\beq
G=2~\dd U~\dd V+\alpha(U,V,\mbf Y)~\dd V^2+2\beta_i
(U,V,\mbf Y)~\dd V~\dd Y^i+C_{ij}(U,V,
\mbf Y)~\dd Y^i~\dd Y^j
\label{GPenrose}\eeq
with $C=(C_{ij})$ a positive definite symmetric matrix. This
coordinate system breaks down at the conjugate points where $C$
becomes degenerate, $\det C=0$. It singles out a twist-free null
geodesic congruence given by constant $V$ and $Y^i$, with $U$ being
the affine parameter along these geodesics. The geodesic $\gamma(U)$
is the one at $V=Y^i=0$.

In string backgrounds one also has to consider generic supergravity $p$-form
gauge potentials $A$ with $(p+1)$-form field strengths $F=\dd A$ in
order to compensate non-trivial spacetime curvature effects. The
G\"uven extension of the Penrose limit to general supergravity fields
shows that any supergravity background has plane wave limits which are
also supergravity backgrounds~\cite{Guven1}. It requires the local
temporal gauge choice
\beq
A_{Ui_1\cdots i_{p-1}}=0
\label{PGgaugechoice}\eeq
in order to ensure well-defined potentials in the limit. With this
gauge choice, which can always be achieved via a gauge transformation
$A\mapsto A+\dd\Lambda$ leaving the flux $F$ invariant, we can write
general potentials and field strengths in the neighbourhood of a null
geodesic $\gamma$ on $M$ as
\bea
A&=&a_{i_1\cdots i_{p-1}}(U,V,\mbf Y)~\dd V\wedge\dd
Y^{i_1}\wedge\cdots\wedge\dd Y^{i_{p-1}}+b_{i_1\cdots i_{p}}
(U,V,\mbf Y)~\dd Y^{i_1}\wedge\cdots\wedge\dd Y^{i_p}
\nn\\&&+\,c_{i_1\cdots i_{p-2}}(U,V,\mbf Y)
{}~\dd U\wedge\dd V\wedge\dd Y^{i_1}\wedge
\cdots\wedge\dd Y^{i_{p-2}} \ , \label{Anull}\\&&{~~~~}_{~~}^{~~}\nn\\
F&=&\frac{\partial b_{i_1\cdots i_p}(U,V,\mbf Y)}
{\partial U}~\dd U\wedge\dd Y^{i_1}\wedge\cdots\wedge\dd Y^{i_p}
+\frac{\partial b_{i_2\cdots i_{p+1}}(U,V,\mbf Y)}
{\partial Y^{i_1}}~\dd Y^{i_1}\wedge\cdots\wedge\dd Y^{i_{p+1}}
\nn\\&&+\,\left(\frac{\partial a_{i_1\cdots i_{p-1}}
(U,V,\mbf Y)}{\partial U}+\frac{\partial c_{i_2\cdots
  i_{p-1}}(U,V,\mbf Y)}{\partial Y^{i_1}}\right)~
\dd U\wedge\dd V\wedge\dd Y^{i_1}\wedge\cdots\wedge\dd Y^{i_{p-1}}
\nn\\&&+\,\left(\frac{\partial b_{i_1\cdots i_p}
(U,V,\mbf Y)}{\partial V}-\frac{\partial a_{i_2\cdots
  i_p}(U,V,\mbf Y)}{\partial Y^{i_1}}\right)~
\dd V\wedge\dd Y^{i_1}\wedge\cdots\wedge\dd Y^{i_p} \ .
\label{Fnull}\eea

The Penrose-G\"uven limit starts with the one-parameter family of
local diffeomorphisms $\psi_\lambda:M\to M$, $\lambda\in\real$ defined
by a rescaling of the Penrose coordinates as
\beq
\psi_\lambda\bigl(U,V,\mbf Y\bigr)=\bigl(u,\lambda^2\,v,
\lambda\,\mbf y\bigr) \ .
\label{psiOmegadef}\eeq
One then defines new fields which are related to the original ones by
a diffeomorphism, a rescaling, and (in the case of potentials) possibly
a gauge transformation, by the well-defined limits
\bea
\widetilde{G}&=&\lim_{\lambda\to0}\,\lambda^{-2}\,\psi^*_\lambda G \ ,
\nn\\\widetilde{A}&=&\lim_{\lambda\to0}\,\lambda^{-p}\,\psi^*_\lambda A
\ , \nn\\\widetilde{F}&=&\lim_{\lambda\to0}\,\lambda^{-p}\,\psi^*_\lambda
F \ .
\label{PGlimitsdef}\eea
Due to (\ref{psiOmegadef}), the only functions in (\ref{GPenrose}),
(\ref{Anull}) and (\ref{Fnull}) which survive this limit are
$C_{ij}(u)=C_{ij}(U,0,\mbf0)$, $b_{i_1\cdots i_p}(u)=b_{i_1\cdots
  i_p}(U,0,\mbf0)$ and $c_{i_1\cdots i_{p-2}}(u)=c_{i_1\cdots
  i_{p-2}}(U,0,\mbf0)$, which are just the pull-backs of the tensor fields $C$,
$b$ and $c$ to the null geodesic $\gamma$. Explicitly, we
obtain a pp-wave metric and supergravity fields in Rosen
coordinates $(u,v,\mbf y)$~\cite{Rosen1} as
\bea
\widetilde{G}&=&2~\dd u~\dd v+C_{ij}(u)~\dd y^i~\dd y^j \ ,
\label{GPGlim}\\&&{~~~~}_{~~}^{~~}\nn\\
\widetilde{A}&=&b_{i_1\cdots i_p}(u)~\dd y^{i_1}
\wedge\cdots\wedge\dd y^{i_p}+c_{i_1\cdots i_{p-2}}(u)~\dd u
\wedge\dd v\wedge\dd y^{i_1}\wedge\cdots\wedge\dd y^{i_{p-2}} \ ,
\label{APGlim}\\&&{~~~~}_{~~}^{~~}\nn\\
\widetilde{F}&=&\frac{\partial b_{i_1\cdots i_p}(u)}
{\partial u}~\dd u\wedge\dd y^{i_1}\wedge\cdots\wedge\dd y^{i_p} \ .
\label{FPGlim}\eea
The physical effect of this limit is to blow up a neighbourhood of the
null geodesic $\gamma$, giving the local background as seen by an
observer moving at the speed of light in $M$. It can be thought of as
an infinite volume limit. We may set $c_{i_1\cdots i_{p-2}}(u)=0$ in
(\ref{APGlim}) via the local gauge transformation
\beq
\widetilde{A}~\longmapsto~\widetilde{A}+\dd\widetilde{\Lambda} \ , ~~
\widetilde{\Lambda}=-\left(\,\mbox{$\int^u$}\,\dd u'~c_{i_1\cdots
    i_{p-2}}(u'\,)\right)~\dd v\wedge\dd y^{i_1}\wedge\cdots\wedge
\dd y^{i_{p-2}} \ .
\label{APGcset0}\eeq

We are now interested in a smooth, local isometric embedding
$\imath:W\subset N\hookrightarrow M$ of a lorentzian manifold $(N,g)$, also
possibly supported by non-trivial $p$-form fields, from an open subset
$W$ of $N$ onto a submanifold of $M$ in codimension $m\geq0$. Thus we
require that $\imath:W\to\imath(W)$ be a diffeomorphism in the induced
${\rm C}^\infty$ structure, and that the derivative map $\dd\imath_x:T_xN\to
T_{\imath(x)}M$ be injective for all $x\in W$. The lorentzian metric
$g$ of $N$ is related to that of $M$ through the pull-back
\beq
g_x\bigl(\,\cdot\,,\,\cdot\,\bigr)=G_{\imath(x)}\bigl(\dd\imath(\,
\cdot\,)\,,\,\dd\imath(\,\cdot\,)\bigr)
\label{gNpullback}\eeq
on $T_xW\otimes T_xW\to\real$, and $p$-form fields $a$ on $N$ are similarly
related to those on $M$ by
\beq
a_x\bigl(\,\cdot\,,\ldots,\,\cdot\,\bigr)=A_{\imath(x)}
\bigl(\dd\imath(\,\cdot\,)\,,\,\ldots,\,\dd\imath(\,\cdot\,)\bigr)
\label{aNpullback}\eeq
on $\otimes^p\,T_xW\to\real$. In most cases of interest in this paper,
$N$ will correspond to
an embedded D-submanifold of the spacetime $M$, but this need not be
the case. In what follows we will usually write $\imath(N)$ for the
injection, with the implicit understanding that it need only be
defined with respect to a local coordinatization of the manifolds.

We shall examine situations in which the Penrose-G\"uven limit will
simultaneously induce the Penrose-G\"uven limits of the ambient spacetime and
of the embedded submanifold. This automatically restricts the types of
embeddings $\imath$ possible. We need to ensure the existence of a
null geodesic on $M$ which starts from a point in $\imath(N)$, whose
initial velocity is tangent to $\imath(N)$, and
which stays on $\imath(N)$ for all time. In certain instances we would
also have to impose further constraints on the geodesic restriction
depending on which target spacetime is desired from the Penrose
limit. A natural restriction is to assume that, in the Penrose
coordinates of (\ref{GPenrose}) adapted to a fixed null geodesic
$\gamma\subset M$ with $\imath(W)$ contained in the set of
conjugate-free points, $\imath$~is the inclusion map defined by fixing
$m$ of the transverse coordinates $Y^{j_1},\dots,Y^{j_m}$ so that
$\imath(N)$ is the intersection of $M$ with the
hypersurface $Y^{j_1},\dots,Y^{j_m}={\rm constant}$. This defines a
null geodesic on $\imath(N)$ which is embedded into the congruence
of null geodesics defined by (\ref{GPenrose}), and the family of such
submanifolds foliates the spacetime $M$ in its local adapted
coordinate system.

In what follows we will simplify matters somewhat by taking the Penrose-G\"uven
limit of $(N,g)$ along the same null geodesic $\gamma$ as that used on
$(M,G)$. The embedded submanifold $\imath(N)$ is then the intersection of
$M$ with the hypersurface $Y^i=0~~\forall i\in{\mathcal
  I}:=\{j_1,\dots,j_m\}$. With the additional requirement that the metric $G$
restricts non-degenerately on $\imath(N)$, there is an orthogonal
tangent space decomposition
\beq
T_xM=T_xN\oplus T_xN^\perp \ .
\label{taudecomp}\eeq
With $\partial_i$ the elements of a local basis of tangent vectors
dual to the one-forms $\dd Y^i$ in an open neighbourhood of $x\in M$,
the fibers of the normal bundle $TN^\perp\to\imath(N)$ over the
embedding are given by $T_xN^\perp=\{\xi^\perp\in
T_xM~|~G(\partial_i,\xi^\perp)=0~~\forall i\notin{\mathcal I}\,\}$. While
these requirements are not the most general ones that one can
envisage, they will suffice for the examples that we consider in this
paper.

Suppose now that we are given another local isometric embedding
$\widetilde{\imath}:\widetilde{W}\subset\widetilde{N}\hookrightarrow
\widetilde{M}$ of lorentzian manifolds $(\,\widetilde{N},\widetilde{g}\,)$ and
$(\,\widetilde{M},\widetilde{G}\,)$, again possibly in the presence of
other supergravity fields. We are interested in the conditions under
which the isometric embedding diagram
\begin{equation}
  \begin{CD}
    M @>\text{PGL}>>                      \widetilde{M}\\
    \text{$\imath$}@AAA @AAA\text{$\widetilde{\imath}$}\\
    N @>\text{PGL}>>                      \widetilde{N}\\
  \end{CD}
\label{isomembdiag}\end{equation}
commutes. The horizontal arrows denote Penrose-G\"uven limits (PGL) of
the respective lorentzian spacetimes, defined in the manner explained
above. In fact such a commutative diagram can only be written down
under very stringent symmetry constraints on the geodesic restrictions
of the transverse plane metric $C_{ij}$ of (\ref{GPenrose}) and
supergravity tensor field $b_{i_1\cdots i_p}$ of
(\ref{Anull}) in the directions normal to the embeddings
$\imath(N)\subset M$ and
$\widetilde{\imath}\,(\,\widetilde{N}\,)\subset\widetilde{M}$.

To formulate these symmetry requirements, let
$\widetilde{\imath}\,(\,\widetilde{N}\,)$ be realized as the
intersection of $\widetilde{M}$ with the hypersurface $y^i=0~~\forall
i\in\widetilde{\mathcal I}:=\{\,\tilde j_1,\dots,\tilde j_m\}$. This
realization of the isometric embedding on the right-hand side of
(\ref{isomembdiag}) is dictated by that on the left-hand side and the
Penrose limit. Consider the submanifold, also denoted
$\widetilde{N}$, defined by the intersection of $M$ with the
hypersurface $Y^i=0~~\forall i\in\widetilde{\mathcal I}$. Denoting the
normal bundle fibers as $T_{\tilde
  x}\widetilde{N}^{\,\perp}:=\{\tilde\xi^\perp\in T_{\tilde
  x}M~|~G(\partial_i,\tilde\xi^\perp)=0~~\forall
i\notin\widetilde{\mathcal I}\,\}$ for $\tilde x\in\widetilde{W}$, there
is an orthogonal tangent space decomposition
\beq
T_{\tilde x}M=T_{\tilde x}\widetilde{N}\oplus T_{\tilde x}
\widetilde{N}^{\,\perp}
\label{tautildedecomp}\eeq
analogous to (\ref{taudecomp}). Along the light-like null geodesic
$\gamma$, where $x=\tilde x=(U,0,\mbf0)$, we fix $p$ tangent vectors
$X,X_1,\dots,X_{p-1}\in T_{(U,0,\mbf0)}M$, and use the lorentzian metric
and $p$-form gauge potentials to define the linear transformations
\bea
G_{(U,0,\mbf0)}(X,\,\cdot\,)\,:\,T_{(U,0,\mbf0)}M&\longrightarrow&
T_{(U,0,\mbf0)}M \ , \label{Gelinmap}\\&&{~~~~}^{~~}_{~~}\nn\\
A_{(U,0,\mbf0)}(X_1,\dots,X_{p-1},\,\cdot\,)\,:\,T_{(U,0,\mbf0)}M&
\longrightarrow&T_{(U,0,\mbf0)}M \ .
\label{Aelinmap}\eea

The isometric embedding diagram (\ref{isomembdiag}) then commutes if,
for every collection of tangent vectors $X,X_1,\dots,X_{p-1}\in
T_{(U,0,\mbf0)}M$, the restrictions of the linear maps (\ref{Gelinmap})
and (\ref{Aelinmap}) to the corresponding orthogonal projections in
(\ref{taudecomp}) and (\ref{tautildedecomp}) agree,
\bea
\bigl.C_{(U,0,\mbf0)}(X,\,\cdot\,)\bigr|_{T_{(U,0,\mbf0)}N^\perp}&=&
\bigl.C_{(U,0,\mbf0)}(X,\,\cdot\,)\bigr|_{T_{(U,0,\mbf0)}
\widetilde{N}^{\,\perp}} \ , \label{Ccommcond}\\&&{~~~~}^{~~}_{~~}
\nn\\\bigl.b_{(U,0,\mbf0)}(X_1,\ldots,X_{p-1},\,\cdot\,)
\bigr|_{T_{(U,0,\mbf0)}N^\perp}&=&
\bigl.b_{(U,0,\mbf0)}(X_1,\ldots,X_{p-1},\,\cdot\,)\bigr|_{T_{(U,0,\mbf0)}
\widetilde{N}^{\,\perp}} \ ,
\label{bcommcond}\eea
in the following sense. The normal subspaces $T_{(U,0,\mbf0)}N^\perp$ and
$T_{(U,0,\mbf0)}\widetilde{N}^{\,\perp}$ of $T_{(U,0,\mbf0)}M$ are
non-canonically
isomorphic as vector spaces. Fixing one such isomorphism, there is
then a one-to-one correspondence between normal vectors $\xi^\perp\in
T_{(U,0,\mbf0)}N^\perp$ and $\tilde\xi^\perp\in
T_{(U,0,\mbf0)}\widetilde{N}^{\,\perp}$, under which we require the
transverse plane metric and $p$-form fields to coincide,
$C_{(U,0,\mbf0)}(X,\xi^\perp)=C_{(U,0,\mbf0)}(X,\tilde\xi^\perp)$ and
$b_{(U,0,\mbf0)}(X_1,\ldots,X_{p-1},\xi^\perp)=b_{(U,0,\mbf0)}
(X_1,\ldots,X_{p-1},\tilde\xi^\perp)$. These symmetry conditions
together ensure that the same supergravity fields are induced on
$\widetilde{\imath}\,(\,\widetilde{N}\,)$ along the two different
paths of the diagram (\ref{isomembdiag}), i.e. that the
Penrose-G\"uven limit of $M$, along the null geodesic described above,
induces simultaneously the Penrose-G\"uven limit of $N$. We stress that
(\ref{Ccommcond}) and (\ref{bcommcond}) are required to simultaneously
hold under only a single such isomorphism of $m$-dimensional vector
spaces, and in all there are $\frac12\,m\,(m+1)$ such commuting
isometric embedding diagrams that can potentially be constructed for
appropriate plane wave profiles.

These symmetry conditions are essentially just the simple statement
that the restrictions of the embeddings $\imath$ and $\widetilde{\imath}$ to
the null geodesic $\gamma(U)$ are equivalent. Nevertheless, there are
many examples whereby the Penrose limit of the metric carries through
in (\ref{isomembdiag}), but not the G\"uven extension to generic
$p$-form supergravity fields, i.e. (\ref{Ccommcond}) can hold with
(\ref{bcommcond}) being violated. Conversely, there can be exotic
isometric embeddings whereby the transverse metric violates the
requirement (\ref{Ccommcond}), leading to target spacetimes with
distinct pp-wave profiles induced by the Penrose limit of essentially
the same lorentzian structure. An interesting example of this would
be a situation wherein the metric is not preserved, but the other
supergravity $p$-form fields are.

\subsection{Hpp-Wave Limits\label{PGLNW}}

Let us now specialize the analysis of the previous subsection to a broad
class of examples that are important to the analysis of this
paper. Consider a real, connected Lie group $\mathcal{G}$ possessing a
bi-invariant metric. On the Lie algebra $\mathfrak g$ of
$\mathcal{G}$, this induces an invariant, non-degenerate inner product
$\langle\,\cdot\,,\,\cdot\,\rangle:\mathfrak{g}\times\mathfrak{g}\to\real$.
We will also write $\mathcal{G}$ for the group manifold.

Symmetric D-branes wrapping submanifolds $D\subset \mathcal{G}$
preserve the maximal (diagonal) symmetry group
$\mathcal{G}\subset\mathcal{G}\times\overline{\mathcal G}$ allowed by
conformal boundary conditions. They are described algebraically by
twisted conjugacy classes of the group~\cite{AS1,FFFS1,FS1,Stanciu2}. Let
$\Omega:\mathfrak{g}\to\mathfrak{g}$ be an outer automorphism of the
Lie algebra of $\mathcal{G}$ preserving its inner product
$\langle\,\cdot\,,\,\cdot\,\rangle$, and let $\omega:\mathcal{G}\to
\mathcal{G}$ be the corresponding Lie group automorphism. The map
$\omega$ is an isometry of the bi-invariant metric on $\mathcal G$, and so it
generates an orbit of any point $g\in \mathcal{G}$ under the
twisted adjoint action of the group as $g\mapsto{\rm
  Ad}_h^\omega(g)=h\,g\,\omega(h^{-1})$, $h\in \mathcal{G}$. We may
thereby identify $D$ with such an orbit as
\beq
D={\mathcal C}_g^\omega=\bigl\{h\,g\,\omega\left(h^{-1}\right)\,
\bigm|\,h\in \mathcal{G}\bigr\} \ .
\label{Dtwistconj}\eeq
This is essentially the generic situation, since additional inner
automorphisms would give rise to twisted conjugacy classes which are
simply translates of one another. In this way, to each such $\omega$
we can associate an equivalence class of D-branes foliating
$\mathcal{G}$.

Since the metric on $\mathcal{G}$ is bi-invariant, the twisted conjugacy class
(\ref{Dtwistconj}) may be exhibited as a homogeneous space
\beq
{\mathcal C}_g^\omega=\mathcal{G}\,/\,{\mathcal Z}_g^\omega
\label{twistconjhom}\eeq
where ${\mathcal Z}_g^\omega\subset \mathcal{G}$ is the stabilizer
subgroup of the point $g$ under the twisted adjoint action of
$\mathcal{G}$ defined by
\beq
{\mathcal Z}_g^\omega=\bigl\{h\in \mathcal{G}\,\bigm|\,h\,g\,\omega\left(h^{-1}
\right)=g\bigr\} \ .
\label{stabdef}\eeq
By this homogeneity, it will always suffice to determine the geometry
(and any $\mathcal{G}$-invariant fluxes) at a single point, as all other points
are related by the twisted adjoint action of the group, which is an
isometry. A natural physical assumption is that the bi-invariant
metric of $\mathcal{G}$ restricts non-degenerately to the twisted conjugacy
classes (\ref{twistconjhom}). The normal bundle over the D-submanifold
then has fibers given by
\beq
\left(T_g\mathcal{C}_g^\omega\right)^\perp=T_g\mathcal{Z}_g^\omega
\label{twistconjnorm}\eeq
with $T_g\mathcal{C}_g^\omega\cap T_g\mathcal{Z}_g^\omega=\{0\}$, so
that there is an orthogonal direct sum decomposition of the tangent
bundle of $\mathcal{G}$ as
\beq
T_g\mathcal{G}=T_g\mathcal{C}_g^\omega\oplus T_g\mathcal{Z}_g^\omega \ .
\label{TgGdecomp}\eeq
The space of normal vectors (\ref{twistconjnorm}) may be identified
with the Lie algebra of the stabilizer subgroup (\ref{stabdef}) given
by
\beq
\mathfrak{z}_g^\omega=\bigl\{X\in\mathfrak{g}\,\bigm|\,g^{-1}\,X\,g=
\Omega(X)\bigr\} \ .
\label{stabLiealg}\eeq

These D-submanifolds are also NS-supported with
$B$-field~\cite{Stanciu1}
\beq
B_{g_0}=-\bigl\langle\dd h\,h^{-1}\,,\,{\rm Ad}_g\circ\Omega\left(
\dd h\,h^{-1}\right)\bigr\rangle
\label{Bfieldg0}\eeq
at $g_0=h\,g\,\omega(h^{-1})\in{\mathcal C}_g^\omega$, and they are
stabilized against decay by the presence of a family of two-form
abelian gauge field strengths $F_{g_0}^{(\zeta)}=\dd
A_{g_0}^{(\zeta)}$, $\zeta\in\mathfrak{z}_g^\omega$ given
by~\cite{BSR1}
\beq
F_{g_0}^{(\zeta)}=\bigl\langle\zeta\,,\,
\left[\dd h\,h^{-1}\,,\,\dd h\,h^{-1}\right]\bigr\rangle \ .
\label{2formfamily}\eeq
If $\zeta$ is a fractional symmetric weight of $\mathcal{G}$, then the
integral of (\ref{2formfamily}) over any two-sphere in the worldvolume
$\mathcal{C}_g^\omega$ of the D-brane is an integer multiple of
$2\pi$. The two-forms
\beq
\mathcal{F}_{g_0}^{(\zeta)}:=B^{~}_{g_0}+F_{g_0}^{(\zeta)}
\label{inv2forms}\eeq
are invariant under gauge transformations $B_{g_0}\mapsto B_{g_0}+\dd\Lambda$,
$F_{g_0}^{(\zeta)}\mapsto F_{g_0}^{(\zeta)}-\dd\Lambda$ of the
$B$-field. They are the gauge invariant combinations (in string units
$\alpha'=1$) that appear in the Dirac-Born-Infeld action governing the
target space dynamics of the D-branes.

There are two ways in which the Penrose limit may be achieved within
this setting~\cite{BFP1}. In both instances it can be understood as an
In\"on\"u-Wigner group contraction of $\mathcal{G}$ whose limit
$\widetilde{\mathcal{G}}$ is a non-compact, non-semisimple Lie group
admitting a bi-invariant metric. In the first case we assume $\mathcal
G$ is simple and consider a one-parameter subgroup
$\mathcal{H}\subset\mathcal{G}$, which is necessarily geodesic
relative to the bi-invariant metric. If it is also null, then it gives
rise to a null geodesic and hence the Penrose limit can be taken as
prescribed in the previous subsection. In the second case we consider
a product group $\mathcal{G}=\mathcal{G}'\times\mathcal{H}$, where
$\mathcal{G}'$ is a simple Lie group with a bi-invariant metric and
$\mathcal{H}\subset \mathcal{G}'$ is a compact subgroup, with Lie subalgebra
$\mathfrak{h}\subset\mathfrak{g}'$, which inherits a bi-invariant
metric from $\mathcal{G}$ by pull-back. The product group
$\mathcal{G}'\times \mathcal{H}$ then carries the bi-invariant product
metric corresponding to the bilinear form
$\langle\,\cdot\,,\,\cdot\,\rangle\oplus
(-\langle\,\cdot\,,\,\cdot\,\rangle|_{\mathfrak{h}})$ on
$\mathfrak{g}'\oplus\mathfrak{h}$. The submanifold $\mathcal{H}\subset
\mathcal{H}\times\mathcal{H}\subset\mathcal{G}'\times \mathcal{H}$
given by the diagonal embedding is a Lie subgroup, and hence it is
totally geodesic and maximally isotropic. The (generalized) Penrose
limit of $\mathcal{G}'\times\mathcal{H}$ along $\mathcal{H}$ thereby
yields a non-semisimple Lie group with a bi-invariant metric. The
related group contraction can also be understood as an infinite volume limit
along the compact directions of $\mathcal{G}'\times \mathcal{H}$,
giving a semi-classical picture of the string dynamics in this
background.

We may now attempt to specialize the isometric embedding diagram
(\ref{isomembdiag}) to the diagram
\begin{equation}
  \begin{CD}
    \mathcal{G} @>\text{PGL}>>                      \widetilde{\mathcal{G}}\\
    \text{$\imath$}@AAA @AAA\text{$\widetilde{\imath}$}\\
    \mathcal{C}_g^\omega @>\text{PGL}>> \mathcal{C}_{\tilde g}^{\tilde\omega}\\
  \end{CD}
\label{twistconjembdiag}\end{equation}
specifying the Penrose-G\"uven limit between twisted D-branes. We have
used the fact that the Penrose limit of a maximally symmetric space is again a
maximally symmetric space~\cite{BFP1}, so that the Penrose-G\"uven isometric
embedding diagrams preserve the symmetry of the embedded spaces and
symmetric D-branes map onto symmetric D-branes in the limit. To
formulate the symmetry conditions (\ref{Ccommcond}) and
(\ref{bcommcond}) within this algebraic setting, we restrict to the
stabilizer algebra (\ref{stabLiealg}).
For any $h\in\mathcal{H}$ and any Lie algebra element
$X\in\mathfrak{g}$, the isometric embedding diagram
(\ref{twistconjembdiag}) then commutes if and only if
\bea
\bigl.\langle X,\,\cdot\,\rangle\bigr|_{\mathfrak{z}_h^\omega}&=&
\bigl.\langle X,\,\cdot\,\rangle\bigr|_{\mathfrak{z}_h^{\tilde\omega}}
\ , \label{innprodtwistcomm}\\{~~~~}_{~~}^{~~}\nn\\\bigl.
B_{h_0}(X,\,\cdot\,)\bigr|_{
\mathfrak{z}_h^\omega}&=&\bigl.B_{h_0}(X,\,\cdot\,)\bigr|_{
\mathfrak{z}_h^{\tilde\omega}} \ , \label{Bfieldtwistcomm}
\\{~~~~}_{~~}^{~~}\nn\\\bigl.F^{(\zeta)}_{h_0}(X,\,\cdot\,)\bigr|_{
\mathfrak{z}_h^\omega}&=&\bigl.F^{(\zeta)}_{h_0}(X,\,\cdot\,)\bigr|_{
\mathfrak{z}_h^{\tilde\omega}} \ ,
\label{Ftwistcomm}\eea
with the analogous restrictions on higher-degree $p$-form fields.

\subsection{NW Limits\label{ApplNW}}

We will now apply these considerations to examine the possible
D-embeddings of Nappi-Witten spacetimes. We start with the
six-dimensional lorentzian manifold $M=\AdS_3\times\Sphere^3$
describing the near horizon geometry of a bound state of fundamental
strings and NS5-branes~\cite{GKS1}. This identification requires that
both factors share a common radius of curvature $R$. We can embed
$\AdS_3\times\Sphere^3$ in the pseudo-euclidean space $\eucl^{2,6}$ as
the intersection of the two quadrics
\bea
\left(x^0\right)^2+\left(x^1\right)^2-\left(x^2\right)^2-\left(x^3\right)^2
&=&R^2\ , \label{AdS3quadric}\\&&{~~~~}^{~~}_{~~}\nn\\
\left(x^4\right)^2+\left(x^5\right)^2+\left(x^6\right)^2+\left(x^7\right)^2
&=&R^2
\label{S3quadric}\eea
with the induced metric. An explicit parametrization is given by
\beq
\begin{matrix}
x^0&=&R\,\sqrt{1+r^2}\,\cos\tau\,\cosh\beta ~~~~ &,& ~~~~
x^4&=&R\,\cos\phi \ , \\x^1&=&R\,\sin\tau
{}~~~~ &,& ~~~~ x^5&=&R\,\chi\,\sin\phi\,\sin\psi \ , \\x^2&=&R\,\sqrt{1+r^2}\,
\cos\tau\,\sinh\beta
{}~~~~ &,& ~~~~ x^6&=&R\,\sqrt{1-\chi^2}\,\sin\phi\,\sin\psi
\ , \\x^3&=&R\,r\,\cos\tau ~~~~ &,& ~~~~ x^7&=&R\,\sin\phi\,\cos\psi \ .
\end{matrix}
\label{AdS3S3param}\eeq
In these coordinates the metric, NS--NS $B$-field and
three-form flux on $\AdS_3\times\Sphere^3$ are given by
\bea
\mbox{$\frac1{R^2}$}\,
G&=&-\dd\tau^2+\cos^2\tau\,\left(\frac{\dd r^2}{1+r^2}+\left(1+r^2
\right)~\dd\beta^2\right)\nn\\&&+\,\dd\phi^2+\sin^2\phi\,\left(
\frac{\dd\chi^2}{1-\chi^2}+\left(1-\chi^2\right)~\dd\psi^2\right) \ ,
\label{AdS3S3metric}\\&&{~~~~}^{~~}_{~~}\nn\\
-\mbox{$\frac1{2R^2}$}\,H&=&\cos^2\tau~\dd\tau\wedge
\dd r\wedge\dd\beta+\sin^2\phi~\dd\phi\wedge\dd\chi\wedge\dd\psi \ ,
\label{AdS3S3flux}\\&&{~~~~}_{~~}^{~~}\nn\\-\mbox{$\frac2{R^2}$}\,
B&=&(\sin2\tau+2\tau)~\dd r\wedge\dd\beta+
(\sin2\phi-2\phi)~\dd\chi\wedge\dd\psi \ .
\label{AdS3S3Bfield}\eea

Viewing $M$ as the group manifold of the Lie group ${\rm
  SU}(1,1)\times{\rm SU}(2)$ with its usual bi-invariant metric
induced by the Cartan-Killing form, its
  embedded submanifolds wrapped by maximally symmetric D-branes are given
  by twisted conjugacy classes of the group. Here we will focus on the
  family of D3-branes which are isometric to $N=\AdS_2\times\Sphere^2$
  and are given by the intersections of the hyperboloids
  (\ref{AdS3quadric},\ref{S3quadric}) with the affine hyperplanes
  $x^3,x^5={\rm constant}$~\cite{BP1}. In this way we may
  exhibit a foliation of $\AdS_3\times\S^3$ consisting of twisted
  D-branes, each of which is isometric to
  $\AdS_2\times\S^2$. Within this family, the intersection of
  $\AdS_3\times\S^3$ with the hyperplane defined by $x^3=x^5=0$ is
  special for a variety of reasons. It
  corresponds to the fixed point set of the reflection isometry of
  $\eucl^{2,6}$ defined by $x^3\mapsto-x^3$, $x^5\mapsto-x^5$ while
  leaving fixed all other coordinates. This isometry preserves the
  embedding (\ref{AdS3quadric},\ref{S3quadric}) and hence induces an
  isometry of $\AdS_3\times\S^3$. The metric (\ref{AdS3S3metric})
  restricts nondegenerately to the corresponding $\AdS_2\times\S^2$
  submanifold, which is thereby totally geodesic and has equal radii of
  curvature $R$.

When we consider such embedded D-submanifolds, we should also add to
the list of supergravity fields
(\ref{AdS3S3metric})--(\ref{AdS3S3Bfield}) the
constant ${\rm U}(1)$ gauge field flux~\cite{BP1,BDS1}
\beq
\mbox{$\frac2{R^2}$}\,F=\mbox{$\frac2{R^2}$}~
\dd A=4\pi\,\kappa_q~\dd r\wedge\dd\beta-
2\pi\,p~\dd\chi\wedge\dd\psi \ ,
\label{AdS3S3U1flux}\eeq
where $p\in\zed$ is the magnetic monopole number through $\S^2$,
while $\kappa_q\in\real$ is related to the quantum number $q\in\zed$ giving
the Dirac-Born-Infeld electric flux through $\AdS_2$. The presence of
this two-form prevents the wrapped D3-branes from collapsing. The quantity
${\mathcal F}=B+F$ is invariant under two-form gauge transformations
of the $B$-field, but it is only the monopole flux and the dual
Dirac-Born-Infeld electric displacement which
are quantized. The addition of such worldvolume electric fields
proportional to the volume form of $\AdS_2$ along with worldvolume
magnetic fields proportional to the volume form of $\S^2$ still
preserves supersymmetry. The sources of such fluxes are $(p,q)$
strings connecting the D3-branes to the NS5/F1 black string background
in the near horizon region~\cite{BP1}. The Dirac-Born-Infeld energy of
the branes wrapping $\AdS_2\times\S^2$ is locally minimized at
$(\tau,\phi)=(2\pi\,\kappa_q,\pi\,p)$~\cite{BP1,BDS1}, and at those
values the gauge-invariant two-form is given by
\beq
-\mbox{$\frac2{R^2}$}\,\mathcal{F}=\sin2\tau~\dd r\wedge\dd\beta+
\sin2\phi~\dd\chi\wedge\dd\psi \ .
\label{gaugeinvAdS3S3}\eeq

Let us now consider the Penrose-G\"uven limit of $M$ which produces the
six-dimensional Nappi-Witten spacetime
$\widetilde{M}=\NW_6$~\cite{BFP1}. As Lie groups, the Penrose limit
can be interpreted as an In\"on\"u-Wigner group contraction of ${\rm
  SU}(1,1)\times{\rm SU}(2)$ onto
$\mathcal{N}_6$~\cite{SF1}. Geometrically, it can be achieved along
any null geodesic which has a non-vanishing velocity component tangent
to the sphere $\S^3$. For this, we change coordinates in the
$(\tau,\phi)$ plane to
\beq
U=\phi+\tau \ , ~~ V=\mbox{$\frac12$}\,(\phi-\tau) \ ,
\label{UVAdS3S3def}\eeq
and relabel the remaining coordinates as
\beq
Y^1=r \ , ~~ Y^2=\beta \ , ~~ Y^3=\chi \ , ~~ Y^4=\psi \ .
\label{Yrelabel}\eeq
This enables us to represent the fields in
(\ref{AdS3S3metric})--(\ref{AdS3S3Bfield}) in the adapted coordinate
forms (\ref{GPenrose}), (\ref{Anull}) and (\ref{Fnull}),  which
thereby exhibits $\frac\partial{\partial
  U}=\frac12\,(\frac\partial{\partial\phi}+\frac\partial{\partial\tau})$
as the null geodesic vector field with $G(\frac\partial{\partial
  U},\frac\partial{\partial U})=0$. After the Penrose-G\"uven limit,
the metric and Neveu-Schwarz fields along the geodesic $\gamma(U)$ are
given by
\bea
\mbox{$\frac1{R^2}$}\,\widetilde{G}&=&2~\dd u~\dd v+\sin^2\mbox{$
\frac u2$}~\dd{\mbf y}^2 \ , \label{NW6tildemetric}\\&&{~~~~}^{~~}_{~~}\nn\\
\mbox{$\frac1{R^2}$}\,\widetilde{H}&=&\cos^2\mbox{$\frac u2$}~\dd u
\wedge\dd y^1\wedge\dd y^2-\sin^2\mbox{$\frac u2$}~\dd u\wedge\dd y^3
\wedge\dd y^4 \ , \label{NW6tildeflux}\\&&{~~~~}^{~~}_{~~}\nn\\
\mbox{$\frac4{R^2}$}\,
\widetilde{B}&=&-(u+\sin u)~\dd y^1\wedge\dd y^2+(u-\sin u)~
\dd y^3\wedge\dd y^4
\label{NW6tildeBfield}\eea
with ${\mbf y}^\top=(y^i)\in\real^4$. At this stage it is convenient
to transform from Rosen coordinates to Brinkman coordinates as
\bea
u&=&\mu\,x^+ \ , \nn\\v&=&\mu^{-1}\,x^-+\mbox{$\frac12$}\,
\bigl(|z|^2+|w|^2\bigr)\,
\cot\mbox{$\frac{\mu\,x^+}2$} \ , \nn\\y^1+\ii y^2&=&w\,
\csc\mbox{$\frac{\mu\,x^+}2$} \ ,
\nn\\y^3+\ii y^4&=&z\,\csc\mbox{$\frac{\mu\,x^+}2$} \ .
\label{BrinkfromAdS3S3}\eea
It is then straightforward to compute that one recovers the standard
NS-supported geometry of $\NW_6$, with supergravity fields
$\frac1{R^2}\,\widetilde{G}=\dd s_6^2$,
$\frac1{2R^2}\,\widetilde{H}=H_6$ and $\frac2{R^2}\,\widetilde{B}=B_6$
given by (\ref{NW6metricBrink},\ref{NW6Bfield}).

The foliating hyperplanes $w=w_0\in\complex$ isometrically embed
$\NW_4$ in $\NW_6$ with its standard geometry
(\ref{NW4metricBrink},\ref{NS2formBrink}) and zero-point energy
$b=-\frac{\mu^2}4\,|w_0|^2$. The Penrose-G\"uven limit of the
worldvolume flux (\ref{AdS3S3U1flux}) vanishes, $\widetilde{F}=0$,
because it is the field strength of a ${\rm U}(1)$ gauge field. On the
other hand, the gauge invariant two-form (\ref{gaugeinvAdS3S3})
transforms as a potential under the Penrose-G\"uven limit and one
finds
\beq
\mathcal{F}_6:=-\mbox{$\frac2{R^2}$}\,\widetilde{\mathcal{F}}=-\ii
\cot\mbox{$\frac{\mu\,x^+}2$}\,\left[\dd\overline{\mz}^{\,\top}
\wedge\dd\mz+\mbox{$\frac\mu2$}\,\cot\mbox{$\frac{\mu\,x^+}2$}~
\dd x^+\wedge\left(\mz^\top~\dd\overline{\mz}-\overline{\mz}^{\,\top}~
\dd\mz\right)\right] \ .
\label{NW6tildegaugeinv}\eeq
Strictly speaking, this field is only defined on the constant time
slices $x^+=\frac\pi\mu\,(2\kappa_q+p)$ of $\NW_6$ induced by the
energy minimizing configurations described above. However, with the
scaling transformations employed in the Penrose-G\"uven limit we can
take the form (\ref{NW6tildegaugeinv}) to be valid at all times. It
induces a worldvolume two-form $\mathcal{F}_4:=\mathcal{F}_6|_{w=w_0}$
on $\NW_4$.

Let us now examine the isometric embedding of the totally geodesic
$\AdS_2\times\S^2$ D-brane in $\AdS_3\times\S^3$, which may be
defined as the hyperplane $Y^1=Y^3=0$. The geodetic property
ensures that the Penrose limit of $\AdS_3\times\S^3$ induces that of
$\AdS_2\times\S^2$, which yields the Cahen-Wallach symmetric space
$\CW_4$~\cite{BFP1}. However, using the general analysis of
Section~\ref{PGLIE}, one can see immediately that the G\"uven
extension breaks down. In this case the normal bundle fiber
$T_{(U,0,\mbf0)}N^\perp$ is locally spanned by the vector fields
$\partial_2,\partial_4$, while $T_{(U,0,\mbf0)}\widetilde{N}^{\,\perp}$ is
spanned by $\partial_3,\partial_4$. From the local form of the
$B$-field (\ref{NW6tildeBfield}), we see for instance that
$b_{12}(u)=-\frac12\,(u+\sin u)$ and $b_{13}(u)=b_{14}(u)=0$ are
clearly distinct. There is thus no gauge transformation such that
(\ref{bcommcond}) is satisfied. On the other hand, since the
transverse plane metric $C_{ij}(u)=\sin^2\frac u2~\delta_{ij}$
in (\ref{NW6tildemetric}) is proportional to the identity,
(\ref{Ccommcond}) is trivially satisfied and so the Penrose limit of
the $\AdS_2\times\S^2$ metric coincides with that of $\CW_4$. This can
also be checked by explicit calculation. Indeed, the restrictions of
the Neveu-Schwarz fields (\ref{AdS3S3flux},\ref{AdS3S3Bfield}) to
constant $Y^1$ and $Y^3$ both vanish, as does that of the two-form
(\ref{gaugeinvAdS3S3}), and hence so do their Penrose-G\"uven limits.

One may think that this problem could be rectified by choosing
an alternative embedding of $\AdS_2\times\S^2$ for which the $B$-field
is non-vanishing, such as that with $\tau,\phi={\rm
  constant}$, as is appropriate for the minimal energy symmetric
D3-branes. However, the induced NS-flux (\ref{AdS3S3flux}) still
vanishes, and indeed this leads to the standard (fuzzy) geometry of
euclidean $\AdS_2\times\S^2$. The Penrose limit of $\AdS_2\times\S^2$ could be
taken now in the adapted coordinates $U=\chi+r$, $V=\frac12\,(\chi-r)$,
$Y^1=\beta$, $Y^2=\psi$, and after another suitable change to
Brinkman coordinates it leads to the anticipated $\CW_4$
geometry. However, now the only non-vanishing components of the
$B$-field are $B_{Ui}$ and $B_{Vi}$, which both necessarily vanish
after the Penrose-G\"uven limit is taken. The same is true of the
gauge invariant two-form (\ref{gaugeinvAdS3S3}), and we are forced to
conclude that there is no way to obtain the fully NS-supported
geometry of $\NW_4$ as a plane wave limit from that of
$\AdS_2\times\S^2$. However, now the $\CW_4$ branes carry a
non-vanishing null worldvolume flux which may be written in Brinkman
coordinates as
\beq
\mbox{$\frac2{R^2}$}\,\widetilde{F}=\mbox{$\frac{\pi\,\mu}2$}\,\csc
\mbox{$\frac{\mu\,x^+}2$}\,\bigl[(2\kappa_q-\ii p)~
\dd x^+\wedge\dd w+(2\kappa_q+\ii p)~\dd x^+\wedge\dd\overline{w}
\,\bigr] \ .
\label{tildefluxalt}\eeq

\subsection{Embedding Diagrams for NW Spacetimes\label{DiagNW}}

Let us now describe two simple and obvious remedies to the problem
which we have raised in the previous subsection. The first one modifies the
embedding $\widetilde{\imath}$ on the right-hand side of the diagram
(\ref{isomembdiag}) to be the intersection of $\widetilde{M}=\NW_6$
with the hyperplane $y^1=y^3=0$, so that
 $T_{(U,0,\mbf0)}N^\perp=T_{(U,0,\mbf0)}\widetilde{N}^{\,\perp}$ and the
conditions (\ref{Ccommcond}) and (\ref{bcommcond}) are always
trivially satisfied. Now the pull-backs of the Neveu-Schwarz fields
(\ref{NW6tildeflux},\ref{NW6tildeBfield}) vanish, as does that of the
two-form (\ref{NW6tildegaugeinv}), while the pull-back
of the metric (\ref{NW6tildemetric}) is still the standard metric on
$\CW_4$. This embedding thereby preserves the basic Cahen-Wallach
structure $\CW_4\subset\NW_6$, and the vanishing of the other
supergravity form fields on $\CW_4$ is indeed induced now by the
Penrose-G\"uven limit from $\AdS_2\times\S^2$. We may thereby write
the commuting embedding diagram
\begin{equation}
  \label{CW4commdiag}
  \begin{CD}
    \AdS_3 \times \S^3             @>\text{PGL}>> \NW_6\\
    \text{$\imath~~$}@AAA @AAA\text{$\widetilde{\imath}$}\\
    \AdS_2 \times \S^2             @>\text{PGL}>> \CW_4\\
  \end{CD}
\end{equation}
which describes the Penrose-G\"uven limit between {\it commutative},
maximally symmetric lorentzian D3-branes in $\AdS_3\times\S^3$ and the
six-dimensional Nappi-Witten spacetime $\NW_6$. Due to the vanishing
of the worldvolume flux and the fact that $\pi_2(\S^3)=0$, these
D-branes can be unstable. They may shrink to zero size completely
corresponding to point-like D-instantons, to euclidean D-strings induced at a
point in $\S^3$ with worldvolume geometries $\AdS_2\subset\AdS_3$, or
to euclidean D-strings sitting at a point in $\AdS_3$ and wrapping
$\S^2\subset\S^3$  on the left-hand side of the
diagram~(\ref{CW4commdiag}). These symmetric decay products will be
generically lost in the Penrose-G\"uven limit, as will become evident
in the ensuing sections.

Alternatively, we may choose to modify the embedding $\imath$ on the
left-hand side of the diagram (\ref{isomembdiag}) to be the
intersection of $M=\AdS_3\times\S^3$ with the hyperplane
$Y^1=Y^2=0$. Again
$T_{(U,0,\mbf0)}N^\perp=T_{(U,0,\mbf0)}\widetilde{N}^{\,\perp}$ and so the
conditions (\ref{Ccommcond}) and (\ref{bcommcond}) are trivially
satisfied. This hyperplane corresponds to the intersection of the
hyperboloid (\ref{AdS3quadric}) with $x^2=x^3=0$, while
(\ref{S3quadric}) is left unchanged. It thereby defines a totally
geodesic embedding of $\S^{1,0}\times\S^3$ in
$\AdS_3\times\S^3$. This does not define a twisted conjugacy class of
the Lie group ${\rm SU}(1,1)\times{\rm SU}(2)$ and so does not
correspond to a symmetric D-brane~\cite{BP1}. Instead, it arises in the near
horizon geometry of a stack of NS5-branes~\cite{GO1}. The pull-backs
of the NS--NS fields (\ref{AdS3S3flux},\ref{AdS3S3Bfield}) are
non-vanishing, and the null geodesic defined by (\ref{UVAdS3S3def})
spins along an equator of the sphere~$\S^3$. The Penrose-G\"uven limit
thus induces the complete NS-supported geometry of
$\NW_4$~\cite{DAK1}, and it can be thought of as a group contraction
of ${\rm U}(1)\times{\rm SU}(2)$ along ${\rm U}(1)$ onto
$\mathcal{N}_4$. We may thereby write the commuting embedding diagram
\begin{equation}
  \label{NW4commdiag}
  \begin{CD}
    \AdS_3 \times \S^3             @>\text{PGL}>> \NW_6\\
    \text{$\imath~~$}@AAA @AAA\text{$\widetilde{\imath}$}\\
    \S^{1,0}\times\S^3         @>\text{PGL}>> \NW_4\\
  \end{CD}
\end{equation}
describing {\it noncommutative} branes. The lorentzian $\NW_4$
D3-brane, supported by non-trivial worldvolume fields, is stabilized
against decay by its worldvolume two-form
$\mathcal{F}_4=\mathcal{F}_6|_{z=0}$.

These results are all consistent with the remark made just after
(\ref{twistconjembdiag}). The brane $\AdS_2\times\S^2$ is a symmetric
D-submanifold of $\AdS_3\times\S^3$. As we discuss in detail in
Section~\ref{TwistedNCBranes}, $\NW_4$ is {\it not} a symmetric
D-brane in $\NW_6$, although $\CW_4$ is. Consistently,
$\S^{1,0}\times\S^3\subset\AdS_3\times\S^3$ is not a maximally
symmetric D-embedding, but rather a member of the hierarchy of
factorizing symmetry-breaking D-branes in ${\rm SU}(1,1)\times{\rm
  SU}(2)$ which preserves the action of an $\real\times{\rm SU}(2)$
subgroup and is localized along a product of images of twisted conjugacy
classes~\cite{Quella1}. Similarly, both the $\AdS_3\times\S^3$ and
$\NW_6$ branes are non-symmetric, while $\CW_6$ is the worldvolume
of a spacetime-filling twisted D5-brane in $\NW_6$.

\newsection{Noncommutative Branes in $\mbf{\NW_6}$: Untwisted
  Case\label{NCBranes}}

In this section and the next we will identify which {\it symmetric}
D-branes of the six-dimensional Nappi-Witten spacetime $\NW_6$ support a
noncommutative worldvolume geometry. After identifying all
supergravity fields living on the respective worldvolumes, we will
proceed to quantize the classical geometries using standard
techniques~\cite{Schom1}. A potential obstruction to this procedure is
that, as in any curved string background, a non-vanishing $H$-flux can lead to
non-associative deformations of worldvolume algebras, typically
giving rise to variants of quantum group algebras that are
deformations of standard noncommutative geometries for which there is
no general notion of quantization. In certain semi-classical
limits the NS-flux vanishes, $H=0$, such as in the conformal field
theory description based on a compact group $\mathcal{G}$ in the limit
of infinite Kac-Moody level $k\to\infty$~\cite{ARS1}. Such limits correspond to
field theory limits of the string theory whereby the Neveu-Schwarz
$B$-field (or more precisely the gauge invariant two-form
(\ref{inv2forms})) induces a symplectic structure on the brane
worldvolume, which can be quantized in principle. In our
situation we can identify $k=\mu^{-2}$, and the semi-classical limit
corresponds to the limit in which the plane wave approaches flat
spacetime. Such limits will always arise for the branes that we
encounter in this paper. We will substantiate our characterizations by
comparison with known results from the boundary conformal field theory
of the Nappi-Witten model~\cite{DK2,Hikida1}. In this section we deal
with branes described by conjugacy classes of the Nappi-Witten group
$\mathcal{N}_6$.

Generally, for the ordinary (untwisted) conjugacy classes
($\omega=\id$ in the notation of (\ref{Dtwistconj})), the
worldvolumes inherit a natural symplectic form via the exponential map
from the usual Kirillov-Kostant form on coadjoint orbits, which
coincides with the symplectic structure induced by the $B$-field when
$H=0$. The conjugacy classes, given as orbits of the adjoint action of the Lie
group $\mathcal{G}$ on itself, are then identified with representations of
$\mathfrak{g}$, obtained from quantization of coadjoint orbits, via
the non-degenerate invariant bilinear form
$\langle\,\cdot\,,\,\cdot\,\rangle$ on $\mathfrak{g}$. If $V_g$ is an
irreducible module over the Lie algebra $\mathfrak{g}$ corresponding
to the D-brane $D=\mathcal{C}^{~}_g:=\mathcal{C}_g^{\id}$, then the
noncommutative algebra of functions on the worldvolume is given
by~\cite{ARS1}
\beq
\mathcal{A}(\mathcal{C}_g)={\rm End}(V_g) \ .
\label{NCalgconjclass}\eeq
The worldvolume algebra (\ref{NCalgconjclass}) carries a natural
(adjoint) action of the group $\mathcal{G}$. In the present case we
will find that these quantized conjugacy
classes carry a noncommutative geometry which is completely
analogous to that carried by D3-branes in flat space $\eucl^4$ with a
uniform magnetic field on their worldvolume~\cite{DNek1,SW1,Sz1,Sz2},
as expected from the harmonic oscillator character of dynamics in
Nappi-Witten spacetime described at length in Section~\ref{NWPW}. We
shall also find an interesting class of commutative null branes whose
quantum geometry generically differs significantly from that of the
classical conjugacy classes, due a transverse ${\rm U}(2)$ rotational
symmetry of the $\NW_6$ background.

\subsection{General Construction\label{GenConstrUntwist}}

It will prove convenient to introduce the doublet
$(\,\underline{\P}^\pm)^\top:=(\P^\pm,\Q^\pm)$ of generators and write
Brinkman coordinates on $\mathcal{N}_6$ as
\beq
g(x^+,x^-,\mz,\overline{\mz}\,)=\e^{\frac\mu2\,x^+\,\J}~
\e^{\mz^\top\,\underline{\P}^++
\overline{\mz}{}^{\,\top}\,\underline{\P}^-}~\e^{\frac\mu2\,x^+\,\J}~
\e^{\mu^{-1}\,x^-\,\T} \ .
\label{NW6globalcoords}\eeq
In these coordinates one can work out the adjoint action ${\rm
Ad}_{g(x^+,x^-,\mz,\overline{\mz}\,)}\,g(x^+_0,x^-_0,\mz^{~}_0,
\overline{\mz}^{~}_0)$ corresponding to a fixed point
$(x_0^+,x_0^-,\mz^{~}_0)\in\NW_6$~\cite{FS1,SF1}, and the conjugacy
classes can be written explicitly as the submanifolds
\bea
&&\mathcal{C}_{(x_0^+,x^-_0,\mz^{~}_0)}\nonumber\\&&~~~~=~
\scriptstyle{\left.\Bigl\{\bigl(x^+_0\,,\,x^-_0-\frac\mu2
\,|\mz|^2\,\sin\mu\,x^+_0+\mu~{\rm Im}\bigl[
\mz_0^\top\,\overline{\mz}~\e^{\frac{\ii\mu}2\,x^+}\,
\cos\frac{\mu\,x_0^+}2\bigr]\,,\,\e^{\ii\mu\,x^+}\,\mz^{~}_0-2\ii
\sin\frac{\mu\,x_0^+}2~\e^{\frac{\ii\mu}2\,x^+}\,\mz
\bigr)~\right|~{}^{x^+\in\S^1}_{\mz\in\complex^2}\Bigr\}}\nonumber\\&&
\label{conjclassgen}\eea
where we have used periodicity to restrict the light-cone time coordinate to
$x^+,x^+_0\in\S^1=\real\,/\,2\pi\,\mu^{-1}\,\zed$. The null planes $x^+=x_0^+$
are thus invariants of the conjugacy classes and can be used
to distinguish the different D-submanifolds of $\NW_6$.
There are two types of euclidean branes generically associated to these
conjugacy classes that we shall now proceed to describe in detail. In
each case we first describe the classical geometry, and then proceed
to quantize the orbits.

\subsection{Null Branes\label{Null}}

We begin with the ``degenerate'' cases where $x_0^+=0$.

\subsubsection*{Classical Geometry}

When $\mz_0={\mbf0}$, the conjugacy class (\ref{conjclassgen})
corresponds to a D-instanton sitting at the point
$(0,x_0^-,\mbf0)\in\NW_6$. When
$\mz_0^\top:=(z^{~}_0,w^{~}_0)\neq\mbf0$ we denote the conjugacy
classes by $\mathcal{C}^{~}_{|z_0|,|w_0|}:=\mathcal{C}_{(0,x_0^-,\mz^{~}_0)}$.
Since
$\mathcal{C}_{|z_0|,|w_0|}=\{(0,x^-,\e^{\ii\mu\,x^+}\,\mz_0~|~
x^-\in\real,x^+\in\S^1\}\cong\real\times\S^1$, the resulting object
may be thought of as a cylindrical brane extended along the null
light-cone direction $x^-$ with fixed radii $|z_0|,|w_0|$ in the two
transverse planes $(z_0,w_0)\in\complex^2$, and will therefore be
refered to as a ``null'' brane. However, the quantization of the
classical worldvolume will generally turn out to depend crucially on
the radii. We can understand this dependence heuristically as
follows. Generally, the Cahen-Wallach metric
(\ref{NW6metricBrink}) possesses an ${\rm SO}(4)$ rotational symmetry of its
transverse space $\mz\in\complex^2\cong\real^4$, which is broken to
${\rm U}(2)$ by the Neveu-Schwarz background (\ref{NW6Bfield}). When
$|z_0|=0$ or $|w_0|=0$, one has
$\mathcal{C}_{|z_0|,0}\cong\mathcal{C}_{0,|w_0|}\cong\real\times\S^1$.
However, when both $|z_0|$ and $|w_0|$ are non-zero, we may use this ${\rm
U}(2)$ symmetry to rotate $w_0\mapsto\e^{\ii\phi}\,w_0$ at fixed $z_0$ by an
arbitrary phase. In this case the cylindrical brane is parametrized effectively
by {\it two} independent periodic coordinates, and one has
\bea
\mathcal{C}_{|z_0|,|w_0|}&=&~\bigl.\bigl\{(0\,,\,x^-\,,\,\e^{\ii\mu\,
  x^+}\,z_0\,,\,\e^{\ii\mu\,y^+}\,w_0)~\bigr|~x^-\in\real\,,\,x^+,y^+
\in\S^1\bigr\}\nn\\&\cong&\left\{\begin{matrix}{\rm point}~~,~~z_0=w_0=0
    \\ \real\times\S^1\times\S^1~~,~~z_0,w_0\neq0 \\
\real\times\S^1~~,~~{\rm otherwise} \ . \end{matrix}\right.
\label{cylnullbrane}\eeq
In each case it carries a degenerate metric
\beq
\dd s_6^2\bigm|_{\mathcal{C}_{|z_0|,|w_0|}}=\mu^2\,|z_0|^2~\left(\dd
  x^+\right)^2+\mu^2\,|w_0|^2~\left(\dd y^+\right)^2 \ .
\label{degmetricnull}\eeq
Such a brane generically has no straightforward interpretation as a
D-brane, as the corresponding Dirac-Born-Infeld action is
ill-defined. Nevertheless, it will play a role in our ensuing analysis
and so we shall analyse it in some detail.

When $\mz_0\neq{\mbf0}$, the centralizer of
$g^{~}_0:=g(0,x^-_0,\mz^{~}_0,\overline{\mz}^{~}_0)$ is the subgroup
of $\mathcal{N}_6$ parametrized as the submanifold
\beq
\mathcal{Z}_{|z_0|,|w_0|}~=~\left.\bigl\{(0\,,\,x^-\,,\,\mz)~\right|~
{\rm Im}~\mz_0^\top\,\overline{\mz}=0\bigr\}~\cong~\real^4
\label{nullcentr}\eeq
with a degenerate metric, while
$\mathcal{Z}_{0,0}\cong\mathcal{N}_6$. The Neveu-Schwarz fields vanish
on the brane,
\beq
H_6\bigm|_{\mathcal{C}_{|z_0|,|w_0|}}=
B_6\bigm|_{\mathcal{C}_{|z_0|,|w_0|}}=0 \ ,
\label{H6B6null0}\eeq
while an elementary calculation using (\ref{NW4algdef}),
(\ref{NW4innerprod}), (\ref{NW4CMform}) and
(\ref{nullcentr}) shows that the abelian gauge field fluxes
(\ref{2formfamily}) also vanish on the null brane worldvolume,
\beq
F^{(\zeta)}_6\bigm|_{\mathcal{C}_{|z_0|,|w_0|}}=0
\label{nullF60}\eeq
for any
$\zeta=\mu^{-1}\,x^-\,\T+\mz^\top\,\underline{\P}^++
\overline{\mz}{}^{\,\top}\,\underline{\P}^-$ in the tangent space to
the centralizer (\ref{nullcentr}). This
suggests that these conjugacy classes should describe {\it
  commutative} branes, and we shall now demonstrate explicitly that
this is indeed the case. Since the $H$-field vanishes on the null
branes, we can use the standard coadjoint orbit method as discussed
earlier.

\subsubsection*{Noncommutative Geometry}

Let us now describe the algebra of functions (\ref{NCalgconjclass}) on
the null brane worldvolume. The Lie algebra $\mathfrak{n}_6$
has three types of unitary irreducible representations
$\mathcal{D}^{p^+,p^-}:U(\mathfrak{n}_6)\to{\rm End}(V^{p^+,p^-})$ labelled by
light-cone momenta $p^\pm\in\real$~\cite{BAKZ1,CFS1,KK1}. On each
module $V^{p^+,p^-}$, elements of the center of the universal
enveloping algebra $U(\mathfrak{n}_6)$ are proportional to the
identity operator $\id$. In particular, the central element $\T$ acts
as
\beq
\mathcal{D}^{p^+,p^-}(\T)=\ii\mu\,p^+\,\id \ .
\label{Tirrepgen}\eeq
According to (\ref{utaudef}), the modules corresponding to the null
branes live in the class $V^{0,p^-}_{\alpha,\beta}$,
$\alpha,\beta\in[0,\infty)$ of continuous representations with
$p^+=0$. In this case, in addition to $\T$ there are {\it two} other
Casimir operators corresponding to the quadratic elements $\P^+\,\P^-$
and $\Q^+\,\Q^-$ of $U(\mathfrak{n}_6)$ with the eigenvalues
\beq
\mathcal{D}^{0,p^-}_{\alpha,\beta}\left(\P^+\,\P^-\right)=-\alpha^2\,\id \ , ~~
\mathcal{D}^{0,p^-}_{\alpha,\beta}\left(\Q^+\,\Q^-\right)=-\beta^2\,\id \ .
\label{otherCasimirs}\eeq

As a separable Hilbert space, the module
$V^{0,p^-}_{\alpha,\beta}$ may be expressed as the linear span
\beq
V^{0,p^-}_{\alpha,\beta}=\bigoplus_{n,m\in\zed}\complex\cdot
\bigl|\,{}^n_\alpha\,{}^m_\beta\,;\,p^-\bigr\rangle
\label{contrepbasis}\eeq
with the non-trivial actions of the generators of $\mathfrak{n}_6$
given by
\bea
\mathcal{D}^{0,p^-}_{\alpha,\beta}\left(\P^\pm\right)
\bigl|\,{}^n_\alpha\,{}^m_\beta\,;\,p^-\bigr\rangle
&=&\ii\alpha\,\bigl|\,{}^{n\mp1}_\alpha\,{}^m_\beta\,;\,p^-\bigr\rangle
\ , \nonumber\\\mathcal{D}^{0,p^-}_{\alpha,\beta}\left(\Q^\pm\right)
\bigl|\,{}^n_\alpha\,{}^m_\beta\,;\,p^-\bigr\rangle
&=&\ii\beta\,\bigl|\,{}^n_\alpha\,{}^{m\mp1}_\beta\,;\,p^-\bigr\rangle
\ , \nonumber\\\mathcal{D}^{0,p^-}_{\alpha,\beta}\bigl(\J\,
\bigr)\bigl|\,{}^n_\alpha\,{}^m_\beta\,;\,p^-\bigr\rangle
&=&\ii\left(\mu^{-1}\,p^--n-m\right)\,\bigl|\,{}^n_\alpha\,{}^m_\beta
\,;\,p^-\bigr\rangle \ .
\label{contrepactions}\eea
The inner product is defined such that the basis (\ref{contrepbasis})
is orthonormal,
\beq
\bigl\langle\,{}^n_\alpha\,{}^m_\beta\,;\,p^-\bigm|\,
{}^{n'}_\alpha\,{}^{m'}_\beta\,;\,p^-\bigr\rangle=
\delta_{nn'}\,\delta_{mm'} \ .
\label{contreportho}\eeq
These representations have no highest or lowest weight states. Note
that shifting the labels $n$ or $m$ by $1$ gives an equivalent
representation
\beq
V^{0,p^-}_{\alpha,\beta}~\cong~V^{0,p^-+\mu}_{\alpha,\beta} \ .
\label{contrepinequiv}\eeq
In other words, only the representations $V^{0,p^-}_{\alpha,\beta}$
with $p^-\in[0,\mu)$ are inequivalent. This is analogous to the
periodicity constraint imposed on the light-cone coordinate $x^+$ in
(\ref{conjclassgen}).

Among these representations is the trivial one-dimensional
representation $V_{0,0}^{0,0}$ which corresponds to the D-instantons
found above. To generically associate them with the null brane worldvolumes
obtained before, we use the correspondence between Casimir elements
and class functions on the group, which are respectively constant in
irreducible representations and on conjugacy classes. Note that the
geodesics in $\NW_6$ obey $p^+=\dot x^+$ and $p^-=\dot
x^--\frac{\mu^2}4\,|\mz|^2\,p^+$
(c.f.~(\ref{masslessLag})). At $p^+=0$, we may thereby identify the
parameters $\alpha$ and $\beta$ with the radii $|z_0|$ and $|w_0|$ in
the two transverse planes $(z_0,w_0)$, while $p^-$ is identified with
a fixed value of the light-cone position $x^-$. The quantized
algebra of functions on the conjugacy class (\ref{cylnullbrane}) is
thus given by
\beq
\mathcal{A}\left(\mathcal{C}_{|z_0|,|w_0|}\right)=
{\rm End}\bigl(V_{|z_0|,|w_0|}^{0,p^-}\bigr) \ ,
\label{quantalgnullbrane}\eeq
where a generic element $\hat f\in{\rm End}(V_{\alpha,\beta}^{0,p^-})$
is a complex linear combination
\beq
\hat
f=\sum_{n,m,n',m'\in\zed}f_{n,m;n',m'}~\bigl|\,{}^n_\alpha\,{}^m_\beta
\,;\,p^-\bigr\rangle\bigl\langle\,{}^{n'}_\alpha\,{}^{m'}_\beta
\,;\,p^-\bigr| \ .
\label{contreplincomb}\eeq
In this context, we may appropriately regard the conjugacy
class (\ref{cylnullbrane}) as the worldvolume of a symmetric
D$p$-brane with $p=-1,1,0$ and volume element
(\ref{degmetricnull}). Because of (\ref{contrepinequiv}), in the
quantum geometry the light-cone position is restricted to a finite
interval $x^-\in[0,\mu)$.

This identification can be better understood by introducing the
coherent states
\beq
\bigl|\,{}^{x^+}_\alpha\,{}^{y^+}_\beta\,;\,p^-
\bigr\rangle:=\sum_{n,m\in\zed}
\e^{\ii n\,\mu\,x^++\ii m\,\mu\,y^+}~\bigl|\,{}^n_\alpha\,{}^m_\beta
\,;\,p^-\bigr\rangle \ , ~~ x^+,y^+\in\S^1 \ ,
\label{contrepcohstates}\eeq
on which the non-trivial symmetry generators are represented as
differential operators
\bea
\mathcal{D}^{0,p^-}_{\alpha,\beta}\left(\P^\pm\right)
\bigl|\,{}^{x^+}_\alpha\,{}^{y^+}_\beta\,;\,p^-\bigr\rangle
&=&\ii\alpha~\e^{\pm\ii\mu\,x^+}\,\bigl|\,{}^{x^+}_\alpha\,{}^{y^+}_\beta
\,;\,p^-\bigr\rangle \ , \nonumber\\
\mathcal{D}^{0,p^-}_{\alpha,\beta}\left(\Q^\pm\right)
\bigl|\,{}^{x^+}_\alpha\,{}^{y^+}_\beta\,;\,p^-\bigr\rangle
&=&\ii\beta~\e^{\pm\ii\mu\,y^+}\,\bigl|\,{}^{x^+}_\alpha\,{}^{y^+}_\beta
\,;\,p^-\bigr\rangle \ , \nonumber\\
\mathcal{D}^{0,p^-}_{\alpha,\beta}\bigl(\J\,\bigr)\bigl|\,
{}^{x^+}_\alpha\,{}^{y^+}_\beta\,;\,p^-\bigr\rangle
&=&\ii\mu^{-1}\,\left(p^--\mbox{$\frac\partial{\partial x^+}-
\frac\partial{\partial y^+}$}\right)\,\bigl|\,{}^{x^+}_\alpha\,
{}^{y^+}_\beta\,;\,p^-\bigr\rangle \ .
\label{contrepcohactions}\eea
The conjugate state to (\ref{contrepcohstates}) is given by
\beq
\bigl\langle\,{}^{x^+}_\alpha\,{}^{y^+}_\beta\bigr|=\sum_{n,m\in\zed}
\e^{-\ii n\,\mu\,x^+-\ii
  m\,\mu\,y^+}~\bigl\langle\,{}^n_\alpha\,{}^m_\beta
\,;\,p^-\bigr| \ ,
\label{contrepconjstate}\eeq
so that the inner product is
\beq
\bigl\langle\,{}^{x_1^+}_\alpha\,{}^{y_1^+}_\beta\,;\,p^-
\bigm|{}^{x_2^+}_\alpha\,{}^{y_2^+}_\beta\,;\,p^-\bigr\rangle=\delta\left(
x_1^+-x_2^+\right)\,\delta\left(y_1^+-y_2^+\right)
\label{contrepinnerprod}\eeq
while the resolution of unity is
\beq
\id~=~\mu^2\,\int\limits_0^{2\pi\,\mu^{-1}}\frac{\dd x^+}{2\pi}~
\int\limits_0^{2\pi\,\mu^{-1}}\frac{\dd y^+}{2\pi}~
\bigl|\,{}^{x^+}_\alpha\,{}^{y^+}_\beta\,;\,p^-\bigr\rangle\bigl
\langle\,{}^{x^+}_\alpha\,{}^{y^+}_\beta\,;\,p^-\bigr| \ .
\label{contrepresunity}\eeq
Thus the states (\ref{contrepcohstates}) form an overcomplete basis
for the Fock space $V^{0,p^-}_{\alpha,\beta}$. The metric
and Kirillov-Kostant symplectic two-form on the orbit can be computed
as the matrix elements in the states (\ref{contrepcohstates}) of the
operators $|\mathcal{D}_{|z_0|,|w_0|}^{0,p^-}(g^{-1}~\dd g)|^2$ and
$\mathcal{D}_{|z_0|,|w_0|}^{0,p^-}([g^{-1}~\dd g\,,\,g^{-1}~\dd g])$,
respectively. They coincide with those computed above as the
pull-backs of the Nappi-Witten geometry to the conjugacy classes
(\ref{cylnullbrane}).

We may now attempt to view the worldvolume algebra
(\ref{quantalgnullbrane}) as a deformation of the classical algebra of
functions on the conjugacy class (\ref{cylnullbrane}). On the null
hyperplanes $x^-={\rm constant}$, we define an isomorphism of
underlying vector spaces
\beq
\Delta\,:\,{\rm C}^\infty\left(\mathcal{C}_{|z_0|,|w_0|}
\right)~\longrightarrow~{\rm End}\bigl(V_{|z_0|,|w_0|}^{0,p^-}\bigr)
\label{Deltanullbranes}\eeq
in the following manner. Decomposing any smooth function $f\in{\rm
  C}^\infty(\mathcal{C}_{|z_0|,|w_0|})$ as a Fourier series over
the torus $\S^1\times\S^1$ given by
\beq
f\left(x^+,y^+\right)=\sum_{n,m\in\zed}f_{n,m}~\e^{\ii n\,\mu\,x^+
+\ii m\,\mu\,y^+} \ ,
\label{Fourierfnull}\eeq
we write
\beq
\hat f=\Delta(f):=\sum_{n,m\in\zed}f_{n,m}~
\Bigl(\mbox{$\frac1{\ii|z_0|}$}\,\mathcal{D}^{0,p^-}_{|z_0|,|w_0|}
\bigl(\P^{\varepsilon(n)}\bigr)\Bigr)^{|n|}\,\Bigl(
\mbox{$\frac1{\ii|w_0|}$}\,\mathcal{D}^{0,p^-}_{|z_0|,|w_0|}
\bigl(\Q^{\varepsilon(m)}\bigr)\Bigr)^{|m|}
\label{hatfDeltanull}\eeq
where the label $\varepsilon(n)=\pm$ corresponds to the sign of
$n\in\zed$. The inverse map is given by
\bea
f\left(x^+,y^+\right)&:=&\Delta^{-1}\bigl(\hat f\,\bigr)\left(x^+,y^+
\right)\nn\\&=&\mu^2\,\int\limits_0^{2\pi\,\mu^{-1}}\frac{\dd\tilde x^+}{2\pi}~
\int\limits_0^{2\pi\,\mu^{-1}}\frac{\dd\tilde y^+}{2\pi}~
\bigl\langle\,{}^{\tilde x^+}_{|z_0|}\,{}^{\tilde y^+}_{|w_0|}\,
;\,p^-\bigr|~\hat f~\bigl|\,{}^{x^+}_{|z_0|}\,{}^{y^+}_{|w_0|}\,;
\,p^-\bigr\rangle \ .
\label{invDeltanull}\eea
As expected, from (\ref{contrepcohactions}) one finds that the
functions corresponding to the generators of $\mathfrak{n}_6$ on
$V_{|z_0|,|w_0|}^{0,p^-}$ coincide with the coordinates of the
conjugacy classes (\ref{cylnullbrane}),
\bea
\Delta^{-1}\Bigl(\mathcal{D}^{0,p^-}_{|z_0|,|w_0|}\left(\P^\pm\right)
\Bigr)\left(x^+,y^+\right)&=&\ii|z_0|~\e^{\pm\ii \mu\,x^+} \ , \nn\\
\Delta^{-1}\Bigl(\mathcal{D}^{0,p^-}_{|z_0|,|w_0|}\left(\Q^\pm\right)
\Bigr)\left(x^+,y^+\right)&=&\ii|w_0|~\e^{\pm\ii \mu\,y^+} \ , \nn\\
\Delta^{-1}\Bigl(\mathcal{D}^{0,p^-}_{|z_0|,|w_0|}\bigl(\J\,\bigr)
\Bigr)\left(x^+,y^+\right)&=&\ii\mu^{-1}\,p^- \ .
\label{nullfnsgens}\eeq
Generically, the classical conjugacy class (\ref{conjclassgen}) in
this case corresponds to the diagonal subspace $x^+=y^+$ of the space
spanned by the coherent states (\ref{contrepcohstates}).

Since the operator (\ref{hatfDeltanull}) is diagonal in the basis
(\ref{contrepcohstates}) with eigenvalue $f(x^+,y^+)$, we easily
establish that in this case the map (\ref{Deltanullbranes})
is in fact an {\it algebra} isomorphism. Namely, the product of two
operators $\hat f$ and $\hat g$ on $V_{|z_0|,|w_0|}^{0,p^-}$
corresponds to the pointwise multiplication of the associated
functions on $\mathcal{C}_{|z_0|,|w_0|}$,
\beq
\Delta^{-1}\bigl(\hat f\,\hat g\,\bigr)\left(x^+,y^+\right)=
f\left(x^+,y^+\right)\,g\left(x^+,y^+\right) \ .
\label{commprodnull}\eeq
Thus the worldvolume algebra (\ref{quantalgnullbrane}) in the present
case describes a {\it commutative} geometry on the null branes, in
agreement with the classical analysis. This is also in accord with the
fact that the $p^+=0$ sector of the dynamics in Nappi-Witten spacetime
does not feel the harmonic oscillator potential of
Section~\ref{Dynamics}, and thereby describes free motion in the
transverse space.

\subsection{D3-Branes\label{CGED3B}}

Let us now turn to the somewhat more interesting cases with
$x_0^+\neq0$, whereby the classical worldvolume geometries are
nondegenerate.

\subsubsection*{Classical Geometry}

In this case the conjugacy class (\ref{conjclassgen}) can be written
after a trivial coordinate redefinition as
\beq
\mathcal{C}_{(x_0^+,x_0^-,\mz_0)}~=~
\bigl\{(x_0^+\,,\,x_0^-+\mbox{$\frac\mu4$}\,\left(|\mz_0|^2-|\mz'\,|^2
\right)\,\cot\mbox{$\frac{\mu\,x_0^+}2$}\,,\,\mz'\,)~\bigm|~\mz'
\in\complex^2\bigr\}~\cong~\eucl^4 \ .
\label{conjpieucl4}\eeq
It may be labelled as $\mathcal{C}_{x_0^+,\chi^{~}_0}$, with
$\chi^{~}_0:=x_0^-+\frac\mu4\,|\mz_0|^2\,\cot\frac{\mu\,x_0^+}2$, so
that the corresponding branes have four-dimensional worldvolumes
located at
\beq
x^+=x_0^+ \ , ~~ x^-=\chi^{~}_0-\mbox{$\frac{\mu}4$}\,
|\mz|^2\,\cot\mbox{$\frac{\mu\,x_0^+}2$} \ .
\label{E4branesloc}\eeq
The worldvolume metric is non-degenerate and given by
\beq
\left.\dd s_6^2\right|_{\mathcal{C}_{x_0^+,\chi_0^{~}}}=|\dd\mz|^2 \ ,
\label{metricpieucl4}\eeq
so that in this case the conjugacy classes are wrapped by flat euclidean
D3-branes. The NS fields on the worldvolume are given by flat space
forms
\beq
H_6\bigl|_{\mathcal{C}_{x_0^+,\chi_0^{~}}}=0 \ , ~~
B_6\bigl|_{\mathcal{C}_{x_0^+,\chi^{~}_0}}=-2\ii\mu\,
x_0^+~\dd\overline{\mz}^{\,\top}\wedge\dd\mz \ .
\label{NSconj}\eeq
The vanishing of the $H$-flux again means that we can apply standard
semi-classical quantization techniques to these D-branes.

If $x_0^+=\frac\pi\mu$ then these branes, like the null branes, are
associated with the conjugate points of the Rosen plane wave geometry
defined by (\ref{Rosen})--(\ref{Cumatrix}). In this case the conjugacy
class $\mathcal{C}_{x_0^-}:=\mathcal{C}_{(\frac\pi\mu,x_0^-,\mz_0)}$ is a
four-plane labelled by the fixed light-cone position $x^-=x_0^-$. When
$x_0^+\neq0,\frac\pi\mu$, the brane worldvolume is a paraboloid
corresponding to a point-like object travelling at the speed of light
while simultaneously expanding or contracting in a three-sphere in the
transverse space $\mz\in\complex^2$ according to (\ref{E4branesloc}). Since
these branes lie in the set of conjugate-free points, we may analyse
them using the Penrose-G\"uven limit of Section~\ref{ApplNW}, which
yields a non-vanishing gauge invariant two-form field via pull-back
of (\ref{NW6tildegaugeinv}) to the conjugacy classes as
\beq
\mathcal{F}_{x_0^+}:=\mathcal{F}_6\bigl|_{\mathcal{C}_{x_0^+,\chi_0^{~}}}
=-\ii\cot\mbox{$\frac{\mu\,x_0^+}2$}~\dd\overline{\mz}^{\,\top}\wedge\dd\mz
\ .
\label{E4F6PGL}\eeq
Thus these D-branes are expected to carry a noncommutative geometry
for $x_0^+\neq0,\frac\pi\mu$. Extrapolating (\ref{E4F6PGL}) to
$x_0^+=\frac\pi\mu$ shows that the conjugacy classes
$\mathcal{C}_{x_0^-}$ are expected to support a commutative
worldvolume geometry like the null branes, despite their non-vanishing
$B$-field.

The Penrose-G\"uven limit can also be used to understand the physical
origin of the euclidean D3-branes. They may be described
through the commuting isometric embedding diagram
\begin{equation}
  \label{E4commdiag}
  \begin{CD}
    \AdS_3 \times \S^3             @>\text{PGL}>> \NW_6\\
    \text{$\imath^{\,\prime}~~$}@AAA
@AAA\text{$\widetilde{\imath}^{~\prime}$}\\
    \AdS_2 \times \S^2             @>\text{PGL$^{\,\prime}$}>> \eucl^4\\
  \end{CD}
\end{equation}
where the primes indicate that the limit is taken along a null geodesic
which does {\it not} pass through the $\AdS_2\times\S^2$ brane
worldvolume, i.e. the embeddings $\imath^{\,\prime}$ and
$\widetilde{\imath}^{~\prime}$ are defined by the constant time
slices $\tau,\phi={\rm constant}$ and $x^+={\rm constant}$, respectively. Thus
the resulting geometry is not of plane wave type. The brane on the
left-hand side of (\ref{E4commdiag}) originates as a flat D3-brane
connected orthogonally to a distant NS5/F1 black string by a stretched
$(p,q)$ string~\cite{BP1}, the origin of the worldvolume flux
(\ref{E4F6PGL}). As the $(p,q)$ string pulls the flat D3-brane, it
deforms its worldvolume geometry, leading to an $\AdS_2\times\S^2$
brane in the near-horizon region of the black string. The
Penrose-G\"uven limit in (\ref{E4commdiag}) pulls the deformed branes
away into the asymptotically flat region of the black string,
decompactifying it as $R\to\infty$ onto the flat euclidean D3-brane
on the right-hand side of the isometric embedding diagram. However,
since the diagram commutes, the standard fuzzy geometry of the
$\AdS_2\times\S^2$ brane induces, through the usual scaling limit~\cite{CMS1},
a
Moyal type noncommutative geometry on the instantonic $\eucl^4$
brane. This argument is consistent with the fact that the (modified)
Penrose-G\"uven limit is also a map between symmetric D-branes. We
shall now substantiate this physical picture through explicit
computation of the quantized worldvolume geometry.

\subsubsection*{Noncommutative Geometry}

To describe the algebra of functions (\ref{NCalgconjclass}) on the
D3-brane worldvolume, we start with the irreducible representations
$\mathcal{D}^{p^+,p^-}:U(\mathfrak{n}_6)\to{\rm End}(V^{p^+,p^-})$
having $p^+>0$. In addition to (\ref{Tirrepgen}), the quadratic
Casimir element ${\sf
  C}_6=2\,\J\,\T+\frac12\,[(\,\Pu^+)^\top\,\Pu^-+(\,\Pu^-)^\top\,\Pu^+]$
acts as a scalar operator
\beq
\mathcal{D}^{p^+,p^-}(\C_6)=-2p^+\,\left(p^-+\mu\right)\,\id \ .
\label{C6irrep}\eeq
This operator is positive for all $p^-\in(-\infty,-\mu)$. As a
separable Hilbert space, the module $V^{p^+,p^-}$ may be
exhibited as the linear span
\beq
V^{p^+,p^-}=\bigoplus_{n,m\in\nat_0}\complex\cdot\bigl|n,m;p^+,p^-
\bigr\rangle
\label{Vhighestex}\eeq
with the non-trivial actions of the Nappi-Witten generators given by
\bea
\mcDp\left(\P^+\right)\bigl|n,m;p^+,p^-\bigr\rangle&=&
2\ii\mu\,p^+\,n\,\bigl|n-1,m;p^+,p^-\bigr\rangle \ , \nonumber\\
\mcDp\left(\P^-\right)\bigl|n,m;p^+,p^-\bigr\rangle&=&
\ii\bigl|n+1,m;p^+,p^-\bigr\rangle \ , \nonumber\\
\mcDp\left(\Q^+\right)\bigl|n,m;p^+,p^-\bigr\rangle&=&
2\ii\mu\,p^+\,m\,\bigl|n,m-1;p^+,p^-\bigr\rangle \ , \nonumber\\
\mcDp\left(\Q^-\right)\bigl|n,m;p^+,p^-\bigr\rangle&=&
\ii\bigl|n,m+1;p^+,p^-\bigr\rangle \ , \nonumber\\
\mcDp\bigl(\J\,\bigr)\bigl|n,m;p^+,p^-\bigr\rangle&=&
\ii\left(\mu^{-1}\,p^--n-m\right)
\bigl|n,m;p^+,p^-\bigr\rangle \ .
\label{highestactions}\eea
The inner product on the basis (\ref{Vhighestex}) is
\beq
\bigl\langle n,m;p^+,p^-\bigm|n',m';p^+,p^-\bigr\rangle=
\left(2\mu\,p^+\right)^{n+m}\,n!\,m!~\delta_{nn'}\,\delta_{mm'} \ .
\label{highestortho}\eeq
This representation admits a highest weight state
$|0,0;p^+,p^-\rangle$ on which $-\ii\mcDp(\J)$ has
weight~$\mu^{-1}\,p^-$.

To associate these representations with euclidean D3-brane
worldvolumes, we note that the constraint on the light-cone position
in (\ref{E4branesloc}) can be written in the semi-classical limit
$\mu\to0$ as
\beq
2x_0^+\,x^-+|\mz|^2=2x_0^+\,\chi^{~}_0 \ .
\label{semiclasspos}\eeq
The relation (\ref{semiclasspos}) agrees with the Casimir eigenvalue
constraint (\ref{C6irrep}) under the identifications of the light-cone
time $x_0^+$ with momentum $p^+$ as before and the class variable
$\chi^{~}_0$ with $p^-+\mu$. We will soon identify the worldvolume coordinates
$\mz\in\complex^2$ with the operators $\mcDp(\,\Pu^+)$. Thus the quantized
algebra of functions on the conjugacy class is given by
\beq
\mathcal{A}\bigl(\mathcal{C}_{x_0^+,\chi^{~}_0}\bigr)=
{\rm End}\bigl(V^{p^+,p^-}\bigr) \ ,
\label{quantalgD3}\eeq
where a generic element $\hat f\in{\rm End}(V^{p^+,p^-})$ is a complex
linear combination
\beq
\hat f=\sum_{n,m,n',m'\in\nat_0}f_{n,m;n',m'}~\bigl|n,m;p^+,p^-
\bigr\rangle\bigl\langle n',m';p^+,p^-\bigr| \ .
\label{genlincombD3}\eeq

As before, it is convenient to work in the conventional coherent state
basis of the Fock module $V^{p^+,p^-}$ defined for
$\mz^\top=(z,w)\in\complex^2$ by
\beq
\bigl|\mz;p^+,p^-\bigr\rangle:=\e^{-\mz^\top\,\Pu^-}
\bigl|0,0;p^+,p^-\bigr\rangle=\sum_{n,m\in\nat_0}
\frac{z^n\,w^m}{\ii^{n+m}\,n!\,m!}~\bigl|n,m;p^+,p^-
\bigr\rangle \ ,
\label{D3cohstates}\eeq
on which the non-trivial symmetry generators are represented by
differential operators
\bea
\mcDp\left(\,\Pu^+\right)\bigl|\mz;p^+,p^-\bigr\rangle&=&
2\mu\,p^+\,\mz\,\bigl|\mz;p^+,p^-\bigr\rangle \ , \nonumber\\
\mcDp\left(\,\Pu^-\right)\bigl|\mz;p^+,p^-\bigr\rangle&=&
-\mdell\bigl|\mz;p^+,p^-\bigr\rangle \ , \nonumber\\
\mcDp\bigl(\J\,\bigr)\bigl|\mz;p^+,p^-\bigr\rangle&=&
\ii\left(\mu^{-1}\,p^--\mz^\top\,\mdell\right)
\bigl|\mz;p^+,p^-\bigr\rangle
\label{symactioncohD3}\eea
with $\mdell^\top:=(\frac\partial{\partial z},\frac\partial{\partial
  w})$. The conjugate state to (\ref{D3cohstates}) is given by
\beq
\bigl\langle\mz;p^+,p^-\bigr|=\bigl\langle0,0;p^+,p^-\bigr|
\e^{\overline{\mz}{}^{\,\top}\,\Pu^+} \ ,
\label{conjD3cohstates}\eeq
so that the inner product of coherent states is
\beq
\bigl\langle\mz;p^+,p^-\bigm|\mz';p^+,p^-\bigr\rangle=
\e^{2\mu\,p^+\,\overline{\mz}{}^{\,\top}\,\mz'}
\label{innprodcohD3}\eeq
while their completeness relation is
\beq
\id~=~\left(2\mu\,p^+\right)^2\,\int\limits_{\complex^2}
\dd\varrho\left(\mz,\overline{\mz}\,\right)~\e^{-2\mu\,p^+\,|\mz|^2}~
\bigl|\mz;p^+,p^-\bigr\rangle\bigl\langle\mz;p^+,p^-\bigr|
\label{D3complrel}\eeq
where $\dd\varrho(\mz,\overline{\mz}\,)=\frac1{\pi^2}~|\dd
z\wedge\dd\overline{z}\wedge\dd w\wedge\dd\overline{w}\,|$ is the
standard flat measure on $\eucl^4\cong\complex^2$. We will now use
these states to describe the worldvolume algebra (\ref{quantalgD3}) as
a deformation of the classical algebra of functions on the conjugacy
class $\mathcal{C}_{x_0^+,\chi^{~}_0}$.

For this, we construct an isomorphism of underlying vector spaces
\beq
\Delta_*\,:\,{\rm C}^\infty\bigl(\mathcal{C}_{x_0^+,\chi^{~}_0}\bigr)
{}~\longrightarrow~{\rm End}\bigl(V^{p^+,p^-}\bigr)
\label{Deltastar}\eeq
by expanding any smooth function $f\in{\rm
  C}^\infty(\mathcal{C}_{x_0^+,\chi^{~}_0})$ as a Taylor series over
$\complex^2$ given by
\beq
f\left(\mz,\overline{\mz}\,\right)=\sum_{n,m,n',m'\in\nat_0}f_{n,m;n',m'}~
z^n\,w^m\,\overline{z}{}^{\,n'}\,\overline{w}{}^{\,m'}
\label{TaylorD3}\eeq
and defining
\bea
\hat f~=~\Delta_*(f)&:=&\sum_{n,m,n',m'\in\nat_0}\,\frac{f_{n,m;n',m'}}
{\left(2\mu\,p^+\right)^{n+m+n'+m'}}~\mcDp\left(\P^-\right)^{n'}\,
\mcDp\left(\Q^-\right)^{m'}\nonumber\\ &&\times\,\mcDp\left(\P^+\right)^{n}\,
\mcDp\left(\Q^+\right)^{m} \ .
\label{DeltafD3}\eea
The inverse map is provided by the normalized matrix elements of
operators as
\beq
f\left(\mz,\overline{\mz}\,\right):=\Delta_*^{-1}\bigl(\hat f\,
\bigr)\left(\mz,\overline{\mz}\,\right)=\e^{-2\mu\,p^+\,|\mz|^2}\,
\bigl\langle\mz;p^+,p^-\bigl|~\hat f~\bigr|\mz;p^+,p^-\bigr\rangle \ .
\label{DeltainvD3}\eeq
As before, the functions corresponding to the generators of the
Nappi-Witten algebra $\mathfrak{n}_6$ coincide with the coordinates of
the conjugacy classes $\mathcal{C}_{x_0^+,\chi^{~}_0}$,
\bea
\Delta_*^{-1}\Bigl(\mcDp\left(\,\Pu^+\right)\Bigr)\left(\mz,
\overline{\mz}\,\right)&=&2\mu\,p^+\,\mz \ , \nonumber\\
\Delta_*^{-1}\Bigl(\mcDp\left(\,\Pu^-\right)\Bigr)\left(\mz,
\overline{\mz}\,\right)&=&2\mu\,p^+\,\overline{\mz} \ , \nonumber\\
\Delta_*^{-1}\Bigl(\mcDp\bigl(\J\,\bigr)\Bigr)\left(\mz,
\overline{\mz}\,\right)&=&
\ii\left(\mu^{-1}\,p^--2\mu\,p^+\,|\mz|^2\right) \ .
\label{DeltainvD3gens}\eea

The map (\ref{Deltastar}) in this case is {\it not} an algebra homomorphism
and it can be used to deform the pointwise multiplication of functions
in the algebra ${\rm C}^\infty(\mathcal{C}_{x_0^+,\chi^{~}_0})$, giving
an associative star-product defined by
\bea
(f*g)\left(\mz,\overline{\mz}\,\right)&:=&
\Delta_*^{-1}\bigl(\hat f\,\hat g\,\bigr)
\left(\mz,\overline{\mz}\,\right)~=~\e^{-2\mu\,p^+\,|\mz|^2}\,
\bigl\langle\mz;p^+,p^-\bigl|~\hat f\,\hat g~\bigr|\mz;p^+,p^-
\bigr\rangle\nonumber\\ &=&\left(2\mu\,p^+\right)^2\,
\int\limits_{\complex^2}\dd\varrho\left(\mz',\overline{\mz}{}^{\,\prime}\,
\right)~\e^{-2\mu\,p^+\,(|\mz'|^2+|\mz|^2)}\nonumber\\
&&\times\,\bigl\langle\mz;p^+,p^-
\bigl|~\hat f~\bigr|\mz';p^+,p^-\bigr\rangle\,\bigl\langle
\mz';p^+,p^-\bigl|~\hat g~\bigr|\mz;p^+,p^-\bigr\rangle \ .
\label{D3stardef}\eea
We can express this star-product more explicitly in terms of a
bi-differential operator acting on ${\rm
  C}^\infty(\mathcal{C}_{x_0^+,\chi^{~}_0})\otimes{\rm
  C}^\infty(\mathcal{C}_{x_0^+,\chi^{~}_0})\to{\rm
  C}^\infty(\mathcal{C}_{x_0^+,\chi^{~}_0})$ by writing the normalized
matrix elements in (\ref{D3stardef}) using translation operators
as~\cite{APS1}
\bea
\e^{-\mz^\top\,\mdell'}~\e^{\mz^{\prime\,\top}\,\mdell}
f\left(\mz,\overline{\mz}\,\right)&=&\e^{-\mz^\top\,\mdell'}
\frac{\bigl\langle\mz;p^+,p^-
\bigl|~\hat f~\bigr|\mz+\mz';p^+,p^-\bigr\rangle}{
\bigl\langle\mz;p^+,p^-\bigm|\mz+\mz';p^+,p^-\bigr\rangle}\nonumber\\ &=&
\frac{\bigl\langle\mz;p^+,p^-
\bigl|~\hat f~\bigr|\mz';p^+,p^-\bigr\rangle}{
\bigl\langle\mz;p^+,p^-\bigm|\mz';p^+,p^-\bigr\rangle} \ .
\label{matrixelttransl}\eea
The translation operator
$\e^{-\mz^\top\,\mdell'}~\e^{\mz^{\prime\,\top}\,\mdell}$, acting on
$\mz'$-independent functions, can be expressed as a normal ordered
exponential $\NO\,\exp(\mz'-\mz)^\top\,\overrightarrow{\mdell}\,\NO$
with derivatives ordered to the right in each monomial of the Taylor
series expansion of the exponential function. In this way we may write
the star-product (\ref{D3stardef}) as
\bea
(f*g)\left(\mz,\overline{\mz}\,\right)&=&\left(2\mu\,p^+\right)^2\,
\int\limits_{\complex^2}\dd\varrho\left(\mz',\overline{\mz}{}^{\,\prime}\,
\right)~f\left(\mz,\overline{\mz}\,\right)~\NO\,\exp
\overleftarrow{\mdell}{}^{\,\top}\,\left(\mz'-\mz\right)\,\NO\nonumber\\
&&\times~\e^{-2\mu\,p^+\,|\mz'-\mz|^2}~\NO\,\exp\left(
\overline{\mz}{}^{\,\prime}-\overline{\mz}\,\right)^\top\,
\overrightarrow{\overline{\mdell}}\,\NO~g\left(\mz,\overline{\mz}\,\right) \ ,
\label{D3starform}\eea
and performing the Gaussian integral in (\ref{D3starform}) leads
to our final form
\beq
(f*g)\left(\mz,\overline{\mz}\,\right)=f\left(\mz,
\overline{\mz}\,\right)~\exp\Bigl(\mbox{$\frac1{2\mu\,p^+}$}\,
\overleftarrow{\mdell}{}^{\,\top}\,\overrightarrow{\overline{\mdell}}
\,\Bigr)~g\left(\mz,\overline{\mz}\,\right) \ .
\label{D3starfinal}\eeq

The star-product (\ref{D3starfinal}) is the Voros
product~\cite{Voros1} on four-dimensional noncommutative euclidean
space $\eucl_\theta^4$ corresponding to the Poisson bi-vector
\beq
\theta=-\mbox{$\frac\ii{2\mu\,p^+}$}~\overline{\mdell}{}^{\,\top}
\wedge\mdell
\label{NCparD3}\eeq
whose components are proportional to the inverse of the magnetic field (or
equivalently harmonic oscillator frequency)
$\omega=\frac12\,\mu\,p^+$ in the effective particle dynamics in
Nappi-Witten spacetime described in Section~\ref{Dynamics}. This
product is {\it not} the same as the standard Moyal product
\beq
(f\star g)\left(\mz,\overline{\mz}\,\right):=f\left(\mz,
\overline{\mz}\,\right)~\exp\left[\mbox{$\frac1{4\mu\,p^+}$}\,
\Bigl(\overleftarrow{\mdell}{}^{\,\top}\,\overrightarrow{\overline{\mdell}}
-\overleftarrow{\overline{\mdell}}{}^{\,\top}\,
\overrightarrow{\mdell}\,\Bigr)\right]~g\left(\mz,\overline{\mz}\,\right)
\label{MoyalD3def}\eeq
which arises from Weyl operator ordering, rather than the normal
ordering prescription employed in (\ref{DeltafD3}). Although
different, these two star-products are cohomologically {\it
  equivalent}~\cite{Voros1}, because the invertible differential
operator $\mathcal{T}:=\exp\frac1{4\mu\,p^+}\,|\mdell|^2$ gives an
algebra isomorphism $\mathcal{T}:({\rm
  C}^\infty(\mathcal{C}_{x_0^+,\chi^{~}_0})\,,\,\star)\to({\rm
  C}^\infty(\mathcal{C}_{x_0^+,\chi^{~}_0})\,,\,*)$,
i.e. $\mathcal{T}(f\star g)=\mathcal{T}(f)*\mathcal{T}(g)~~\forall
f,g\in{\rm C}^\infty(\mathcal{C}_{x_0^+,\chi^{~}_0})$.

Finally, to deal with the geometry in the case $p^+<0$, we construct a
lowest-weight module $\widetilde{V}^{p^+,p^-}$ which defines the
representation conjugate to $V^{p^+,-p^-}$ above by interchanging the
roles of the generators $\Pu^+\leftrightarrow\Pu^-$ and replacing the
generators $\J,\T$ with their reflections $-\J,-\T$. The two
representations have the same quadratic Casimir
eigenvalue (\ref{C6irrep}) and are dual to each other as
\beq
\widetilde{V}^{p^+,p^-}~\cong~\bigl(V^{p^+,p^-}\bigr)^* \ .
\label{dualreps}\eeq
The star-product is generically given as (\ref{D3starfinal}) or
(\ref{MoyalD3def}) with $p^+$ replaced by $|p^+|$, and the worldvolume
algebra (\ref{quantalgD3}) is canonically isomorphic as a vector space
to
\beq
\mathcal{A}\bigl(\mathcal{C}_{x_0^+,\chi^{~}_0}\bigr)=V^{p^+,p^-}
\otimes\widetilde{V}^{p^+,p^-} \ , ~~ p^+>0 \ .
\label{quantalgD3duals}\eeq
The $\mathcal{N}_6$-module structure is then determined by the
Clebsch-Gordan decomposition of (\ref{quantalgD3duals}) into the
irreducible continuous representations of the Nappi-Witten algebra as
\beq
\mathcal{A}\bigl(\mathcal{C}_{x_0^+,\chi^{~}_0}\bigr)=~{\int\limits_0^\infty
\!\!\!\!\!\!\!\!\!\!\!\!~\,\bigodot\!\!\!\!\!\!\!\!\!\!\mbf-\!\!
\mbf-}~\dd\alpha~\alpha~~{\int\limits_0^\infty
\!\!\!\!\!\!\!\!\!\!\!\!~\,\bigodot\!\!\!\!\!\!\!\!\!\!\mbf-\!\!
\mbf-}~\dd\beta~\beta~~V_{\alpha,\beta}^{0,0} \ .
\label{D3module}\eeq
The organization of the quantized worldvolume algebra into irreducible
representations associated with null branes owes to the fact that
the isometry generators of the noncommutative D3-branes in $\NW_6$ are
given by the Killing vectors $P^\pm$, $Q^\pm$ and $J+\overline{J}$ in
(\ref{6CWKilling},\ref{extraHKilling}) with $T=0$, corresponding to
translations in each transverse plane $z,w\in\complex$ along with
simultaneous rotations of the two planes. These isometries generate
the subgroup (\ref{S5subgp}) which coincides with the group ${\rm
Inn}(\mathfrak{n}_6)$ of inner automorphisms of the Nappi-Witten Lie algebra.
The symmetric untwisted D3-brane breaks the generic rotational symmetry ${\rm
U}(2)={\rm SU}(2)\times{\rm U}(1)$ to ${\rm U}(1)\cong{\rm SO}(2)$, leaving the
overall isometry subgroup (\ref{S5subgp}) which is precisely the symmetry group
of noncommutative euclidean space in four dimensions with equal magnetic fields
through each parallel plane~\cite{AlVaz1,CMNTV1}. Thus the embedding
of flat branes into $\NW_6$ realizes explicitly the usual breaking of
${\rm ISO}(4)$ invariance in passing to the noncommutative space
$\eucl_\theta^4$. The geometrical role of the broken ${\rm SU}(2)$
symmetry of the background will be discussed in
Section~\ref{D3Fuzzy}.

\subsection{Open String Description\label{D3OSD}}

Let us now compare the semi-classical results obtained above with the
predictions from the open string dynamics on the NS supported D-branes
in the Seiberg-Witten decoupling limit. Generally, let $G$ be the
closed string metric on a D-brane and $\mathcal{F}=B+F$ the
gauge-invariant two-form which we assume is non-degenerate. Whenever
$\dd(\mathcal{F}-G\,\mathcal{F}^{-1}\,G)=0$, the momentum of an open string
attached to the D-brane is small in the low-energy limit~\cite{HY1}.
This is just the requirement $H=0$ of a vanishing NS flux in the limit
of large $B$-field. The strings are then very short and see only a
small portion of the worldvolume, which is approximately flat. The
same expressions that apply to flat backgrounds can thereby be applied
in these instances~\cite{ARS1,HY1}. In particular, the
noncommutativity parameters and the open string metric may be computed
from the usual Seiberg-Witten formulas~\cite{SW1}
\bea
\Theta&=&\frac1{\mathcal{F}-G\,\mathcal{F}^{-1}\,G}~=~-\frac1{G+\mathcal{F}}
\,\mathcal{F}\,\frac1{G-\mathcal{F}} \ , \label{SWThetagen}
\\&&{~~~~}^{~~}_{~~}\nn\\
G_{\rm o}&=&G-\mathcal{F}\,G^{-1}\,\mathcal{F} \ .
\label{SWopenmetgen}\eea
For the flat euclidean D3-branes of the previous subsection, we
substitute (\ref{metricpieucl4}) and (\ref{E4F6PGL}) in
(\ref{SWThetagen}) to get the bi-vector
\beq
\Theta=-\mbox{$\frac\ii{(\mathcal{F}_{x_0^+})^{~}_{z\,\overline{z}}+
(\mathcal{F}_{x_0^+})^{-1}_{z\,\overline{z}}}$}~
\overline{\mdell}{}^{\,\top}\wedge\mdell=
-\mbox{$\frac\ii2$}\,\sin\mu\,x_0^+~\overline{\mdell}{}^{\,\top}
\wedge\mdell \ .
\label{ThetaD3}\eeq
This does not agree with (\ref{NCparD3}) in the limit $\mu\to0$, as
one would have naively expected, but the reason for the discrepancy is
simple. The semi-classical analysis of the previous subsection is
strictly speaking valid only in the limit of large $B$-field, for
which the formula (\ref{SWThetagen}) reduces to $\Theta=B^{-1}$ and
coincides with (\ref{NCparD3}) on the D3-branes. This is equivalent to
the zero-slope field theory limit ($\alpha'\to0$) of the open string
dynamics, and it yields the Kirillov-Kostant Poisson bi-vector on the
coadjoint orbits corresponding to the symplectic two-form $B$. This
situation is characteristic of branes in curved
backgrounds~\cite{ARS1,HY1}, and we will regard the $\mu\to0$ limit of
(\ref{ThetaD3}), for which
$\Theta=-\frac{\ii\mu\,p^+}2~\overline{\mdell}{}^{\,\top}\wedge\mdell$,
as ``dual'' to the field theoretic quantity (\ref{NCparD3}) with
respect to the inner product $\langle\,\cdot\,,\,\cdot\,\rangle$ on
$\mathfrak{n}_6$.

We can gain more insight into this description by appealing to the
exact boundary conformal field theory description of the conjugacy
classes~\cite{DK2,Hikida1}, which can be done on the limiting planes
$w=0$ whereby the D3-branes restrict to the symmetric euclidean
D-strings of the four-dimensional Nappi-Witten spacetime
$\NW_4$~\cite{FS1}. With this restriction understood everywhere, the
zero-mode Hilbert space for open strings which start and end on a
D1-brane labelled by light-cone momenta $p^\pm$ coincides as a vector
space with the worldvolume algebra
(\ref{quantalgD3duals},\ref{D3module}) (at $\beta=0$). To each open
string boundary condition corresponding to an irreducible module
$V_{\alpha,0}^{0,0}$ in (\ref{D3module}), there is associated a
collection of boundary vertex operators $\Psi_\alpha^{p^+,p^-}(x)$,
$x\in\real$ of conformal dimension $h_\alpha=\alpha^2$ (the quadratic
Casimir eigenvalue) in one-to-one correspondence with the vectors of
$V_{\alpha,0}^{0,0}$. The operator product expansion of two such ${\rm
  U}(1)$ boundary primary fields reads~\cite{DK2}
\bea
\Psi_{\alpha_1}^{p^+,p^-}(x_1)\,\Psi_{\alpha_2}^{p^+,p^-}(x_2)&=&
\int\limits_0^\infty\dd\alpha~\alpha~(x_1-x_2)^{h_\alpha-
h_{\alpha_1}-h_{\alpha_2}}~\delta\left(\alpha^2-
\alpha_1^2+\alpha_2^2+2\alpha^{~}_1\,\alpha^{~}_2\,\cos\phi\right)
\nn\\ &&\times~ {\rm F}^{p^+,p^-;\alpha}_{\alpha_1,\alpha_2}~
\Psi_{\alpha}^{p^+,p^-}(x_2)+\dots
\label{OPEbdryfields}\eea
for $x_1<x_2$ and up to contributions involving descendant fields. The
structure constants ${\rm F}^{p^+,p^-;\alpha}_{\alpha_1,\alpha_2}$
coincide with the (quantum) Racah coefficients of $\mathcal{N}_4$ and
are given explicitly by
\beq
{\rm F}^{p^+,p^-;\alpha}_{\alpha_1,\alpha_2}=\frac{\e^{\frac{\ii\pi}2\,
\alpha_1\,\alpha_2\,\sin\phi\,\cot\frac{\mu\,p^+}2}}{
\pi\,\alpha_1\,\alpha_2\,\sin\phi}
\label{Racahquantum}\eeq
with
$\alpha^2:=\alpha_1^2+\alpha_2^2-2\alpha^{~}_1\,\alpha^{~}_2\,\cos\phi$
expressing momentum conservation in the two-dimensional transverse
plane to the $\NW_4$ wave.

An element $\psi_{\mbf\alpha}^{p^+,p^-}$, $\mbf\alpha\in
T\mathcal{C}_{x_0^+,\chi^{~}_0}$ of the vector space
$V_{\alpha,0}^{0,0}$ is an eigenfunction of the momentum operators
(\ref{6CWKilling}) pulled back to the brane worldvolume. It can be
represented as a plane wave whose radial momentum in the transverse
plane is $\alpha=|\mbf\alpha|$, and the association
$\Psi_{\alpha}^{p^+,p^-}:={\rm V}[\psi_{\mbf\alpha}^{p^+,p^-}]$ yields
the standard, flat space open string tachyon vertex
operators~\cite{Schom1}. The operator product expansion
(\ref{OPEbdryfields}) written in the form
\beq
{\rm V}\bigl[\psi_{\mbf\alpha_1}^{p^+,p^-}\bigr](x_1)~{\rm V}
\bigl[\psi_{\mbf\alpha_2}^{p^+,p^-}\bigr](x_2):={\rm V}
\bigl[\psi_{\mbf\alpha_1}^{p^+,p^-}*\psi_{\mbf\alpha_2}^{p^+,p^-}
\bigr](x_2)+\dots
\label{OPEstarform}\eeq
can then be used to define a star-product in the zero-slope limit by
\beq
\psi_{\mbf\alpha_1}^{p^+,p^-}*\psi_{\mbf\alpha_2}^{p^+,p^-}=
\e^{\frac{\ii\pi}2\,\mbf\alpha_1\wedge\mbf\alpha_2\,\cot\frac{\mu\,p^+}2}
{}~\psi_{\mbf\alpha_1+\mbf\alpha_2}^{p^+,p^-} \ \def\R{{\sf R}}
{}.
\label{psistardef}\eeq
The star-product of any two functions on the classical worldvolume algebra
is then given by expressing them, according to (\ref{D3module}), as
expansions $f=\int_0^\infty\dd\alpha~\alpha~\tilde
f_\alpha~\psi_{\mbf\alpha}^{p^+,p^-}$. According to the standard
formulas for D-branes in flat space with a magnetic field on their
worldvolume~\cite{SW1}, the conformal weight power in
(\ref{OPEbdryfields}) should be identified as
$\alpha^{~}_1\,\alpha^{~}_2\,\cos\phi=\mbf\alpha_1^\top\,\mbf\alpha_2^{~}:=
G_{\rm o}^{-1}(\mbf k_1,\mbf k_2)$. In the present case the open
string metric (\ref{SWopenmetgen}) is given by $G_{\rm
  o}=\csc^2\frac{\mu\,x_0^+}2~|\dd\mz|^2$, yielding
$\mbf\alpha_i=\sin\frac{\mu\,x_0^+}2~\mbf k_i$ for
$i=1,2$. Identifying now the phase factor in (\ref{psistardef}) as
$\e^{\frac{\ii\pi}2\,\mbf\alpha^{~}_1\wedge\mbf\alpha^{~}_2\,
\cot\frac{\mu\,p^+}2}:=
\e^{-\frac{\ii\pi}2\,\Theta(\mbf k_1,\mbf k_2)}$ leads
immediately to the anticipated result (\ref{ThetaD3})~\cite{DK2}.

\subsection{Foliation by Fuzzy Spheres\label{D3Fuzzy}}

We will now clarify the role of the broken ${\rm U}(2)={\rm
  SU}(2)\times{\rm U}(1)$ symmetry on the noncommutative D3-branes. The ${\rm
U}(1)$ factor is generated by the vector
field $J+\overline{J}$ producing simultaneous rotations in the two
transverse planes, while the ${\rm SU}(2)$ generators can be
represented on each irreducible module $V^{p^+,p^-}$ over
$U(\mathfrak{n}_6)$ as
\beq
\hat I^{~}_a:=\mbox{$\frac1{(2\mu\,p^+)^2}$}~
\mcDp\left(\bigl(\,\Pu^-\bigr){}^\top\,
\sigma_a^\top\,\Pu^+\right)
\label{SU2gensdef}\eeq
where $\sigma_a$, $a=1,2,3$ are the standard Pauli spin matrices. The
operators (\ref{SU2gensdef}) generate, up to a rescaling, the ${\rm
  su}(2)$ Lie algebra
\beq
\bigl[\hat I_a\,,\,\hat I_b\bigr]=
\mbox{$\frac\ii{2\mu\,p^+}$}\,\epsilon_{ab}^{~~~c}\,\hat I_c \ ,
\label{su2liealg}\eeq
and under their adjoint action the transverse space operators
transform in the fundamental and anti-fundamental representations of
${\rm SU}(2)$ as
\bea
\left[\hat I^{~}_a\,,\,\mcDp\bigl(\,\Pu^+\bigr)
\right]&=&\mbox{$\frac1{2\mu\,p^+}$}~
\sigma^{~}_a~\mcDp\bigl(\,\Pu^+\bigr) \ , \nn\\
\left[\hat I^{~}_a\,,\,\mcDp\bigl(\,\Pu^-\bigr)
\right]&=&-\mbox{$\frac1{2\mu\,p^+}$}~
\sigma_a^\top~\mcDp\bigl(\,\Pu^-\bigr) \ .
\label{Pfundreps}\eea
The quadratic Casimir operator $\hat R^2:=\sum_a\hat I_a^2\in U({\rm
  su}(2))\subset{\rm End}(V^{p^+,p^-})$ may be written using
(\ref{C6irrep}) as
\beq
\hat R=\mbox{$\frac1{2\mu\,p^+}$}\,\Bigl(\ii\mcDp\bigl(\J\,\bigr)
+\mu^{-1}\,p^-\,\id\Bigr) \ .
\label{su2quadCas}\eeq

This ``hidden'' algebraic structure has the following geometric
interpretation. Corresponding to these operators, the functions on the
conjugacy classes $\mathcal{C}_{x_0^+,\chi^{~}_0}$ are given by
\beq
x_a:=\Delta_*^{-1}\bigl(\hat I^{~}_a\bigr)\left(\mz,
\overline{\mz}\,\right)=-\overline{\mz}{}^{\,\top}
\,\sigma_a^\top\,\mz \ , ~~ r^2:=
\Delta_*^{-1}\bigl(\hat R\bigr)\left(\mz,
\overline{\mz}\,\right)=|\mz|^2 \ .
\label{su2classfns}\eeq
The D3-brane worldvolume $\eucl^4$ admits a foliation into three-spheres
$\S^3$ defined by $|\mz|=r={\rm constant}$. The orbits are all
parallel and lie on two-spheres $\S^2\cong\complex{\rm
  P}^1\subset\eucl^4$ of radii $(\,\sum_ax_a^2)^{1/2}=r^2$. The corresponding
maps
$\S^3\to\S^2$ defined by (\ref{su2classfns}) are Hopf fibrations over
the coadjoint orbits ${\rm SU}(2)\,/\,{\rm U}(1)$ of the ${\rm SU}(2)$
symmetry group.

The ${\rm SU}(2)$ symmetry of the Nappi-Witten plane wave $\NW_6$
persists as well on the quantized conjugacy classes, giving a
noncommutative version of the Hopf fibration via the Jordan-Schwinger
map~\cite{G-BLMV1,HLS-J1}. This is the residual symmetry ${\rm
  USp}(2)\times{\rm USp}(2)\,/\,{\rm U}(1)$ of the reduction of
$\eucl_\theta^4$. The corresponding
two-spheres $\S_{p^+,j}^2\subset\eucl_\theta^4$ are fuzzy spheres and
can be constructed for each $j\in\frac12\,\nat_0$ by defining a
$(2j+1)$-dimensional module $W^{p^+}_j\subset V^{p^+,p^-}$ over
$U({\rm su}(2))$ as the linear span of the Schwinger basis vectors
\beq
\bigl|j,\ell\,\bigr\rangle\!\!\!
\bigm\rangle_{p^+}:=\bigl|j+\ell,j-\ell;p^+,p^-\bigr\rangle
\label{fuzzybasisdef}\eeq
for $\ell\in\{-j,-j+1,\dots,j-1,j\}$. The actions of the ${\rm su}(2)$
generators are given by
\beq
\hat I_\pm\bigl|j,\ell\,\bigr\rangle\!\!\!
\bigm\rangle_{p^+}=-\mbox{$\frac{j\pm\ell}
{\mu\,p^+}$}\,\bigl|j,\ell\mp1\,\bigr\rangle\!\!\!\bigm\rangle_{p^+} \ , ~~
\hat I_3\bigl|j,\ell\,\bigr\rangle\!\!\!\bigm\rangle_{p^+}
=-\mbox{$\frac\ell{\mu\,p^+}$}\,
\bigl|j,\ell\,\bigr\rangle\!\!\!\bigm\rangle_{p^+}
\label{su2irrepaction}\eeq
where $\hat I_\pm:=\hat I_1\pm\ii\hat I_2$, while the operator
$\mcDp(\J\,)$ and hence the Casimir (\ref{su2quadCas}) are scalar
operators on $W_j^{p^+}$ with
\beq
\hat R\bigm|_{W_j^{p^+}}=R\,\id:=\mbox{$\frac j{\mu\,p^+}$}\,\id \ .
\label{quadCasscalar}\eeq
This module is the irreducible spin-$j$ representation of ${\rm
  SU}(2)$. The algebra of functions on the fuzzy sphere is given by
\beq
\mathcal{A}\bigl(\S^2_{p^+,j}\bigr)={\rm End}\bigl(W^{p^+}_j\bigr)
\label{fuzzyfns}\eeq
and it is isomorphic to the finite-dimensional algebra of
$(2j+1)\times(2j+1)$ matrices. It admits a natural action of the group
${\rm SU}(2)$ by conjugation with group elements evaluated in the
$(2j+1)$-dimensional representation of ${\rm SU}(2)$. Under this
action, the ${\rm SU}(2)$-module structure is determined by its
decomposition into irreducible representations as
\beq
\mathcal{A}\bigl(\S^2_{p^+,j}\bigr)=\bigoplus_{k=0}^{2j}W_k^{p^+} \ .
\label{fuzzyfnsmodule}\eeq

Since the basis (\ref{Vhighestex}) can be presented in terms of ${\rm
  SU}(2)$ module vectors as
$|n,m;p^+,p^-\rangle=|\frac{n+m}2,\frac{n-m}2\,\rangle\!\rangle_{p^+}$, the
noncommutative worldvolume algebra may be expressed in terms of a
foliation by fuzzy spheres as
\beq
\mathcal{A}\bigl(\mathcal{C}_{x_0^+,\chi^{~}_0}\bigr)=
\bigoplus_{j\in\frac12\,\nat_0}\Biggl(\mathcal{A}\bigl(\S^2_{p^+,j}\bigr)
{}~\oplus~\bigoplus_{\stackrel{\scriptstyle j'\in\frac12\,\nat_0}
{\scriptstyle j'\neq j}}W_j^{p^+}\otimes\bigl(W_{j'}^{p^+}\bigr)^*
\Biggr) \ .
\label{D3algS2fol}\eeq
The first summand in (\ref{D3algS2fol}) represents the foliation of
noncommutative $\S^3\cong\real^3\cup\infty$, defined as the representation of
the universal enveloping algebra $U({\rm su}(2))$ in ${\rm End}(V^{p^+,p^-})$,
by fuzzy spheres of increasing quantized radii (\ref{quadCasscalar}). With
$\mbf x^\top=(x_a)\in\real^3$, the three-dimensional noncommutative space
$\mathcal{A}(\real_\theta^3):=
\bigoplus_{j\in\frac12\,\nat_0}\mathcal{A}(\S^2_{p^+,j})$ may be
viewed as a deformation of the algebra ${\rm C}^\infty(\real^3)$ by
using (\ref{su2classfns}) to reduce the Voros product
(\ref{D3starfinal}) on $\eucl_\theta^4$ to the
star-product~\cite{HLS-J1}
\beq
(f*_{\real^3}g)(\mbf x)=f(\mbf x)~\exp\Bigl[\mbox{$\frac1{4\mu\,p^+}$}~
\mbox{$\frac{\overleftarrow{\partial}}{\partial x_a}$}\,\left(
\delta_{ab}\,r^4+\ii\epsilon_{ab}^{~~~c}\,x_c\right)\,
\mbox{$\frac{\overrightarrow{\partial}}{\partial x_b}$}\,\Bigr]~
g(\mbf x) \ .
\label{starprodR3}\eeq
The reduction to the fuzzy sphere (\ref{fuzzyfns}) is achieved explicitly by
introducing the orthogonal projection
\beq
\hat P_j=\mbox{$\frac1{(2j+1)\,\left(2\mu\,p^+\right)^{2j}}$}~
\sum_{\ell=-j}^j\mbox{$\frac1{(j+\ell)!\,(j-\ell)!}$}~
\bigl|j,\ell\,\bigr\rangle\!\!\!
\bigm\rangle_{p^+}\,{}^{~}_{p^+}\bigl\langle\!
\bigm\langle j,\ell\,\bigr|
\label{projspinj}\eeq
onto the spin-$j$ module $W_{p^+}^j$, so that
\beq
\mathcal{A}\bigl(\S^2_{p^+,j}\bigr)=\hat P_j\mathcal{A}\bigl
(\real_\theta^3\bigr)=\mathcal{A}\bigl
(\real_\theta^3\bigr)\hat P_j=\hat P_j\mathcal{A}\bigl(
\mathcal{C}_{x_0^+,\chi^{~}_0}\bigr)\hat P_j \ .
\label{fuzzyalgproj}\eeq
This projection onto $\S^2_{p^+,j}$ may be translated into the star-product
formalism on ordinary functions~\cite{HLS-J1}, with the coordinate
generators of the deformation of ${\rm C}^\infty(\S^2)$ given by
$x_a*_{\real^3}P_j$. The function $P_j$ corresponding to the projection
operator
(\ref{projspinj}) is radially symmetric on $\eucl^4$ and may
be straightforwardly computed using (\ref{DeltainvD3}) to get
\beq
P_j\bigl(|\mz|\bigr):=\Delta_*^{-1}\bigl(\hat P_j\bigr)
=\mbox{$\frac{\left(2\mu\,p^+\right)^{2j}}{(2j+1)!}$}~
|\mz|^{4j}~\e^{-2\mu\,p^+\,|\mz|^2} \ .
\label{projspinjfn}\eeq
A similar foliation will be exploited in Section~\ref{TDNC} to construct the
full noncommutative geometry of Nappi-Witten spacetime.

The second summand in (\ref{D3algS2fol}) is the foliation of $\eucl_\theta^4$
by noncommutative three-spheres of increasing radius, and the direct sum
decomposition (\ref{D3algS2fol}) reflects the ${\rm SU}(2)$ symmetry inherent
on the noncommutative D3-branes. It can be compared with the module
decomposition (\ref{D3module}) which suggests that the euclidean D3-branes in
$\NW_6$ are composed of elementary constituents with the
$\mathcal{N}_6$ quantum numbers of null branes. In the present case,
the algebra isomorphism (\ref{D3algS2fol}) suggests that the
elementary branes originate as objects expanded into fuzzy spheres in
$\eucl_\theta^4$. It is interesting at this stage to compare these
configurations with those which arise in the RR-supported pp-wave obtained as
the Penrose-G\"uven limit of $\AdS_5\times\S^5$~\cite{BFHP1}. Because
of the RR fluxes, giant gravitons, i.e. massless particles in
anti-de~Sitter space which polarize into massive
branes, correspond to D-branes. In the strongly-coupled Type~IIA plane wave,
degenerate vacua corresponding to fundamental strings blown up into fuzzy
spheres provide an exact microscopic description of giant gravitons in
$\AdS_7\times\S^4$ corresponding to M2-branes polarized along $\S^4$. These
latter objects become the fuzzy sphere configurations of the BMN matrix model
in the Penrose-G\"uven limit~\cite{BMN1}. In the Type~IIB case, the
gravitational waves expand into spherical D3-branes wrapping a fuzzy
three-sphere determined by the noncommutative Hopf fibration described
above~\cite{JLR-G1,S-J1}. In the present case the
situation is a bit different. The intrinsic ${\rm SU}(2)$ symmetry of the NS
background implies that long strings, of light-cone momenta $p^+>1$, can move
freely in the two transverse planes to the Nappi-Witten pp-wave and correspond
to spectral-flowed null brane states~\cite{BAKZ1}. The long strings in
$\NW_6$ thus correspond instead to fundamental string states. It would
be interesting to understand their origin more precisely through the
Penrose-G\"uven limit of the corresponding giant graviton
configurations in $\AdS_3\times\S^3$ which wrap fuzzy
cylinders~\cite{JLR-G2}. A general definition of fuzzy spheres in
diverse dimensions is given in~\cite{S-J1} and applied to the
construction of fuzzy three-spheres and four-spheres in~\cite{S-JT1}
by embedding them into $\complex^4_\theta$, generalizing the
construction outlined above.

\subsection{Flat Space Limits\label{FlatLim}}

The spectral flow of long string states also implies that, unlike the
RR-supported pp-waves, the strong NS-field limit $\mu\to\infty$ gives a flat
space theory~\cite{DAK1}, just like the semiclassical limit $\mu\to0$
does. This flat space limit is described in terms of the fuzzy sphere
foliations above, with radii $R=\frac j{\mu\,p^+}$, as follows. If we
send the spin $j\to\infty$, along with $\mu\to\infty$ at fixed
light-cone momentum $p^+$, then the operators $\hat I_a$ in
(\ref{su2liealg}) commute and generate the commutative algebra ${\rm
  C}^\infty(\S^2)$ of functions on an ordinary sphere $\S^2$. This is
the intuitive expectation, as the noncommutativity parameter of
$\eucl_\theta^4$ vanishes in the limit $\mu\to\infty$. However,
although a flat space theory is attained, noncommutative dynamics
still persist signalling the remnant of the ${\rm SU}(2)$ symmetry of
the NS background, because the $j\to\infty$ limit of (\ref{su2liealg})
describes a noncommutative plane $\eucl_\vartheta^2$ under
stereographic projection~\cite{APS1,CMS1}. This limit is completely
analogous to the scaling of the volume and flux of $\AdS_2\times\S^2$ to
infinity that obtains the noncommutative space $\eucl_\theta^4$ in the
Penrose-G\"uven limit of (\ref{E4commdiag}).

The analog of stereographic projection for fuzzy coordinates is
defined by the operators~\cite{APS1,CMS1}
\beq
\hat Z:=\hat I_-\,\left(\id-\mbox{$\frac{\mu\,p^+}j$}\,\hat I_3
\right)^{-1} \ , ~~ \hat Z^\dag:=\left(\id-\mbox{$\frac{\mu\,p^+}j$}\,
\hat I_3\right)^{-1}\,\hat I_+
\label{fuzzystereo}\eeq
which for large $j$ have the commutator
\beq
\left[\hat Z\,,\,\hat Z^\dag\,\right]=\mbox{$\frac j{\left(\mu\,p^+
\right)^2}$}\,\left(\id-\mbox{$\frac{\mu\,p^+}j$}\,\hat I_3
\right)^{-2}+O\left(\mbox{$\frac1j$}\right) \ .
\label{Zcommj}\eeq
In the limit $j\to\infty$, the whole $\hat Z$-plane is covered by
projecting all operators onto the subspace of the eigenspace of the
hermitian operator $(\frac j{\mu\,p^+})^2\,(\id+\frac{\mu\,p^+}j\,\hat
I_3)$ corresponding to its eigenvalues which lie in the range
$[0,\frac{j^{3/2}}{(2\mu\,p^+)^2}]$. Thus for $\mu,j\to\infty$ with
$j/(\mu\,p^+)^2$ fixed, the commutation relation (\ref{Zcommj})
becomes
\beq
\left[\hat Z\,,\,\hat Z^\dag\,\right]=\vartheta:=
\mbox{$\frac j{\left(2\mu\,p^+\right)^2}$} \ .
\label{Zcommtheta}\eeq
Therefore, even though the foliation of $\eucl^4\subset\NW_6$ becomes
that by ordinary commutative spheres $\S^2$ as $\mu,j\to\infty$, the
quantization of the D3-brane worldvolumes persists in the
neighbourhoods of points on the two-spheres which can be
stereographically projected onto the noncommutative planes
$\eucl_\vartheta^2$.

\newsection{Noncommutative Branes in $\mbf{\NW_6}$: Twisted
  Case\label{TwistedNCBranes}}

The quantization of twisted conjugacy classes is more
intricate~\cite{AFQS1}. The $\omega$-twisted D-branes in the Lie group
$\mathcal{G}$ are labelled by representations of the invariant
subgroup
\beq
\mathcal{G}^\omega:=\bigl\{g\in\mathcal{G}\,\bigm|\,\omega(g)=g
\bigr\} \ .
\label{Gomegagen}\eeq
In a neighbourhood of the identity of $\mathcal G$, a twisted
conjugacy class $\mathcal{C}_g^\omega$ may be regarded as a fibration
\begin{equation}
  \label{conjfibre}
  \begin{CD}
    \breve{\mathcal C}^{~}_g ~~     @>>> \mathcal{C}_g^\omega\\
    @.    @VVV\\ @.
     \mathcal{G}\,/\,\mathcal{G}^\omega  \\
  \end{CD}
\end{equation}
where $\breve{\mathcal C}_g$ is an ordinary conjugacy class of
$\mathcal{G}^\omega$ and $\mathcal{G}^\omega$ acts on $\mathcal{G}$ by
right multiplication. In particular, $\breve{\mathcal C}_g$ can
be identified with a coadjoint orbit of the subgroup (\ref{Gomegagen}),
with the standard linear Poisson structure coinciding with that induced
by pull-back $B|_{\mathcal{G}^\omega}$ of the Neveu-Schwarz
$B$-field when $H=0$. Semiclassically, after quantization
$\mathcal{C}_g^\omega=\mathcal{G}\times_{\mathcal{G}^\omega}\breve{\mathcal
  C}^{~}_g$ becomes a trivial bundle with noncommutative fibers labelled by
irreducible modules $\breve{V}_g^\omega$ over the group
$\mathcal{G}^\omega$ but with a classical base space
$\mathcal{G}\,/\,\mathcal{G}^\omega$~\cite{AFQS1}. The associative
noncommutative algebra of functions on the worldvolume in this case is
thus given by
\beq
\mathcal{A}\left(\mathcal{C}_g^\omega\right)=\Bigl({\rm C}^\infty\bigl(
\mathcal{G}\bigr)\otimes{\rm End}\bigl(\breve{V}_g^\omega\bigr)
\Bigr)^{\mathcal{G}^\omega} \ ,
\label{NCtwistalg}\eeq
where the superscript denotes the $\mathcal{G}^\omega$-invariant
part and $\mathcal{G}^\omega$ acts on ${\rm C}^\infty(\mathcal{G})$
through the induced derivative action of right isometries of
$\mathcal{G}$. Again, the worldvolume algebra (\ref{NCtwistalg})
carries a natural action of the group $\mathcal{G}$, now
through the induced action on ${\rm C}^\infty(\mathcal{G})$ by right
isometries.

If $\omega=\id$, then
$\mathcal{G}^\omega=\mathcal{G}$ and (\ref{NCtwistalg}) reduces to the
definition (\ref{NCalgconjclass}). If the module $\breve{V}_g^\omega$ is
one-dimensional (e.g. the trivial representation of
$\mathcal{G}^\omega$), then ${\rm
  End}(\breve{V}_g^\omega)\cong\complex$ and
the algebra (\ref{NCtwistalg}) becomes the commutative algebra of
functions $\mathcal{A}(\mathcal{C}_g^\omega)\cong{\rm
  C}^\infty(\mathcal{G}\,/\,\mathcal{G}^\omega)$ on $\mathcal{G}$
invariant under the action of the subgroup
$\mathcal{G}^\omega\subset\mathcal{G}$ by right isometries. It is this latter
situation that we shall discover in the present case. All twisted
D-branes in $\NW_6$ support {\it commutative} worldvolume geometries,
given by the algebra of functions on the classical twisted conjugacy
class, consistent with the remarks made in
Section~\ref{DiagNW}. This result will be supported by the fact that all
twisted branes have vanishing NS flux $H=0$, so that again semi-classical
quantization applies. However, for certain lorentzian D-branes that we shall
encounter, there are some important subtleties hidden in the structure of the
worldvolume two-form fields that are invisible in the field theoretic analysis
outlined here.

\subsection{General Construction\label{GenConstrTwist}}

The group of outer automorphisms of the Lie algebra $\mathfrak{n}_6$ is given
by ${\rm Out}(\mathfrak{n}_6)=\zed_2\ltimes{\rm
  SU}(2)\,/\,\zed_2$~\cite{SF1}, where the ${\rm SU}(2)$ factor is the
transverse space rotational symmetry described in the previous
section. There are thus two families of outer automorphisms
$\omega_\pm^S$ parametrized by a matrix
\beq
S=\begin{pmatrix}a&b\\-\overline{b}&\overline{a}\end{pmatrix}~\in~
{\rm SU}(2) \ ,
\label{SU2matrixouter}\eeq
with $a,b\in\complex$ obeying $|a|^2+|b|^2=1$ and with the identification
$S\sim-S$. Corresponding to the identity element of $\zed_2$, the automorphism
$\omega_+^S$ acts on a group element (\ref{NW6globalcoords}) via a rotation in
the transverse space as
\beq
\omega_+^S(g)=g\left(x^+,x^-,S\mz,\overline{S}\,\overline{\mz}\,\right)
\label{omegapdef}\eeq
with $S\neq\id$ (or else the automorphism is trivial). As a consequence, the
invariant subgroup (\ref{Gomegagen}) in this case is given by
\beq
\mathcal{N}_6^{\,\omega_+^S}~=~\bigl\{g\left(x^+,x^-,\mbf0,\mbf0\right)
\in\mathcal{N}_6~\bigm|~x^\pm\in\real\bigr\}~\cong~\real^2 \ .
\label{invsubgpp}\eeq
This is the two-dimensional abelian group of translations of the light-cone in
$\NW_6$, and its irreducible representations are all one-dimensional. Thus the
quantized worldvolume algebra (\ref{NCtwistalg}) in this case is
$\mathcal{A}(\mathcal{C}_g^{\omega_+^S})\cong{\rm C}^\infty(\eucl^4)$
corresponding to {\it commutative} euclidean D3-branes. We shall explicitly
construct these branes in what follows, showing that the group
(\ref{invsubgpp}) is isomorphic to the stabilizer of the twisted
conjugacy class and hence its points (corresponding to its conjugacy
classes) label the locations of the various D3-branes in $\NW_6$. This
characterization is consistent with the fact that only the (trivial)
one-dimensional subspace of $V^{p^+,p^-}$ spanned by the ground state
vector $|0,0;p^+,p^-\rangle$ is invariant under the action of the
automorphism (\ref{omegapdef}).

The other class of automorphisms $\omega_-^S$ combines the transverse space
rotations with the non-trivial element of $\zed_2$ which acts by charge
conjugation on the generators of $\mathfrak{n}_6$ to give
\beq
\omega_-^S(g)=g\left(-x^+,-x^-,S\,\overline{\mz},\overline{S}\,\mz
\right) \ .
\label{omegamdef}\eeq
It defines an isomorphism
$V^{p^+,p^-}\leftrightarrow\widetilde{V}{}^{p^+,p^-}$ on irreducible
representations. The invariant subgroup (\ref{Gomegagen}) in this case
is determined by the equation $S\,\overline{\mz}=\mz$ for
$\mz\in\complex^2$. The only solution of this equation is $\mz=\mbf0$,
unless the parameter $b$ in (\ref{SU2matrixouter}) is purely imaginary
in which case the subgroup reduces to the two-dimensional abelian
group generated by the elements
\beq
\P_S=a\,\P^++\P^-+b\,\Q^+ \ , ~~ \Q_S=
\overline{a}\,\Q^++\Q^-+b\,\P^+ \ .
\label{2Dabgp}\eeq
Thus
\beq
\mathcal{N}_6^{\,\omega_-^S}~=~\bigl\{g\left(0,0,\mz,\overline{\mz}\,\right)
\in\mathcal{N}_6~
\bigm|~\mz=S\,\overline{\mz}\in\complex^2\bigr\}~\cong~\left\{\begin{matrix}
\real^2~~,~~b\in\ii\real\\\{\id\}~~,~~{\rm otherwise}\end{matrix}
\right.
\label{invsubgp}\eeq
is again always an abelian group (of transverse space translations) with only
one-dimensional irreducible representations, corresponding to commutative
worldvolume geometries (\ref{NCtwistalg}). As we show explicitly below, the
first instance corresponds to a class of commutative, lorentzian D3-branes
wrapping $\CW_4\subset\NW_6$, i.e.
$\mathcal{A}(\mathcal{C}_g^{\omega_-^S})\cong{\rm C}^\infty(\CW_4)$,
with the points of (\ref{invsubgp}) again parametrizing the locations
of the branes in $\NW_6$. The second case represents a family of
commutative, spacetime filling lorentzian D5-branes each isometric to
$\CW_6$, i.e. $\mathcal{A}(\mathcal{C}_g^{\omega_-^S})\cong{\rm
  C}^\infty(\mathcal{N}_6)$. These results are consistent with the
fact that only the self-dual null brane modules
$V_{\alpha,\beta}^{0,0}$ and $V_{\alpha,\beta}^{0,\frac\mu2}$ are
invariant under the action of the automorphism (\ref{omegamdef}).

As in Section~\ref{GenConstrUntwist}, one can now straightforwardly work
out the twisted adjoint actions ${\rm
Ad}^{\omega_\pm^S}_{g(x^+,x^-,\mz,\overline{\mz}\,)}\,g(x^+_0,x^-_0,\mz^{~}_0,
\overline{\mz}^{~}_0)$ corresponding to a fixed point
$(x_0^+,x_0^-,\mz^{~}_0)\in\NW_6$~\cite{FS1,SF1}. For the automorphism
(\ref{omegapdef}) the twisted conjugacy
classes can be written explicitly as the submanifolds
\bea
\mathcal{C}^{\omega_+^S}_{(x_0^+,x^-_0,\mz^{~}_0)}&=&
\scriptstyle{\Bigl\{\bigl(x^+_0\,,\,x^-_0+\frac\mu2
{}~{\rm
  Im}\bigl[\overline{\mz}{}_0^{\,\top}\,(\e^{-\frac{\ii\mu}2\,x_0^+}
\,S+\e^{\frac{\ii\mu}2\,x_0^+}\,\id)\mz~\e^{-\frac{\ii\mu}2\,x^+}+
\frac12\,\mz^\top\,(\e^{-\ii\mu\,x_0^+}\,S-\e^{\ii\mu\,x_0^+}\,\id)
\overline{\mz}\,\bigr]}\,,\,\bigr.\Bigr.\nonumber\\&&~~~~~
\scriptstyle{\left.\Bigl.\bigl.
\e^{\frac{\ii\mu}2\,(x^+-x_0^+)}\,(S-\e^{\ii\mu\,x_0^+}\,\id)\mz+
\e^{\ii\mu\,x^+}\,\mz^{~}_0
\bigr)~\right|~x^+\in\S^1\,,\,\mz\in\complex^2\Bigr\}} \ .
\label{conjclassgenp}\eea
Again $x_0^+$ is an orbit invariant, and the corresponding branes are
thus euclidean. In fact, except for the nature of their noncommutative
worldvolume geometry, these branes are completely analogous to the
branes described in the previous section. For the automorphism
(\ref{omegamdef}) one finds instead the submanifolds
\bea
\mathcal{C}^{\omega_-^S}_{(x_0^+,x^-_0,\mz^{~}_0)}&=&
\scriptstyle{\Bigl\{\bigl(x^+_0+2x^+\,,\,x^-_0+2x^--\mu
{}~{\rm
  Im}\bigl[\overline{\mz}{}_0^{\,\top}\,\mz~\e^{\frac{\ii\mu}2\,
(x_0^++x^+)}-(\,\overline{\mz}{}_0^{\,\top}+
\e^{\frac{\ii\mu}2\,(x^++x_0^+)}\,\overline{\mz}^{\,\top})\,S
\,\overline{\mz}~\e^{\frac{\ii\mu}2\,(x^++x_0^+)}
\bigr]\,,\,\bigr.\Bigr.}\nonumber\\&&~~~~~\scriptstyle{\left.\Bigl.\bigl.
\mz^{~}_0-S\,\overline{\mz}~\e^{-\frac{\ii\mu}2\,(x^++x_0^+)}+\mz~
\e^{\frac{\ii\mu}2\,(x^++x_0^+)}
\bigr)~\right|~x^+\in\S^1\,,\,x^-\in\real\,,\,\mz\in\complex^2\Bigr\}}
\label{conjclassgenm}\eea
wrapped by lorentzian branes. We will now briefly describe the
supergravity fields supported by each of these classes of branes.

\subsection{Euclidean D3-Branes\label{EuclD3Twist}}

The geometry of the twisted conjugacy classes (\ref{conjclassgenp}) is
determined by the complex map on $\complex^2\to\complex^2$ defined by
\beq
\mz':=\e^{\frac{\ii\mu}2\,(x^+-x_0^+)}\,\left(S-\e^{\ii\mu\,x_0^+}\,\id
\right)\mz+\e^{\ii\mu\,x^+}\,\mz^{~}_0 \ .
\label{complexmapeucl}\eeq
Whenever the $2\times2$ matrix $S-\e^{\ii\mu\,x_0^+}\,\id$ is
invertible, this map defines an isomorphism of linear spaces and
one has
$\mathcal{C}^{\omega_+^S}_{(x_0^+,x^-_0,\mz^{~}_0)}\cong\eucl^4$ with
the same worldvolume fields as in (\ref{metricpieucl4}) and
(\ref{NSconj}), the $x^+$-dependence of (\ref{conjclassgenp})
cancelling out. However, although they are qualitatively the same as the
euclidean D3-branes of Section~\ref{CGED3B}, the worldvolume two-forms
on these branes are different. The stabilizer of the twisted conjugacy
class in this case is parametrized as the cylindrical submanifold
\beq
\mathcal{Z}^{\omega_+^S}_{(x_0^+,x^-_0,\mz^{~}_0)}=\scriptstyle{\left.\Bigl\{
\bigl(x^+,x^-,\e^{\frac{\ii\mu}2\,(x_0^+-x^+)}\,
(1-\e^{\ii\mu\,x^+})\,(S-\e^{\ii\mu\,x_0^+}\,
\id)^{-1}\mz^{~}_0\bigr)~\right|~x^+\in\S^1\,,\,x^-\in\real\Bigr\}} \ ,
\label{stabE4twist}\eeq
from which we may compute the abelian gauge field fluxes
(\ref{2formfamily}) to be
\beq
F^{(\zeta)}_6\bigm|_{\mathcal{C}^{\omega_+^S}_{(x_0^+,x^-_0,\mz^{~}_0)}}=
2\ii x_0^+~\dd\overline{\mz}{}^{\,\top}\wedge\dd\mz \ .
\label{E4F6twist}\eeq
Thus the gauge-invariant two-forms $\mathcal{F}_6^{(\zeta)}$ vanish on these
twisted conjugacy classes, leading to the anticipated commutative
worldvolume geometry.

We can also understand the origin of these
branes through the Penrose-G\"uven limit of Section~\ref{ApplNW} by
considering the commuting isometric embedding diagram
\begin{equation}
  \label{E4commdiagH2}
  \begin{CD}
    \AdS_3 \times \S^3             @>\text{PGL}>> \NW_6\\
    \text{$\imath^{\,\prime}~~$}@AAA
@AAA\text{$\widetilde{\imath}^{~\prime}$}\\
           \Hyp^2\times\S^2    @>\text{PGL$^{\,\prime}$}>> \eucl^4\\
  \end{CD}
\end{equation}
analogous to (\ref{E4commdiag}). The hyperbolic plane $\Hyp^2$ is
wrapped by symmetric instantonic D-strings in $\AdS_3$~\cite{Stanciu3},
corresponding to conjugacy classes of the group ${\rm SU}(1,1)$, and
it is obtained from the intersection of the three-dimensional hyperboloid
(\ref{AdS3quadric}) with the affine hyperplane $x^1=\tau=0$. Now the
pull-back of the $B$-field (\ref{AdS3S3Bfield}) to the $\Hyp^2$
worldvolume vanishes, leading to a vanishing gauge-invariant
two-form.

\subsection{Spacetime Filling D5-Branes\label{LorD5}}

Let us now examine the twisted conjugacy classes
(\ref{conjclassgenm}). By defining $y^+:=x_0^++2x^+$ and
$\mw:=\e^{\frac{\ii\mu}4\,(y^++x_0^+)}\,\mz$ we may express them as the
submanifolds
\beq
\mathcal{C}_{\mz_0}^S=\left.\left\{(y^+,y^-,\mz_0-S\,\overline{\mw}+\mw)~
\right|~y^+\in\S^1\,,\,y^-\in\real\,,\,\mw\in\complex^2\right\} \ .
\label{conjmrew}\eeq
The geometry of the twisted conjugacy class (\ref{conjmrew}) is
thereby determined by the {\it real} linear transformation on
$\real^4\to\real^4$ defined by
\beq
\mw':=S\,\overline{\mw}-\mw \ ,
\label{reallinmapm}\eeq
whose determinant is straightforwardly worked out to be $4({\rm
  Re}~b)^2$. As long as the parameter $b$ in (\ref{SU2matrixouter}) has
a non-zero real part, the map (\ref{reallinmapm}) is a linear
isomorphism and we can write the induced geometry in terms of the new
transverse space coordinates $\mw'\in\complex^2$. After an irrelevant
shift, the pull-back of the plane wave metric (\ref{NW6metricBrink})
to the twisted conjugacy class (\ref{conjmrew}) is given by
\beq
\dd s_6^2\bigm|_{\mathcal{C}_{\mz_0}^S}=2~\dd y^+~\dd y^-+\left|\dd
\mw'\,\right|^2-\mbox{$\frac{\mu^2}4$}\,\left|\mw'\,
\right|^2~\left(\dd y^+\right)^2 \ ,
\label{D5metricpull}\eeq
and the corresponding worldvolumes are therefore wrapped by spacetime
filling D-branes isometric to $\CW_6$. From the identity
\beq
\dd\overline{\mw}{}^{\,\prime\,\top}\wedge\dd\mw'=\dd\overline{\mw}{}^{\,\top}
\wedge\left(\id-\overline{S}{}^{\,\top}\,S\right)~\dd\mw
\label{dwidentity}\eeq
and the unitarity of the matrix $S\in{\rm SU}(2)$ it follows that the
pull-back of the NS background (\ref{NW6Bfield}) is trivial,
\beq
H_6\bigm|_{\mathcal{C}_{\mz_0}^S}~=~0~=~
B_6\bigm|_{\mathcal{C}_{\mz_0}^S} \ .
\label{D5Bpull}\eeq
Furthermore, since the corresponding stabilizer in this case is
euclidean, having $x^+=0$, it is elementary to see that the pull-backs
of the gauge-invariant two-forms also vanish,
\beq
\mathcal{F}^{(\zeta)}_6\bigm|_{\mathcal{C}_{\mz_0}^S}~=~0 \ .
\label{D5F6pull}\eeq
In perfect agreement with the general analysis of
Section~\ref{GenConstrTwist}, these D5-branes thus all carry
commutative worldvolume geometries.

As described in Section~\ref{ApplNW}, these D-branes originate through
the Penrose limit of D5-branes isometric to $\AdS_3\times\S^3$. The
latter do {\it not} correspond to symmetric D-branes of the ${\rm
  SU}(1,1)\times{\rm SU}(2)$ group and, unlike the symmetric
D5-branes obtained here which are spacetime filling branes of the
six-dimensional plane wave, they do not fill the whole
$\AdS_3\times\S^3$ target space. They wrap only a three-dimensional
submanifold of the $\S^3$ component given by the complement in $\S^3$
of the disjoint union of two circles~\cite{Stanciu3}. While the Penrose limit
decompactifies the branes to spacetime filling ones, the Neveu-Schwarz
background vanishes due to the non-trivial embedding $\imath$
represented through the isometric embedding diagram
\begin{equation}
  \label{CW6commdiag}
  \begin{CD}
    \AdS_3 \times \S^3             @>\text{PGL}>> \NW_6\\
    \text{$\imath~~$}@AAA @AAA\text{$\widetilde{\imath}$}\\
    \AdS_3 \times \S^3             @>\text{PGL}>> \CW_6\\
  \end{CD}
\end{equation}
with $\widetilde{\imath}$ constructed as above.

\subsection{Lorentzian D3-Branes\label{LorD3}}

We now describe the situations wherein the linear maps defined above
are not of maximal rank, beginning with (\ref{reallinmapm}). In this
case the parameter $b\in\ii\real$ is purely imaginary, and the map has
real rank~$2$ so that the variable $\mw^{\prime\,\top}=(w_1',w_2')$
lives in a one-dimensional complex subspace of $\complex^2$. When
${\rm Re}~a=a=1$, one has $b=0$ and $\mw'\in\ii\real^2$. Defining
$z\in\complex$ by $z:=2~{\rm Im}~w_1+2\ii~{\rm Im}~w_2$, it follows that
$|\mw'\,|=|z|$ and the metric (\ref{D5metricpull}) coincides with the
usual Cahen-Wallach geometry (\ref{NW4metricBrink}) of $\CW_4$. When
${\rm Re}~a<1$, we may coordinatize the twisted conjugacy class
(\ref{conjmrew}) by $\mw\in\real^2$. Then by defining
\beq
z:=\sqrt{2(1-{\rm Re}~a)}~\left(w_1-\mbox{$\frac b{2(1-{\rm Re}~a)}$}
\,w_2\right)+\ii\sqrt{2(1-{\rm Re}~a)-\mbox{$\frac{b^2}
{2(1-{\rm Re}~a)}$}}~w_2
\label{zD3def}\eeq
one finds $|\mw'\,|=|z|$ and (\ref{D5metricpull}) again reduces to the
form (\ref{NW4metricBrink}). The identity (\ref{dwidentity}) once more
implies the vanishing (\ref{D5Bpull}) of the NS background, so that
the twisted conjugacy classes (\ref{conjmrew}) are wrapped by
curved lorentzian D3-branes isometric to $\CW_4$. All of this is in
perfect harmony with the general findings of
Section~\ref{GenConstrTwist}.

The $\AdS_3\times\S^3$ origin of these D3-branes is described by
(\ref{CW4commdiag}). However, now an apparent discrepency
arises. Using (\ref{dwidentity}) and the fact that the matrix
(\ref{SU2matrixouter}) is symmetric whenever $b\in\ii\real$, one finds
a non-vanishing pull-back of the worldvolume flux
(\ref{NW6tildegaugeinv}) given by
\beq
\mathcal{F}_6\bigm|_{\mathcal{C}_{\mz_0}^S}=-\mbox{$\frac{\ii\mu}2$}\,
\cot^2\mbox{$\frac{\mu\,y^+}2$}~\dd y^+\wedge\left(\mz_0^\top~
\dd\overline{\mw}{}^{\,\prime}-\overline{\mz}{}_0^{\,\top}~
\dd\mw'\,\right) \ .
\label{non0D3flux}\eeq
This worldvolume two-form depends on the locations of the branes in
$\NW_6$, and it should induce a spacetime noncommutative geometry on
the D3-brane worldvolume. However, this flux corresponds to a worldvolume {\it
  electric} field, rather than a magnetic field, and D-branes in
electric backgrounds do not exhibit a well-defined decoupling of the
massive string states. Instead of field theories, such backgrounds
lead to noncommutative open string theories~\cite{GMMS1,SST1} which
lie beyond the scope of the semi-classical analysis at the beginning
of this section. Such a noncommutativity agrees with the non-trivial
boundary three-point couplings~\cite{DK2} on the symmetric lorentzian
D-membranes~\cite{FS1} wrapping the pp-waves
$\CW_3\hookrightarrow\NW_4$ defined by the volumes ${\rm Im}~z={\rm
  constant}$ in the above. Given the origin of the null worldvolume
flux (\ref{non0D3flux}) from Section~\ref{ApplNW} and the mechanism
described after (\ref{E4commdiag}), this noncommutative open string
theory could provide a curved space analog of the expected flat space
S-duality with $(p,q)$ strings.

\subsection{D-Strings\label{DStrings}}

Let us now turn to the degenerate cases of the map
(\ref{complexmapeucl}). The vanishing of the determinant of the
$2\times2$ matrix $S-\e^{\ii\mu\,x_0^+}\,\id$ shows that these branes
arise when the fixed light-cone time coordinate obeys the equation
\beq
\cos\mu\,x_0^+={\rm Re}~a \ .
\label{nullcoordeq}\eeq
Since $|a|<1$ in this case, these worldvolumes lie in the set of
conjugate free points $x_0^+\neq0,\frac\pi\mu$ of the Rosen plane wave
geometry and $S-\e^{\ii\mu\,x_0^+}\,\id$ is of complex
rank~$1$. Explicitly, with $\mz^\top=(z,w)$ and $\xi\in\complex$
defined by $\xi:=b\,w-\ii(\sin\mu\,x_0^+-{\rm Im}~a)\,z$, one finds
using (\ref{nullcoordeq}) that
\beq
\left(S-\e^{\ii\mu\,x_0^+}\,\id\right)\mz=\begin{pmatrix}
\xi\\[2mm]\,\frac{\ii\overline{b}}{\sin\mu\,x_0^+-{\rm Im}~a}~\xi
\end{pmatrix} \ .
\label{Sdegexpl}\eeq
If $\mz_0^{~}=(S-\e^{\ii\mu\,x_0^+}\,\id)\mw_0^{~}$ for some fixed
$\mw_0^{~}\in\complex^2$, then a simple coordinate redefinition in
(\ref{complexmapeucl}) enables us to parametrize the twisted conjugacy
class by a single complex variable. The pull-back of the $\NW_6$
metric is non-degenerate and the classes in this case are wrapped by
euclidean D1-branes. As in Section~\ref{EuclD3Twist}, all worldvolume
fields in this case are easily found to be trivial. These branes
do not explicitly appear in the general analysis of
Section~\ref{GenConstrTwist}, and like the null branes of
Section~\ref{Null} their quantized worldvolume geometry, although
commutative, differs from the classical one. They can nevertheless be
regarded as subclasses of the twisted euclidean D3-branes constructed in
Section~\ref{EuclD3Twist}. In particular, they originate from
symmetric euclidean D-strings in $\AdS_3\times\S^3$, either wrapping
$\Hyp^2\subset\AdS_3$ and sitting at a point in $\S^3$ or wrapping
$\S^2\subset\S^3$ and sitting at a point in $\AdS_3$.

\subsection{D-Membranes\label{DMem}}

Finally, if $\mz_0^{~}\notin{\rm im}(S-\e^{\ii\mu\,x_0^+}\,\id)$ on
$\complex^2$, then from (\ref{complexmapeucl}) it follows that the
twisted conjugacy class (\ref{conjclassgenp}) is isometric to
$\S^1\times\eucl^2$. The metric $\dd s_6^2$ restricts
non-degenerately, and once again all worldvolume form fields are
trivial. The orbits are thus wrapped by euclidean D2-branes, which can
likewise be regarded as subclasses of the D3-branes in
Section~\ref{EuclD3Twist}. Like the spacetime filling D-branes of
Section~\ref{LorD5}, these branes do not originate from symmetric
D-branes in $\AdS_3\times\S^3$, but rather from either of the
trivially embedded $\AdS_3$ or $\S^3$ submanifolds.

\newsection{Noncommutative Geometry of $\mbf{\NW_6}$\label{TDNC}}

In this final section we will construct a noncommutative deformation
of the six-dimensional Nappi-Witten spacetime $\NW_6$. We will do so
by using the fact that the conjugacy classes foliate the
$\mathcal{N}_6$ group with respect to the standard Kirillov-Kostant
symplectic structure. Generally, from (\ref{twistconjhom}) it follows
that in a neighbourhood of $g\in\mathcal{G}$ one has
$\mathcal{G}=\mathcal{C}_g^\omega\times\mathcal{Z}_g^\omega$ and thus
the group manifold is foliated by hyperplanes corresponding to twisted
conjugacy classes. This construction is completely analogous to the
foliation by fuzzy spheres that we described in Section~\ref{D3Fuzzy}, which
exhibits a foliation of the ${\rm SU}(2)$ group by its conjugacy
classes. In this way we will exhibit $\NW_6$ as a foliation by the
noncommutative D3-branes constructed in Section~\ref{CGED3B}. We will
also compare this noncommutative geometry with that which arises in
the open string decoupling limit of Nappi-Witten spacetime, showing
that our foliation realizes an explicit quantization of the Lie
algebra $\mathfrak{n}_6$. Let us remark that in principle a
noncommutative deformation of $\NW_6$ can be induced via group
contraction of ${\rm SU}(1,1)\times{\rm SU}(2)$ (or of
$\real\times{\rm SU}(2)$ for $\NW_4$). However, the corresponding
star-products on $\real^{1,2}\times\real^3$~\cite{G-BLMV1,HLS-J1,HNT1}
are not well-defined under the contraction, which does not extend to
the corresponding universal enveloping algebras.

\subsection{Foliation by Noncommutative D3-Branes\label{D3Fol}}

Our starting point is the Peter-Weyl theorem which gives the
linear decomposition of the algebra of functions on the group $\mathcal{N}_6$
into its irreducible representations. With respect to the regular
action of $\mathcal{N}_6\times\overline{\mathcal{N}}_6$ given by left
and right multiplication of the group $\mathcal{N}_6$ itself, one has
\beq
{\rm C}^\infty\left(\mathcal{N}_6\right)=~{\int\limits_0^\infty
\!\!\!\!\!\!\!\!\!\!\!\!~\,\bigodot\!\!\!\!\!\!\!\!\!\!\mbf-\!\!
\mbf-}~\dd q^+~~{\int\limits_{-\infty}^\infty
\!\!\!\!\!\!\!\!\!\!\!\!~\bigodot\!\!\!\!\!\!\!\!\!\!\mbf-\!\!
\mbf-}~\dd q^-~\left[\bigl(V^{q^+,q^-}\otimes\widetilde{V}
{}^{q^+,q^-}\bigr)\oplus\bigl(\widetilde{V}
{}^{q^+,q^-}\otimes V^{q^+,q^-}\bigr)\right]
\label{PeterWeylthm}\eeq
where the $q^+=0$ contributions contain an implicit integration over all null
brane representations described in Section~\ref{Null}. The explicit
decomposition is obtained by expanding any function in the complete
system of eigenfunctions of the scalar laplacian
$\Box_6=2\,\partial_+\,\partial_-+\frac{\mu^2}4\,|\mz|^2\,
\partial_-^2+|\mdell|^2$ in the plane wave background. The spectrum of
$\Box_6$ is organized into representations of
$\mathcal{N}_6\times\overline{\mathcal{N}}_6$ for $p^+\neq0$ as
\beq
\Box_6\Phi_{\mell,\mm}^{p^+,p^-}\left(x^+,x^-,\mz,\overline{\mz}\,\right)=
E_{\mell,\mm}^{p^+,p^-}\,\Phi_{\mell,\mm}^{p^+,p^-}\left(x^+,x^-,\mz,
\overline{\mz}\,\right)
\label{Box6eigeneq}\eeq
with $\mell^\top=(\ell_1,\ell_2)\in\nat_0^2$, $\mm^\top=(m_1,m_2)\in\zed^2$ and
$\mz^\top=(z_1,z_2)\in\complex^2$. As follows from the analysis of
Section~\ref{Dynamics}, the eigenfunctions are given by
\beq
\Phi_{\mell,\mm}^{p^+,p^-}\left(x^+,x^-,\mz,\overline{\mz}\,\right)=
\e^{\ii p^+x^-+\ii
  p^-x^+}~\varphi^{p^+}_{\mell,\mm}\left(\mz,\overline{\mz}\,\right)
\label{Box6eigenfns}\eeq
where $\varphi^{p^+}_{\mell,\mm}(\mz,\overline{\mz}\,)$ are the Landau
wavefunctions for a particle moving in four dimensions, with equal
magnetic fields $\omega=\frac12\,\mu\,|p^+|$ through each transverse
plane, given by
\bea
\varphi^{p^+}_{\mell,\mm}\left(\mz,\overline{\mz}\,\right)&=&
\mbox{$\frac{\mu\,\left|p^+\right|}{2\pi}$}~
\prod_{a=1,2}\sqrt{\mbox{$\frac{\ell_a!}{(\ell_a+|m_a|)!}$}}~
\e^{\ii m_a~{\rm arg}~z_a}~\e^{-\frac{\mu\,|p^+|}4\,|z_a|^2}\nonumber\\&&
\times\,\left(\mbox{$\frac{\mu\,\left|p^+\right|}2$}\,|z_a|^2\right)^{|m_a|/2}~
L_{\ell_a}^{|m_a|}\left(\mbox{$\frac{\mu\,\left|p^+\right|}2$}\,|z_a|^2\right)
\label{Landauwavefns}\eea
with $L_\ell^m(x)$, $\ell,m\in\nat_0$ the associated Laguerre
polynomials.

The corresponding eigenvalues are given in terms of the energies of
Landau levels by
\beq
E_{\mell,\mm}^{p^+,p^-}=2p^+p^--\mu\,\left|p^+\right|\,
\sum_{a=1,2}\bigl(2\ell_a+|m_a|+1\bigr) \ .
\label{Landaulevels}\eeq
They are matched with the quadratic Casimir eigenvalues
(\ref{C6irrep}) of the representation $V^{q^+,q^-}$ by relating the
light-cone momenta as
\beq
q^+=\left|p^+\right| \ , ~~ q^-=p^--\mu\,\sum_{a=1,2}\bigl(2\ell_a+
|m_a|\bigr) \ ,
\label{momid}\eeq
while the quantum numbers of a basis state
$|n,m;q^+,q^-\rangle\otimes|\widetilde{n},\widetilde{m}\,;q^+,q^-\rangle$
of (\ref{PeterWeylthm}) are related through
\beq
\mell=\begin{pmatrix}\min\left(n,\widetilde{n}\,\right)\\
\min\left(m,\widetilde{m}\,\right)\end{pmatrix} \ , ~~
\mm=\begin{pmatrix}n-\widetilde{n}\,\\m-\widetilde{m}\,
\end{pmatrix} \ .
\label{quantnumid}\eeq
For $p^+=0$, the null brane representations of
Section~\ref{Null} correspond to the zero modes of the laplacian
$\Box_6$ and are given as a product of Bessel functions yielding the
decomposition of a plane wave whose radial momentum in the two
transverse planes is $\alpha^2$ and $\beta^2$. They will play no
explicit role in the analysis of this section. Any function $f\in{\rm
  C}^\infty(\mathcal{N}_6)$ can be thereby expanded as
\beq
f\left(x^+,x^-,\mz,\overline{\mz}\,\right)=\int\limits_{-\infty}^\infty
\frac{\dd p^+}{2\pi}~\int\limits_{-\infty}^\infty
\frac{\dd p^-}{2\pi}~\e^{\ii p^+x^-+\ii
  p^-x^+}\,\sum_{\mell\in\nat_0^2}~\sum_{\mm\in\zed^2}f_{\mell,\mm}^{p^+,p^-}~
\varphi^{p^+}_{\mell,\mm}\left(\mz,\overline{\mz}\,\right)
\label{fNW6exp}\eeq
with the appropriate integration over the $p^+=0$ zero modes again
implicitly understood throughout.

To quantize the algebra of functions on the $\mathcal{N}_6$ group
manifold, we use the Peter-Weyl decomposition of the group algebra
$\complex(\mathcal{N}_6)$ into matrix elements of irreducible
representations regarded as functions on $\mathcal{N}_6$. Thus we use
the canonical isomorphism (\ref{quantalgD3duals}) to
interpret (\ref{PeterWeylthm}) as an algebra decomposition into
the noncommutative D3-brane worldvolumes (\ref{quantalgD3}) given by
\beq
\mathcal{A}\left(\mathcal{N}_6\right)=~{\int\limits_0^\infty
\!\!\!\!\!\!\!\!\!\!\!\!~\,\bigodot\!\!\!\!\!\!\!\!\!\!\mbf-\!\!
\mbf-}~\dd q^+~~{\int\limits_{-\infty}^\infty
\!\!\!\!\!\!\!\!\!\!\!\!~\bigodot\!\!\!\!\!\!\!\!\!\!\mbf-\!\!
\mbf-}~\dd q^-~\left[{\rm End}\bigl(V^{q^+,q^-}\bigr)\oplus
{\rm End}\bigl(\widetilde{V}^{q^+,q^-}\bigr)\right] \ ,
\label{NW6quantalgdef}\eeq
so that any element $\hat f\in\mathcal{A}(\mathcal{N}_6)$ admits an
expansion
\beq
\hat f=\int\limits_{-\infty}^\infty
\frac{\dd q^+}{2\pi}~\int\limits_{-\infty}^\infty\frac{\dd q^-}
{2\pi}~\sum_{n,\widetilde{n},m,\widetilde{m}\in\nat_0}
f_{n,m;\widetilde{n},\widetilde{m}}^{q^+,q^-}~
\bigl|n,m;q^+,q^-\bigr\rangle\bigl\langle
\widetilde{n},\widetilde{m}\,;q^+,q^-\bigr|
\label{hatfNW6exp}\eeq
with $|n,m;q^+,q^-\rangle$ the Fock space basis of
$\widetilde{V}^{|q^+|,q^-}$ when $q^+<0$ and
$f_{n,m;\widetilde{n},\widetilde{m}}^{-q^+,q^-}=
\overline{f_{n,m;\widetilde{n},\widetilde{m}}^{q^+,q^-}}$.
We can view the algebra
(\ref{NW6quantalgdef}) as a deformation of the algebra of functions
${\rm C}^\infty(\mathcal{N}_6)$ by using the fact that, with the
identifications (\ref{momid}) and (\ref{quantnumid}),
the images of the Landau wavefunctions (\ref{Landauwavefns}) under the
quantization map of Section~\ref{CGED3B} coincide with rank~$1$
operators on the Fock module (\ref{Vhighestex}) as~\cite{LSZ1}
\beq
\Delta_\star\bigl(\varphi^{p^+}_{\mell,\mm}\bigr)=
\frac1{16\pi}\,\sqrt{\mbox{$\frac{\left(2\mu\,\left|p^+\right|
\right)^{2-n-\widetilde{n}-m-\widetilde{m}}}{n!\,\widetilde{n}!\,m!\,
\widetilde{m}!}$}}~\bigl|n,m;q^+,q^-\bigr\rangle\bigl\langle
\widetilde{n},\widetilde{m}\,;q^+,q^-\bigr| \ ,
\label{LandauWigner}\eeq
where the linear isomorphism $\Delta_\star:{\rm
  C}^\infty(\mathcal{C}_{x_0^+,\chi_0^{~}})\to{\rm End}(V^{q^+,q^-})$
is defined via symmetric operator ordering. The implicit dependence on
the light-cone momentum $p^-$ occurs in (\ref{LandauWigner}) through
the level sets of the class function exhibiting the conjugacy classes,
as explained in (\ref{semiclasspos}), or equivalently through
(\ref{momid}). We extend this map to the foliation of $\NW_6$ by the
conjugacy classes of the $\mathcal{N}_6$ group via the identification
\beq
f_{n,m;\widetilde{n},\widetilde{m}}^{q^+,q^-}=
\frac1{16\pi}\,\sqrt{\mbox{$\frac{\left(2\mu\,\left|p^+\right|
\right)^{2-n-\widetilde{n}-m-\widetilde{m}}}{n!\,\widetilde{n}!\,m!\,
\widetilde{m}!}$}}~f_{\mell,\mm}^{p^+,p^-} \ ,
\label{fextid}\eeq
giving an isomorphism of underlying vector spaces
$\Delta_\star:{\rm
  C}^\infty(\mathcal{N}_6)\to\mathcal{A}(\mathcal{N}_6)$ which maps the
two expansions (\ref{fNW6exp}) and (\ref{hatfNW6exp}) into one another
as
\beq
\hat f~=~\Delta_\star(f) \ .
\label{DeltastarNW6}\eeq

At the classical level, the pull-back of a function (\ref{fNW6exp}) to
a conjugacy class is obtained by restricting its light-cone
coordinates as in (\ref{E4branesloc}). In the quantum geometry, it is
achieved via an orthogonal projection which maps operators $\hat
f\in\mathcal{A}(\mathcal{N}_6)$ onto functions $\hat f_{q^+,q^-}\in{\rm
  End}(V^{q^+,q^-})$ on the quantized conjugacy classes as
\beq
\hat f_{q^+,q^-}=\hat f\,\hat P_{q^+,q^-}=\hat P_{q^+,q^-}\,\hat f \ ,
\label{orthoprojfns}\eeq
where the hermitian projector
\beq
\hat P_{q^+,q^-}=\sum_{n,m\in\nat_0}\mbox{$\frac1{\left(2\mu\,\left|q^+
\right|\right)^{n+m}\,n!\,m!}$}~\bigl|n,m;q^+,q^-\bigr\rangle
\bigl\langle n,m;q^+,q^-\bigr|
\label{projqdef}\eeq
is the identity operator on the Fock module (\ref{Vhighestex}). It
obeys
\beq
\hat P_{q^+,q^-}\,\hat P_{s^+,s^-}=\delta\left(q^+-s^+\right)\,
\delta\left(q^--s^-\right)~\hat P_{q^+,q^-}\ , ~~
\int\limits_{-\infty}^\infty\frac{\dd q^+}{2\pi}~
\int\limits_{-\infty}^\infty\frac{\dd q^-}{2\pi}~\hat P_{q^+,q^-}
=\id
\label{projqprops}\eeq
and has (uncountably) infinite rank. From (\ref{Landauwavefns}),
(\ref{momid}) and (\ref{quantnumid}) it follows that the function
corresponding to (\ref{projqdef}) is radially symmetric in each
transverse plane and can be expressed in terms of Laguerre polynomials
as
\bea
P_{q^+,q^-}\left(x^+,x^-,|z_1|,|z_2|\right)&:=&\Delta_\star^{-1}\bigl(
\hat P_{q^+,q^-}\bigr)\left(x^+,x^-,\mz,\overline{\mz}\,\right)
\nonumber\\ &=&4~\e^{-\ii q^+x^--\ii q^-x^+}~\e^{-\frac{\mu\,|q^+|}4\,
|\mz|^2}~\sum_{n=0}^\infty L_n\left(\mbox{$\frac{\mu\,\left|q^+\right|}2$}
|z_1|^2\right)\nonumber\\ &&\times~
\sum_{m=0}^\infty L_m\left(\mbox{$\frac{\mu\,\left|q^+\right|}2$}
|z_2|^2\right) \ .
\label{projqLaguerre}\eea
This construction is completely analogous to that sketched in
Section~\ref{D3Fuzzy}. In particular, the star-projectors
(\ref{projspinjfn}) are special instances of (\ref{projqLaguerre}).

\subsection{Star-Product\label{NW6Star}}

Using the construction of the previous subsection, we can define an
associative star-product of fields on $\NW_6$ in the usual way by $f\star
g:=\Delta_\star^{-1}(\hat f\,\hat g\,)$. Written in terms of the
expansion (\ref{fNW6exp}), one has
\bea
(f\star g)\left(x^+,x^-,\mz,\overline{\mz}\,\right)&=&
\int\limits_{\real^4}\dd\varrho\left(p^+,p^-,r^+,r^-\right)~
\e^{\ii(p^++r^+)x^-+\ii(p^-+r^-)x^+}\nonumber\\ &&\times\,
\sum_{\mell,\mell'\in\nat_0^2}~\sum_{\mm,\mm'\in\zed^2}
f_{\mell,\mm}^{p^+,p^-}\,
g_{\mell',\mm'}^{r^+,r^-}~\bigl(\varphi_{\mell,\mm}^{p^+}
\star\varphi_{\mell',\mm'}^{r^+}\bigr)\left(\mz,\overline{\mz}\,
\right) \ .
\label{starNW6def}\eea
{}From (\ref{highestortho}), (\ref{momid}), (\ref{quantnumid}) and
(\ref{LandauWigner}) it follows that the Landau wavefunctions obey the
star-product projector relation~\cite{LSZ1}
\beq
\varphi_{\mell,\mm}^{p^+}\star\varphi_{\mell',\mm'}^{r^+}=
\mbox{$\frac{\mu\,\left|p^+\right|}{8\pi}$}~\delta_{\widetilde{n},n'}~
\delta_{\widetilde{m},m'}~\delta\left(p^+-r^+
\right)~\delta\left(p^--r^--\mu\,\tilde m_1'-\mu\,\tilde m_2'
\right)~\varphi^{p^+}_{\tilde\mell,\tilde\mm'}
\label{Landaustarproj}\eeq
with
\beq
\tilde\mell=\begin{pmatrix}\min\left(n,\widetilde{n}^{\,\prime}\,\right)\\
\min\left(m,\widetilde{m}^{\,\prime}\,\right)\end{pmatrix} \ , ~~
\tilde\mm'=\begin{pmatrix}n-\widetilde{n}^{\,\prime}\,
\\m-\widetilde{m}^{\,\prime}\,\end{pmatrix} \ .
\label{quantnumtilde}\eeq
{}From (\ref{Landaustarproj}) it follows that the expansion coefficients
of the star-product (\ref{starNW6def}) vanish at
$p^+=0$. Thus the zero mode wavefunctions do not contribute to the
star-product, as expected from the commutativity of the null brane
worldvolume geometries. In particular, if either $f$ or $g$ is
independent of the light-cone position $x^-$, then $f\star
g=0$. Furthermore, from (\ref{Box6eigeneq}) it
follows that noncommutative scalar field theory on $\NW_6$ with this
star-product is of a similar type as the flat space noncommutative field
theories studied in~\cite{LSZ1} which can be reformulated as exactly
solvable matrix models.

We can express this star-product in a form which does not relate to
the particular basis used to expand functions on $\NW_6$. For this, we
write the star-product of Landau wavefunctions in (\ref{starNW6def})
in terms of the Moyal bi-differential operator in (\ref{MoyalD3def})
to get
\bea
(f\star g)\left(x^+,x^-,\mz,\overline{\mz}\,\right)&=&
\int\limits_{\real^4}\dd\varrho\left(p^+,p^-,r^+,r^-\right)~
\e^{\ii(p^++r^+)x^-+\ii(p^-+r^-)x^+}
\nonumber\\ &&\times~\delta\left(p^+-r^+\right)~\delta\bigl(p^--r^-+\mu\,
\mbox{$\sum\limits_{a=1,2}$}(2\ell_a'+|m_a'|-2\ell^{~}_a-|m^{~}_a|)\bigr)
\nonumber\\ &&\times~\sum_{\mell,\mell'\in\nat_0^2}~
\sum_{\mm,\mm'\in\zed^2}f_{\mell,\mm}^{p^+,p^-}~
g_{\mell',\mm'}^{r^+,r^-}~\varphi_{\mell,\mm}^{p^+}
\left(\mz,\overline{\mz}\,\right)\nonumber\\ &&
\times~\exp\left[\mbox{$\frac1{4\mu\,
\left|p^+\right|}$}\,\Bigl(\overleftarrow{\mdell}{}^{\,\top}\,
\overrightarrow{\overline{\mdell}}
-\overleftarrow{\overline{\mdell}}{}^{\,\top}\,
\overrightarrow{\mdell}\,\Bigr)\right]~
\varphi_{\mell',\mm'}^{r^+}\left(\mz,\overline{\mz}\,\right) \ .
\label{starNW6bidiff}\eea
We have exploited the occurence of the Dirac delta-functions in
(\ref{Landaustarproj}) to write the star-product (\ref{MoyalD3def})
between representations of different light-cone momenta. For us to be
able to do so, it is important to use the Poisson bi-vector
(\ref{NCparD3}) associated with the field theory limit of large
$B$-field, not the string theoretic one of Section~\ref{D3OSD}, as
(\ref{Landaustarproj}) holds {\it only} when $\theta$ is the inverse
of the magnetic field appearing in the Landau wavefunctions
(\ref{Landauwavefns}). The full stringy deformation of $\NW_6$
is much more complicated and it would not produce the nice explicit
formulas for the star-product that we derive here. It is also
important that the Moyal product is used, and not the Voros
product which leads to the same Hochschild cohomology of the
noncommutative algebra of functions.

Resolving the delta-functions in (\ref{starNW6bidiff}) and replacing
the light-cone momenta in the bi-differential operator by
bi-derivatives in the light-cone position $x^-$, we find
\bea
(f\star g)\left(x^+,x^-,\mz,\overline{\mz}\,\right)&=&
\int\limits_{\real^4}\dd\varrho\left(p^+,p^-,r^+,r^-\right)~
\int\limits_{\real^2}
\dd\varrho\left(\lambda^+,\lambda^-\right)~\e^{\ii\lambda^-(p^+-r^+)+
\ii\lambda^+(p^--r^-)}\nonumber\\ &&\times~
\sum_{\mell,\mell'\in\nat_0^2}~\sum_{\mm,\mm'\in\zed^2}
f_{\mell,\mm}^{p^+,p^-}~g_{\mell',\mm'}^{r^+,r^-}~
\e^{\ii\mu\,\lambda^+\,\sum\limits_{a=1,2}
(2\ell_a'+|m_a'|-2\ell^{~}_a-|m^{~}_a|)}\nonumber\\ &&\times~
\e^{\ii p^+x^-+\ii p^-x^+}\,\varphi_{\mell,\mm}^{p^+}
\left(\mz,\overline{\mz}\,\right)\nonumber\\ &&\times~
\exp\left[\mbox{$\frac\ii{2\mu}$}~
\mbox{$\frac{\overleftarrow{\mdell}{}^{\,\top}\,
\overrightarrow{\overline{\mdell}}
-\overleftarrow{\overline{\mdell}}{}^{\,\top}\,
\overrightarrow{\mdell}}{|\overleftarrow{\partial_-}|+
|\overrightarrow{\partial_-}|}$}\right]~\e^{\ii r^+x^-+\ii r^-x^+}\,
\varphi_{\mell',\mm'}^{r^+}\left(\mz,\overline{\mz}\,\right) \ .
\label{starNW6resolve}\eea
For the exponentials of the Landau level quantum numbers in
(\ref{starNW6resolve}), we use the eigenvalue problem
(\ref{Box6eigeneq},\ref{Landaulevels}) to replace
$\mu\,|p^+|\,\sum_{a=1,2}(2\ell_a+|m_a|)$ with the differential
operator $\Box_6-2\,\partial_+\,\partial_-$ when acting on the
eigenfunctions (\ref{Box6eigenfns}). The
light-cone momentum integrals can then be performed explicitly to
recover the original functions, leading to the desired basis
independent form of the star-product in terms of an
integro-bidifferential operator as
\bea
&&(f\star g)\left(x^+,x^-,\mz,\overline{\mz}\,\right)~=~
\int\limits_{\real^2}\dd\varrho\left(\lambda^+,\lambda^-\right)~
f\left(x^++\lambda^+,x^-+\lambda^-,\mz,\overline{\mz}\,\right)
\nonumber\\ &&~~~~~~~~~~~~~~~\times\,\exp\left[\lambda^+~
\mbox{$\frac{\frac{\mu^2}4\,|\mz|^2\,
|\overleftarrow{\partial_-}|^{2}+|\overleftarrow{\mdell}|^2}
{|\overleftarrow{\partial_-}|}$}\right]
\exp\left[\mbox{$\frac\ii{2\mu}$}~
\mbox{$\frac{\overleftarrow{\mdell}{}^{\,\top}\,
\overrightarrow{\overline{\mdell}}
-\overleftarrow{\overline{\mdell}}{}^{\,\top}\,
\overrightarrow{\mdell}}{|\overleftarrow{\partial_-}|+
|\overrightarrow{\partial_-}|}$}\right]\,
\exp\left[-\lambda^+~\mbox{$\frac{\frac{\mu^2}4\,|\mz|^2\,
|\overrightarrow{\partial_-}|^{2}+|\overrightarrow{\mdell}|^2}
{|\overrightarrow{\partial_-}|}$}\right]
\nonumber\\ &&~~~~~~~~~~~~~~~\times\,
g\left(x^+-\lambda^+,x^--\lambda^-,\mz,\overline{\mz}\,\right) \ .
\label{starNW6final}\eea
Note the particular ordering of bi-differential operators in this
expression. To recover the usual Moyal star-product on the quantized
coadjoint orbits, one uses the projector functions
(\ref{projqLaguerre}) to map functions $f\in{\rm C}^\infty(\NW_6)$ to
functions $f_{q^+,q^-}$ on conjugacy classes as
\beq
f_{q^+,q^-}=f\star P_{q^+,q^-}=P_{q^+,q^-}\star f \ .
\label{fconjclassprojq}\eeq
With this definition one has $(f\star g)_{q^+,q^-}=f_{q^+,q^-}\star
g_{q^+,q^-}$. The quantum space $\mathcal{A}(\mathcal{N}_6)$ does {\it
  not} reduce to the classical Nappi-Witten spacetime in the
commutative limit, as (\ref{starNW6final}) is not a deformation
quantization of the pointwise product of functions in ${\rm
  C}^\infty(\NW_6)$ (an analogous statement applies to the foliation of
$\real_\theta^3$ by fuzzy spheres in
Section~\ref{D3Fuzzy}~\cite{HLS-J1}).

For generic functions on $\NW_6$ the star-product
(\ref{starNW6final}) will be divergent, and so it is only well-defined
on a (relatively small) subalgebra of ${\rm
  C}^\infty(\NW_6)$. Formally, we may remove this divergence by
multiplying the right-hand side of (\ref{starNW6def}) by the rank
$\Tr\,\hat P_{q^+,q^-}$ of the projector (\ref{projqdef}). Using
zeta-function regularization on the sums over Landau levels, this rank
is given in terms of the volume of the light-cone momentum space as
$({\rm vol}~\real^2)^{-1}$, which will cancel against the divergences
coming from (\ref{starNW6final}). With this regularization understood,
a simple set of non-vanishing star-products is given as
\beq
\left(z_a\,x^-\right)\star\left(\,\overline{z}_b\,x^-\right)=
-\left(\,\overline{z}_a\,x^-\right)\star\left(z_b\,x^-\right)=
\mbox{$\frac{\ii\mu}2$}~\delta_{ab}
\label{simplestarprods}\eeq
for $a,b=1,2$. By changing to the new transverse space coordinates
$\mw:=x^-\,\mz$, we see that one of the new features of the
noncommutative geometry of the pp-wave, compared to the flat space
case, is the dependence of the noncommutativity on the light-cone
position $x^-$.

\subsection{The Dolan-Nappi Model\label{DNModel}}

The noncommutativity described by the star-product
(\ref{starNW6final}) does not depend on the light-cone time $x^+$, in
contrast to other low-energy effective field theories in
time-dependent backgrounds~\cite{DRRS1,HS1}. At a field theoretic
level, this is not entirely surprising, given that the dynamics in
Nappi-Witten spacetime is described by time independent harmonic
oscillators. At the string theoretic level, this property can be
understood by introducing the one-form
\beq
\Lambda:=-\ii\left(\mu^{-1}\,x_0^-+\mu\,x^+\right)\,\left(
\mz^\top~\dd\overline{\mz}-\overline{\mz}{}^{\,\top}~\dd\mz
\right)
\label{Lambdadef}\eeq
on the null hypersurfaces of constant $x^-=x_0^-$, and computing the
corresponding two-form gauge transformation of the $B$-field in
(\ref{NW6Bfield}) to get
\beq
B_6^\Lambda:=B^{~}_6+\dd\Lambda=-\ii\mu~\dd x^+\wedge\left(
\mz^\top~\dd\overline{\mz}-\overline{\mz}{}^{\,\top}~\dd\mz
\right)+2\ii\mu^{-1}\,x_0^-~\dd\overline{\mz}{}^{\,\top}
\wedge\dd\mz \ .
\label{B6gaugeequiv}\eeq
With $x_0^+=0$ and restricted to the four-dimensional hypersurface
defined by $\mz^\top=(z,0)$, the metric (\ref{NW4metricBrink}) and NS
potential (\ref{B6gaugeequiv}) coincide with those of the Dolan-Nappi
model~\cite{DN1} describing a (non-symmetric) D3-brane with the complete
NS-supported geometry of $\NW_4$. In~\cite{HT1} this geometry is
realized as a null Melvin twist of a {\it flat} commutative D3-brane
with twist parameter $\frac\mu2$ (in string units $\alpha'=1$),
leading to the Melvin universe with a boost. Despite the
non-vanishing null NS three-form flux of $\NW_4$, it can be argued
from this realization that the usual flat space Seiberg-Witten formulas
(\ref{SWThetagen},\ref{SWopenmetgen}) hold in this closed string
background with $F=0$ a consistent solution to the corresponding
Dirac-Born-Infeld equations of motion. The
isometry with respect to which the background is twisted corresponds
to the R-symmetry of the D3-brane worldvolume field theory, which
thereby becomes a non-local theory of dipoles whose length is
proportional to the R-charge. The open string metric
(\ref{SWopenmetgen}) correctly captures the non-local dipole-like open
string dynamics on the D3-brane.

Extrapolating this argument to $x_0^-\neq0$ and to the full
six-dimensional spacetime $\NW_6$, a straightforward calculation gives
the Seiberg-Witten bi-vector (\ref{SWThetagen}) with $F=0$ for the
background (\ref{NW6metricBrink},\ref{B6gaugeequiv}) as
\beq
\Theta^\Lambda=-\mbox{$\frac{2\ii\mu}{\mu^2+\left(x_0^-\right)^2}$}\,
\left[\mu^2~\partial_-\wedge\left(\mz^\top~\mdell-
\overline{\mz}{}^{\,\top}~\overline{\mdell}\,\right)+4x_0^-~
\mdell^\top\wedge\overline{\mdell}\,\right] \ ,
\label{ThetaLambda}\eeq
while the corresponding open string metric (\ref{SWopenmetgen}) is
given by
\beq
G_{\rm o}^\Lambda=2~\dd x^+~\dd x^-+\mbox{$\frac{\mu^2+\left(x_0^-
\right)^2}{\mu^2}$}~|\dd\mz|^2+2\ii x_0^-\,\left(\mz^\top~\dd
\overline{\mz}-\overline{\mz}{}^{\,\top}~\dd\mz\right)~\dd x^+ \ .
\label{GopenLambda}\eeq
Since (\ref{ThetaLambda}) is degenerate on the whole $\NW_6$
spacetime, it does not define a symplectic structure. Generally, if
the components of a bi-vector
$\theta:=\theta^{ij}~\partial_i\wedge\partial_j$ obey
\beq
\theta^{il}\,\partial_l\theta^{jk}+\theta^{jl}\,\partial_l\theta^{ki}+
\theta^{kl}\,\partial_l\theta^{ij}=0
\label{thetaPoissoncondn}\eeq
for all $i,j,k$, then $\theta$ defines a Poisson structure, i.e. it is
a Poisson bi-vector and (\ref{thetaPoissoncondn}) is equivalent to the Jacobi
identity for the corresponding Poisson brackets. If in addition
$\theta$ is invertible, then (\ref{thetaPoissoncondn}) is equivalent
to the required closure condition $\dd(\theta^{-1})=0$ for a
symplectic two-form. It is easily checked that (\ref{ThetaLambda})
satisfies (\ref{thetaPoissoncondn}) and hence that it defines a
Poisson bi-vector. In the flat space limit $x_0^-\to0$ of
(\ref{GopenLambda}), the corresponding quantization of $\NW_6$ is
given by the associative Kontsevich star-product~\cite{Kont1} in this
case.

The important feature of the noncommutativity parameter
(\ref{ThetaLambda}) is that it is time independent, though
non-constant. If we think of the light-cone position $x^-$ as being
dual to the Nappi-Witten generator $\J$, then the form of
(\ref{ThetaLambda}) agrees with its representation in
(\ref{nullfnsgens}) and (\ref{DeltainvD3gens}). On the other hand, the
calculation of~\cite{DN1}
provides evidence for a time-dependent Poisson bi-vector in the
original closed string background (\ref{NW6Bfield}). To make this precise,
however, one would require a detailed understanding of the worldvolume
stabilizing flux on the $\NW_6$ brane, which is difficult to determine
for non-symmetric D-branes. The noncommutativity parameter and open
string metric in the decoupling limit of D5-branes in Nappi-Witten
spacetime $\NW_6$ are thereby presently given by (\ref{ThetaLambda})
and (\ref{GopenLambda}). In particular, at the special value
$x_0^-=\mu$ and with the rescaling $\mz\to\sqrt{2/\mu\,\tau}~\mz$, the
metric (\ref{GopenLambda}) becomes that of $\CW_6$ in global
coordinates analogous to (\ref{NW4metricNW}), while the non-vanishing
Poisson brackets corresponding to (\ref{ThetaLambda}) read
\beq
\left\{z_a\,,\,\overline{z}_b\right\}&=&2\ii\mu\,\tau~\delta_{ab} \ ,
\nonumber\\ \left\{x^-\,,\,z_a\right\}&=&-\ii\mu\,z_a \ ,
\nonumber\\ \left\{x^-\,,\,\overline{z}_a\right\}&=&\ii\mu\,
\overline{z}_a
\label{Poissonspecial}\eeq
for $a,b=1,2$. The Poisson algebra thereby coincides with the
Nappi-Witten Lie algebra $\mathfrak{n}_6$ in this case and the metric
on the branes with the standard curved geometry of the pp-wave. In the
semi-classical flat space limit $\mu\to0$, the
quantization of the brackets (\ref{Poissonspecial}) thereby yields a
noncommutative worldvolume geometry on D5-branes wrapping $\NW_6$
which can be associated with a quantization of $\mathfrak{n}_6$ (or
more precisely of its dual $\mathfrak{n}_6^*$). The foliation of the
previous subsection, resulting in the star-product
(\ref{starNW6final}) on $\NW_6$, yields but one possible realization
of this deformation. Other realizations, based directly on the
Kontsevich formula, will be given elsewhere~\cite{inprep}.

\subsection*{Acknowledgments}

We thank B.~Dolan, J.~Figueroa-O'Farrill, L.~Freidel, J.~Gracia-Bond\'{\i}a,
P.-M.~Ho, G.~Landi, F.~Lizzi, R.~Myers, N.~Obers, S.~Philip,
V.~Schomerus, B.~Schroers and K.~Zarembo for helpful discussions and
correspondence. The work of S.H. was supported in part by an EPRSC
Postgraduate Studentship. The work of R.J.S. was supported in part by a
PPARC Advanced Fellowship, by PPARC Grant PPA/G/S/2002/00478, and by
the EU-RTN Network Grant MRTN-CT-2004-005104.

\end{document}